\documentclass[11pt]{article}

\usepackage{fullpage}
\usepackage[pdftex]{graphicx}
\usepackage{url, float, verbatim, bigints}
\usepackage{xcolor, dsfont, pdfpages, sectsty, MnSymbol}

\usepackage{amsthm,amsmath,MnSymbol,bbm}
\usepackage{psfrag,epsf}
\usepackage{latexsym}
\usepackage{calc, mathtools, array}
\usepackage[ragged, symbol]{footmisc}
\usepackage{grffile}
\usepackage{booktabs}
\usepackage{algorithmic}
\usepackage{algorithm}
\usepackage{caption}
\usepackage{comment}
\usepackage{color}

\DeclarePairedDelimiter\ceil{\lceil}{\rceil}

\usepackage{titlesec,pdflscape}

\titleformat{\paragraph}
{\normalfont\normalsize\bfseries}{\theparagraph}{1em}{}
\titlespacing*{\paragraph}
{0pt}{3.25ex plus 1ex minus .2ex}{1.5ex plus .2ex}

\usepackage[nonamebreak, sort&compress, round, authoryear]{natbib}

\usepackage{breakcites}
\usepackage{etoolbox}
\makeatletter
\patchcmd{\@citex}{,}{;}{}{}
\makeatother

\usepackage[inline]{enumitem}
\makeatletter
\newcommand{\inlineitem}[1][]{%
	\ifnum\enit@type=\tw@
	{\descriptionlabel{#1}}
	\hspace{\labelsep}
	\else
	\ifnum\enit@type=\z@
	\refstepcounter{\@listctr}\fi
	\quad\@itemlabel\hspace{\labelsep}
	\fi}
\makeatother

\newcommand\pN{\mathcal{N}}
\newcommand{\blind}{0}

\usepackage[left=1.1in, right = 1.1in, top = 1.1in, bottom = 1.1in]{geometry}
\renewcommand{\baselinestretch}{1.3}

\begin{document}

	\def\spacingset#1{\renewcommand{\baselinestretch}%
		{#1}\small\normalsize} \spacingset{1}

		\title{\bf }

		\if0\blind
		{
			\title{\bf Functional Variable Selection for EMG-based Control of a Robotic Hand Prosthetic}
			\author{Md Nazmul Islam\thanks{
					Department of Statistics, North Carolina State University (Email: \textit{mnislam@ncsu.edu})}
				\and
				Jonathan Stallings\thanks{
					Department of Statistics, North Carolina State University (Email: \textit{jwstalli@ncsu.edu}); corresponding author}
				\and
				Ana-Maria Staicu\thanks{
					Department of Statistics, North Carolina State University (Email: \textit{astaicu@ncsu.edu})}
				\and
				Dustin Crouch\thanks{
					Department of Mechanical, Aerospace, and Biomedical Engineering, University of Tennessee (Email: \textit{dcrouch3@utk.edu})}
				\and
				Lizhi Pan\thanks{
					Department of Biomedical Engineering, North Carolina State University (Email: \textit{lpan3@ncsu.edu})}
				\and
				He Huang \thanks{
					Department of Biomedical Engineering, North Carolina State University (Email: \textit{hhuang11@ncsu.edu})}
			}			
			\date{}
			\maketitle
		} \fi

	\baselineskip=16pt

\section*{Abstract}
State-of-the-art robotic hand prosthetics generate finger and wrist movement through pattern recognition (PR) algorithms using features of forearm electromyogram (EMG) signals, but requires extensive training and is prone to poor predictions for conditions outside the training data \citep{Scheme2010,peerdeman2011myoelectric}.  We propose a novel approach to develop a dynamic robotic limb by utilizing the recent history of EMG signals in a model that accounts for physiological features of hand movement which are ignored by PR algorithms.  We do this by viewing EMG signals as functional covariates and develop a functional linear model that quantifies the effect of the EMG signals on finger/wrist velocity through a bivariate coefficient function that is allowed to vary with current finger/wrist position.  
The model is made parsimonious and interpretable through a two-step variable selection procedure, called Sequential Adaptive Functional Empirical group LASSO (SAFE-gLASSO).  Numerical studies show excellent selection and prediction properties of SAFE-gLASSO compared to popular alternatives. For our motivating dataset, the method correctly identifies the few EMG signals that are known to be important for an able-bodied subject with negligible false positives and the model can be directly implemented in a robotic prosthetic.

\textbf{\underline{Keywords:}} Electromyography signal; Varying functional regression; Functional variable selection; Group LASSO; Post-selection predictive inference

\section{Introduction}

More than 160,000 Americans are transradial (i.e. below-elbow) amputees, henceforth TRAs, and must relearn how to perform tasks that typically require an intact hand \citep{ziegler2008estimating}.  Passive hand prostheses and related devices are useful, but cannot emulate the full functionality of an intact limb.  Multifunctional robotic prosthetics, such as the FDA-approved DEKA arm system 
\citep{linda2011development, resnik2011using}, have become very popular with recent advancements in their mechanical systems.  The available software that provides user control of the hardware, however, is often nonintuitive to operate.  For example, one approach is to have the user control the prosthetic with their foot.  Forearm muscle contractions are known to cause hand movement for an able-bodied subject, henceforth AB subject.  A TRA's residual forearm, which once caused hand movement, will still contract and these contractions exhibit measurable EMG signals that are somewhat consistent with hand movement even though they no longer use that hand.  A reliable mapping of a TRA's forearm electromyogram (EMG) signals to hand movement would then provide a more intuitive approach for prosthesis control.

\begin{figure}[h]
	\centering
	\caption{Visualization of biomechanical system for hand movement for AB subject and TRA. (A) Internal limb representation in the motor cortex. (B) Neural signal sends motor commands to forearm muscles. (C) Forearm muscles contract to perform desired movement, which, for an AB subject, results in direct hand movement through tendon connections (red lines) to the hand. (D) For a TRA, a prosthesis controller projects real time hand movements using forearm EMG information, and (E) the robotic limb performs the intended movement.}
	
	\includegraphics[width=0.80\textwidth]{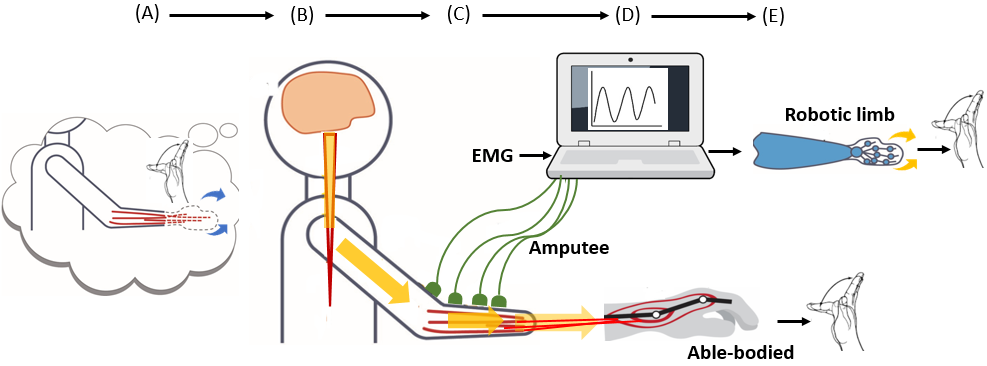} 
	\label{brain_process} 
\end{figure}

Figure \eqref{brain_process} shows the process of movement intention to movement production for both an AB subject and TRA.  For an AB subject, tendons are attached to the forearm muscles, travel through the wrist and connect to bones in the hand.  The muscle contractions transmit force through tendons to the bones, generating finger or wrist movement.  For a TRA, this physical connection among muscles, tendons, and bones no longer exists, but they may still sense movement in their phantom limb, accompanied by observable muscle contractions in their residual limb.  Their generated EMG signal data feed into a prosthesis controller that processes the data and predicts movement, which is then produced by the robotic limb.

Direct myoelectric control, a traditional approach for prosthesis control, uses EMG signals from two antagonistic muscles and assigns to each a specific type of hand movement, such as finger flexion or extension.  The EMG signal can only control that class of movement and, if activated for any reason, will produce that movement with speed proportional to the magnitude of the signal.  Direct control is limited in its ability to direct multiple degrees of freedom of hand movement.
The user must switch manually between functions (e.g. from wrist pronation/supination to hand open/close), but then the user's attempted movement may not match with the prosthesis action.  
Since it is cumbersome for controlling multiple limb functions, direct control prostheses have shown a 75\% rejection rate by users \citep{biddiss2007upper,peerdeman2011myoelectric}.  State-of-the-art prosthetics overcome the restrictions of direct myoelectric control using data-driven pattern recognition (PR) algorithms capable of synthesizing a large number of EMG signals (potentially over 100) into classes of intended movement \citep{Englehart2003, Zhou2007, Huang2008}.  

PR is more flexible than direct myoelectric control, but has its own set of limitations that prohibit effective, intuitive user control.  PR does not incorporate knowledge of human neurophysiology and biomechanics and relies heavily on the representativeness of the data used to train the algorithm in identifying EMG patterns associated with classes of hand movements.  This leads to poor prediction performance for conditions not directly observed in the training data.  It assumes that a constant, repeatable pattern of muscle contractions leads to a specific movement class and users may find it difficult to perfectly repeat this pattern. PR also significantly reduces the EMG information, using summarizing features of the EMG signals such as the mean absolute value and number of slope sign changes instead of the original signals.  The loss of important information due to data reduction as well as redundancy of predictor information leads to overfitting to the training data, as indicated by \cite{Scheme2010}.

Recently, \cite{crouch2016lumped} and \cite{crouchhuang2017} proposed and implemented an EMG-based controller using a planar link-segment dynamic model that directly incorporates features of the neurophysiological and biomechanical system.  Using EMG data from only four forearm muscles, they were able to accurately predict wrist and finger movement.  The model they use, however, is nonlinear and it is not clear how well it will translate to a TRA, since their model requires the subset of muscles to be pre-specified.  Their results do show that incorporating the neurophysiological and biomechanical properties can yield movement predictions that are more typical of biological movement than movements predicted by PR. This suggests that incorporating such properties in a data-driven approach may help improve movement predictions.

In this paper, we propose to model two degrees of freedom for hand movement, finger and wrist flexion/extension, using recent past behavior of the EMG signals.  Our statistical model uses either finger or wrist velocity as its response and allows the effect of each EMG signal on the response to vary with the current finger or wrist position.  This model retains much of the EMG information and respects known biomechanical constraints, thus allowing it to explain a broader range of hand movements using fewer EMG signals.  We fit the model using a novel algorithm for functional regression which we call Sequential Adaptive Functional Empirical group LASSO (SAFE-gLASSO) that selects only a subset of the EMG to be included in the model while simultaneously requiring that the estimated functional coefficients are smooth and interpretable.  SAFE-gLASSO pairs a generalization of the adaptive sparse-smoothness penalty from \cite{gertheiss2013variable} with an efficient, sequential variable selection and fitting approach inspired by the relaxed LASSO \citep{meinshausen2007relaxed}.  We also extend the ideas of \cite{Lei2017} to generate post-selection prediction confidence bands.

The paper is organized as follows.  Section \ref{Biomechanics} reviews the neurophysical and biomechanical relationships involved in hand movement that we address in our proposed statistical model.  Section \ref{data} describes the collection process for collecting EMG and hand movement data. Section~\ref{Proposed} details the analysis methodology starting with the proposed functional linear model and explains how it accounts for known underlying biomechanical relationships.  We then describe the variable selection and post-selection fit procedures. 
Section \ref{Data analysis} presents the analysis results from data collected from an AB subject, where the underlying truth is known.  An extensive simulation study is performed in sections \ref{Simulation} and \ref{Mathematical_simul}, showing the robustness of our method to varying assumptions.  Section \ref{Discuss} concludes the paper with a discussion of limitations and possible avenues of extension.

\section{Biomechanics of hand movement}\label{Biomechanics}

This section describes the underlying biomechanical process that generates intentional hand movement for an AB subject and highlights the unique challenges that will need to be addressed by the proposed model.  Incorporating features of this process in the statistical analysis and the robotic prosthetic software should significantly aid user control of robotic prosthetics by TRAs.  In particular, this section motivates the use of recent past behavior of the EMG signals as functional predictors and justifies the need for position-dependent effects to map the relationship between these EMG signals and velocity.  Velocity was chosen as the desired response instead of position because almost all the upper limb prostheses on the market have direction and velocity control as their outputs \citep{SchemeEnglehart2011}.

The biological process of hand movement begins with initiation of action potentials that are first conducted along motor neurons from the central nervous system.  At the neuromuscular junction, the action potential causes the release of neurotransmitters from the nerve that initiates an action potential on the muscle fiber membrane. That motor unit action potential (MUAP) conducts along the surface of the muscle fiber membrane that extends over and within the muscle fiber.  This stimulates the release of calcium ions within myofibrils and initiates cross-bridge cycling of overlapping protein chains that cause the entire muscle fiber to shorten (contract), generating mechanical energy.  The output force of an entire muscle varies depending on the proportion of fibers in the muscle that are contracted and the frequency at which they are stimulated.  The EMG signal measured from surface electrodes represents the sum of individual MUAPs and thus conveys information about the magnitude and duration of muscle contraction.

The mechanical aspect of hand movement explains how the mechanical energy generated by the muscle contractions lead to specific hand movements.  For AB subjects, tendons connect forearm muscles to bones in the hand and it is through this connection that movements are generated.  That is, a muscle contraction will pull its tendon which in turn moves the hand depending on the path of the tendon relative to the joint(s) that it crosses.  As mentioned earlier, AB subjects have an internal representation that maps muscle contractions to hand movements, and this mapping is maintained for TRAs following surgery, although it may become distorted.  Our goal is to build and implement a statistical model that decodes this internal biomechanical representation, specifically for relating finger and wrist movement to EMG signal data.  

\begin{figure}[h]
	\centering
	\caption{Top panel corresponds to finger and wrist movement: finger extension, finger flexion, wrist extension, and wrist flexion (from left to right). Bottom panel illustrates the arm postures at which data is collected during the experiment.}
	
	\includegraphics[width=0.95\textwidth]{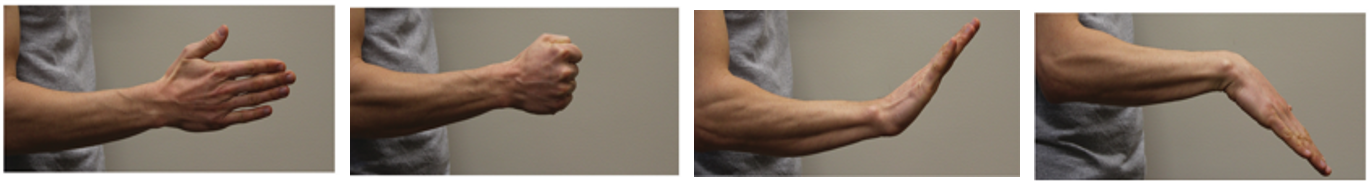} \\
	\includegraphics[width=0.95\textwidth]{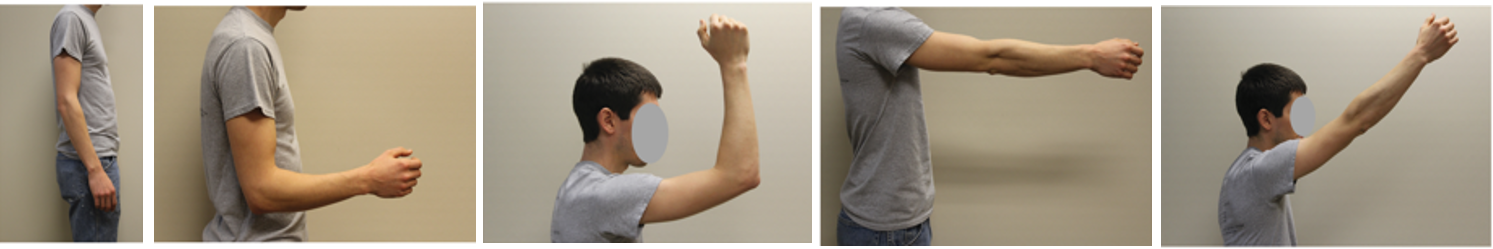}
	
	\label{type} 
\end{figure}

The top panel of Figure~\ref{type} demonstrates what is meant by finger and wrist movement.  For example, finger flexion and extension refers to the simultaneous opening and closing of all finger digits, respectively, except for the thumb.  The bottom panel of Figure~\ref{type} shows a range of arm postures considered in this paper.  A PR-based controller trained on one posture setting would likely perform poorly in one of the other postures because EMG patterns change with limb posture due to, for example, a shift in electrode locations relative to the underlying muscles \citep{Scheme2010}.  A biomechanically motivated model should behave consistently across postures because the biomechanical process does not change significantly with posture.

The known biomechanical process of AB subjects suggests that these two movement degrees of freedom can be explained by relatively few forearm muscles.  The nonlinear model of \cite{crouch2016lumped} is capable of accurate predictions with few EMG signals but does not clearly admit a variable selection procedure to decide which muscles to use.  This presents a problem for its use by a TRA for whom the intended biomechanical actions of the EMG are not directly observable and may be altered following amputation.  Thus we require a statistical model that is capable of accurate predictions using few EMG signals and the model must lend itself to a variable selection procedure.

There are many challenges and constraints that the proposed statistical model must accommodate to successfully approximate this biomechanical process.  First, the fingers and the wrist are limited in how far they can be extended/flexed. Second, muscles generate passive movement forces when stretched, which may produce movement in the absence of EMG information. For instance, if one relaxes their forearm muscles following a contraction, the tendons will return back to their resting length, generating a passive force, and the hand will return to a neutral configuration.  Therefore, we may observe movement without observing concurrent EMG activation.  While this tendon connection no longer exists for a TRA, they may still anticipate these passive forces.

\begin{figure}[h]
	\centering
	\caption{Depicted are the joint finger angle (radians) for flexion and extension movements, in black, and two normalized EMG signals, in red and green.  Left and right Y-axes correspond to finger angle and EMG signals, respectively. Event (a) demonstrates restricted movement due to a physical constraint, in this case maximal finger extension. Event (b) demonstrates movement due to passive forces, lacking active EMG signals.}
	\includegraphics[width=0.70\textwidth]{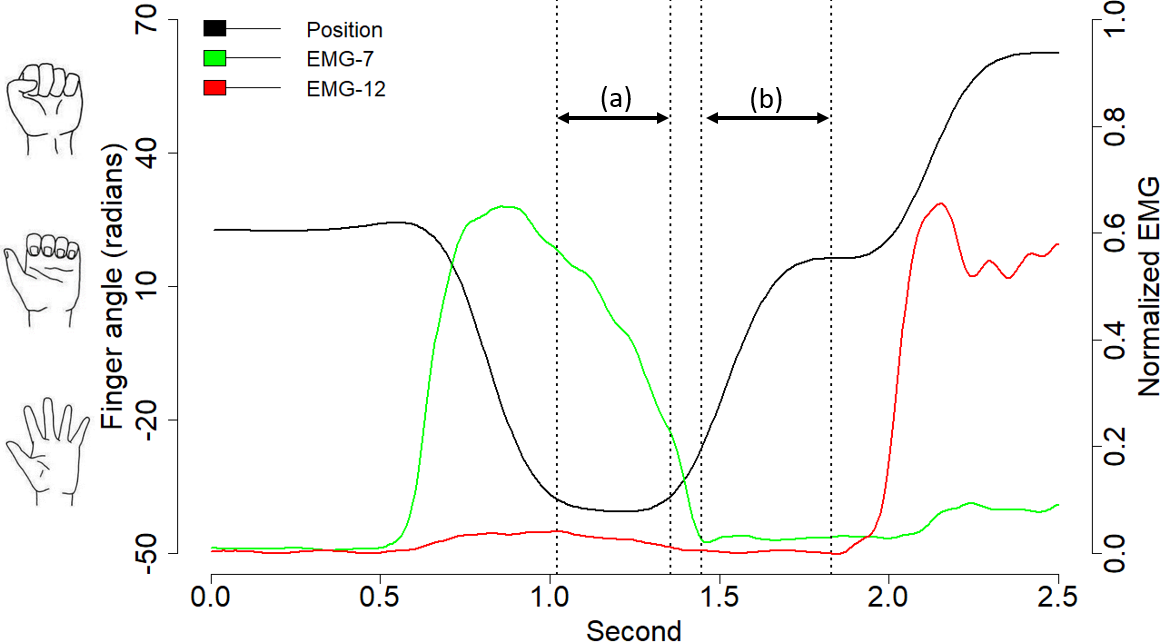}
	\label{PASSIVE} 
\end{figure}

Figure \ref{PASSIVE} illustrates some finger position data (in radians, shown in black) taken from an AB subject for a short time-window and overlays two concurrent EMG signals, labeled EMG 7 (green) and EMG 12 (red), known to contribute to finger movement.  Figure \ref{PASSIVE}(a) is an event window with an active EMG 7 signal but no movement; the result of a physical constraint.  Figure \ref{PASSIVE}(b) demonstrates finger movement in the absence of active, concurrent EMG signals, due to passive forces.  This figure demonstrates the opportunity for past EMG behavior to predict passive movement, as seen by the declining green line preceding Figure \ref{PASSIVE}(b).

\section{Data collection and processing}\label{data}

This section and the analysis in section~\ref{Data analysis} focuses on the collection and analysis of data from an AB subject.  The process of calibrating a robotic arm is subject-specific and we are not interested in estimating a population model.  The data collection procedure described here and the analysis methodology described in Section~\ref{Proposed} could be modified to perform inference across multiple subjects.  A data collection strategy is given at the end of Section~\ref{subsec:datacollect} for TRAs consistent with our analysis approach, requiring measured hand movements.

Surface EMG electrodes (Biometrics, Newport, UK) were placed at different locations of the forearm.  The subject was then asked to perform specific types of hand movement and the EMG signals were recorded continuously at 960 Hz.  Electrodes were placed over 16 specific forearm muscles considered to be potentially important muscles for generating hand movements of interest.  Raw EMG data were high-pass filtered at 40 Hz, rectified, and low-pass filtered at 6 Hz using a 4th order Butterworth zero-phase filter \citep{crouch2016lumped}. Prior to data collection, the subject was asked to perform movements exhibiting maximal muscle contraction, allowing normalization of the subject's EMG signals to be between 0 (no contraction) and 1 (maximal contraction).

Movement data were collected by reflective markers placed on 9 anatomical locations on the forearm, wrist, and hand. Three-dimensional marker positions were recorded at 120 Hz using an infrared motion capture system (Vicon Motion Systems Ltd., UK), and processed (filtered) at 6 Hz using a 4th order Butterworth filter \citep{butterworth1930theory}. The joint angles (in radians), which we call positions, were then calculated from filtered marker data through a musculoskeletal model \citep{holzbaur2005model} in OpenSim \citep{delp2007opensim}. EMG and joint angle data were collected synchronously for a given period of time during which a subject was asked to perform a set of finger and/or wrist movement.  Figure~\ref{PASSIVE} provides a snapshot of this data while a subject is performing finger flexion and extension with consistent movements in a fixed posture.

\subsection{Data collection procedure}\label{subsec:datacollect}
\label{sec:data collection}

The subject was asked to perform basic finger and wrist movements in each arm posture as shown in the bottom panel of Figure \ref{type}.  For a given posture, the subject was asked to perform single degree-of-freedom movements for their fingers/wrist following either a consistent or random pattern.  For example, the subject performed consistent finger movement that continuously alternated between extending (opening) and flexing (closing) their fingers, while keeping their wrist in a neutral position.  Such a protocol has been followed by others; see for example \cite{kawashima2016psychophysical}. The subject was also asked to perform a series of random movements, which was chosen solely by the subject, providing a challenging benchmark to determine how robust our fitted models are to different movements.

For a TRA, there would be no movement data for the corresponding hand.  Instead, movement data can be collected from the other, intact limb while the TRA performs (attempts) mirrored movements with both limbs; see \cite{SchemeEnglehart2011}.

\subsection{Post-collection data processing for analysis}\label{sec:process}
To accommodate the biomechanical constraints described in section~\ref{Biomechanics}, we utilize recent past behavior of EMG signals, as well as current finger/wrist position, to predict finger/wrist velocity. For example, a concurrent EMG value of 0.50 leads to different movements depending on the historical EMG trend.   We briefly describe here the data processing that was done that allowed us to fit the model described in the next section.

The velocity values were estimated from a penalized smoother of the entire series of recorded position data across time using the \texttt{R} package {\tt fda}.  In particular, we used a second-order smooth regularization penalty to control the goodness of fit and smoothness of the fitted curve, where the smoothing parameter is selected by cross-validation. We then generated a velocity estimate for each observed position data point.  For each velocity data point, we extracted the previous EMG observations that ended with the concurrent EMG value to the velocity value.  In this application, we chose a past time window of roughly 1/3 seconds.  The value was chosen based on observed passive force movement (see Figure~\ref{PASSIVE}).  This was done for all EMG signals, so each velocity estimate had associated with it a position value and  equally-spaced past measurements across 1/3 seconds for all $16$ EMG signals collected. A visualization of this data restructuring may be found in the Supplementary Materials, Section \ref{AppendixChp3}.  We propose to model the current velocity of the finger/wrist movement as a function of the recent history of the EMG signals which is to be discussed next.

\section{Variable selection and inferential framework}\label{Proposed} 

Denote the observed data by $\left[y_i, z_i, \{X_{k,i}(s_r); r = 1, \ldots, R\}, k = 1, \cdots, K, i=1,\dots,N\right]$  where $i$ indexes the instance at which data are collected, $y_i$ is the scalar response, $z_i$ is the continuous scalar covariate, $k$ indexes the functional predictors, and $X_{k,i}(s_r)$ is the $k$th functional predictor observed at point $s_r$ such that $s_r \in \mathcal{S}$.   It is assumed that  $z_i \in \mathcal{Z},$ and both $\mathcal{S}$ and $\mathcal{Z}$ are closed compact sets. In our application,
$y_i$ is the current velocity, $z_i$ is the current position,  and $X_{k,i}(\cdot)$ is the recent history of the $k$th EMG signal and $\mathcal{S}$ is a window of time that depicts the ``recent" past. 

We consider a functional linear model with varying smooth effects:
\begin{equation}
\begin{aligned}
\label{FLM}
E[y_{i} | X_{1,i}, \cdots, X_{K,i}] =  \alpha + \sum^{K}_{k=1}  \int_{\mathcal{S}} X_{k,i}(s) \gamma_{k}(s, z_{i}) ds,
\end{aligned}
\end{equation}
where $\alpha$ is an intercept and $\gamma_{k}(\cdot, \cdot)$ is an unknown bivariate function defined on $\mathcal{S} \times \mathcal{Z}$ that quantifies the effect of the $k$th functional predictor on the mean response of $y_i$ conditional on $z_i$. Model \eqref{FLM} is a direct extension of the functional linear model (FLM) described in \cite{ramsay2005springer, cardot2003spline,   james2009functional,goldsmith2011penalized,ferraty2012presmoothing,mclean2014functional} who assume nonvarying functional coefficients. In our application accounting for positions in the statistical model is crucial for modeling of passive forces since the effect of the EMG signals on finger/wrist movement heavily depends on the current state of the hand as argued in section~\ref{Biomechanics}.  Varying coefficients for scalar covariates characterize the effect of a covariate on response through another covariate and have been studied extensively in nonparametric and semiparametric literature; see, for example, \cite{fan2008statistical,maity2012partially,fan2014nonparametric, davenport2015parametrically,bandyopadhyay2017asymptotic}. Recently, the varying-coefficient model has been extended to analyze functional data in \cite{cardot2008varying, wu2010varying,davenport2013semiparametric}. 

Our primary objective is to select the functional predictors whose effects on the mean response as described by \eqref{FLM} is non-zero. For our application this would allow us to select the most important forearm  muscles in explaining finger and wrist movements. Let $\mathcal{K}\subseteq \{1, \cdots, K\}$ denote the true index set where $\gamma_k (\cdot,\cdot) \neq 0,$ and let $\mathcal{B} = \{\gamma_{k}(\cdot,\cdot); k \in \mathcal{K} \}$ denote the set of true effects.  Established fitting approaches focus on the estimation of the smooth coefficients rather than variable selection. Variable selection in scalar-on-function regression with invariant smooth coefficients, $\gamma_k(\cdot)$, has been discussed in \cite{matsui2011variable, gertheiss2013variable,  fan2015functional, pannu2017robust}.  In this paper, we extend these approaches to the varying functional linear model \eqref{FLM} that is motivated by our data application.

\subsection{Approximation to linear model} \label{model_strategy1}

Following the procedure described in \cite{wood2006low} and \cite{eilers2003multivariate}, we approximate  $\gamma_{k}(\cdot, \cdot)$ using a tensor product of two univariate basis functions. Let $\{\omega_{l} (s)\}^{L}_{l}$ and $\{\tau_{m} (s)\}^{M}_{m}$ be two truncated univariate bases defined in $\mathcal{S}$ and $\mathcal{Z},$ respectively; and let $\gamma_{k}(s, z) \approx \sum_{l=1}^{L} \sum_{m=1}^{M} \omega_{l} (s) \tau_{m} (z) \beta_{klm}$, where $\beta_{klm}$'s are the unknown basis coefficients associated with the $k$th functional predictor. For simplicity of exposition, we use the same bases for all $\gamma_{k}(\cdot, \cdot); k = 1, \ldots, K.$ Generalization to different bases for different predictors is straightforward. 

It is more convenient to rewrite the tensor product in matrix form $\gamma_{k}(s,z) \approx \boldsymbol {\omega}^{T}(s) \boldsymbol {B_k} \boldsymbol { \tau} (z)$ where $\boldsymbol {\omega}(s) = \big( \omega_{1}(s), \cdots, \omega_{L}(s) \big)^{T}$, $\boldsymbol {\tau}(z) = \big( \tau_{1}(z), \cdots, \tau_{M}(z) \big)^{T}$, and $\boldsymbol {B_k} = (\beta_{klm})_{l,m}$ is the $L \times M$ matrix of basis coefficients, which is our target for inference.  
The generic summand in \eqref{FLM} can be approximated by Riemann sum approximation as  
\begin{equation}
\begin{aligned}
\int_{\mathcal{S}} X_{k,i}(s) \gamma_{k}(s,z_i)ds
\approx \sum^{L}_{l=1} \big\{  \sum_{r=1}^R \Delta_r  X_{k,i}(s_{r}) \omega_{l}(s_r) \big\} \boldsymbol {B_{lk}} \boldsymbol { \tau} (z_i)  = \boldsymbol {X}_{ki\omega}^T\boldsymbol {B_k} \boldsymbol { \tau} (z_i),
\end{aligned}
\end{equation}
where $\boldsymbol {X}_{ki\omega}^T = \big( \sum_{r=1}^R  \Delta_r   X_{k,i}(s_{r}) \omega_{1}(s_r), \ldots,  \sum_{r=1}^R  \Delta_r   X_{k,i}(s_{r}) \omega_{L}(s_r)\big)$ is a $1 \times L$ row vector and $\Delta_r = s_r - s_{r - 1},$ . By an abuse of notation, define the $1 \times LM$ vector ${\boldsymbol{X}}_{ki\omega\tau}^T= \boldsymbol {X}^{T}_{ki\omega} \otimes \boldsymbol {\tau}(z_i),$ where $\otimes$ indicates the Kronecker product. For notational simplicity, we omit indices $\omega$ and $\tau$ in the subscript of  ${\boldsymbol{X}}_{ki\omega\tau}^{T}$ and denote it by $\widetilde{\boldsymbol{X}}_{ki}^T$. Vectorize $\boldsymbol B_k$ by appropriately stacking its $M$ columns vertically and denote the $LM \times 1$ vector of basis coefficients $\beta_{klm}$'s  by $\boldsymbol {\beta_k}$.  
This yields an approximating linear model to \eqref{FLM} with unknown regression coefficients $\boldsymbol {\beta}_{k}$'s 
\begin{equation}
\label{approx_FLM}
E[y_{i}|\widetilde{\boldsymbol {X}}_{1i}, \ldots, \widetilde{\boldsymbol {X}}_{Ki}] \approx  \alpha + \sum_{k=1}^{K} \widetilde{\boldsymbol {X}}_{ki}^T \boldsymbol {\beta}_{k}.
\end{equation}

\subsection{Penalized criterion for variable selection} \label{model_strategy2}	

We allow $L$ and $M$ to be sufficiently large to capture the complexity of the regression surfaces and penalize the degree of smoothness. We adopt the penalized least squares approach for estimation that simultaneously induces group sparsity and controls smoothness of the corresponding regression surfaces.

Let $|| \gamma_{k}||^{2} = \int_{\mathcal{S}} \int_{\mathcal{Z}} \{\gamma_{k}(s, z) \}^{2} dz ds$ be the $L^{2}$ norm of $\gamma_k$, and let $\gamma''_{k,s}$ and $\gamma''_{k,z}$ be the second partial derivatives of $\gamma_{k}$ with respect to $s$ and $z,$ respectively. Prior to analysis, we center the response $y_i$ and functional predictors $X_{k,i}$'s, and remove $\alpha$ from the model. By an abuse of notation, we use the same notation for the centered model. The estimates of the $\gamma_{k}$'s are then selected to be the minimizers of the following penalized criterion 
\begin{equation}
\label{penalized_estimation}
\sum^{N}_{i=1} \left\{ y_{i} -  \sum^{K}_{k=1} \int_{\mathcal{S}} X_{k,i}(s) \gamma_{k}(s, z_{i}) ds \right\}^{2} + \lambda \sum^{K}_{k=1} 
\left(|| \gamma_{k}||^{2} + \phi_{1} || \gamma''_{k,s} ||^2 + \phi_{2} || \gamma''_{k,z} ||^2\right)^{1/2}; 
\end{equation}	
where $\lambda > 0$ controls the model sparsity, and $\phi_{1}, \phi_{2} > 0$ control the smoothness of $\gamma_{k}(\cdot, \cdot)$ in the $s$ and $z$ dimension, respectively. Let $\phi = \{\phi_{1}, \phi_{2}\}$ and $P_{\phi}(\gamma_{k})=( || \gamma_{k}||^{2} + \phi_{1} || \gamma''_{k,s} ||^2 + \phi_{2}|| \gamma''_{k,z} ||^2)^{1/2}.$ The proposed penalty function $P_{\phi}(\gamma_{k})$ was first introduced by  \cite{meier2009high} for variable selection in high dimensional additive model. As $\lambda$ increases, more penalty weight is placed on both the magnitude and smoothness of the surface which results in $||\gamma_{k}||^2 = 0$, excluding the corresponding functional predictor from the model. Although $\gamma_{k}(\cdot, \cdot)$ is assumed to be a smooth function, the fitted regression surfaces may be rough if the penalty on the second order derivatives $|| \gamma''_{k,s} ||^2$ and $||\gamma''_{k,z} ||^2$ are not included. The extent of smoothing of the non-zero regression surfaces is controlled by the vector of smoothing parameters $\phi$. For example, as $\phi$ increases, the model more heavily penalizes the departure from linearity, producing linear effects in both directions. The proposed penalty  $P_{\phi}(\gamma_{k})$  ensures smoothness and sparseness in two dimensions; it can be easily extended for multiple dimensions without loss of generality.

We now turn our attention to efficient calculation of $P_\phi(\gamma_k)$ for a given function $\gamma_k$. For simplicity, we further assume that the bases $\{\omega_{l}(\cdot) \}^{L}_{l=1}$ and $\{\tau_{m}(\cdot) \}^{M}_{m=1}$ are orthogonal B-splines. Using the orthogonal property of the bases, define by $\boldsymbol {\Omega}_{s}= \boldsymbol {I}_{L}$  the $L \times L$ identity matrix with the $(l,l')$th element as $\int_{\mathcal{S}}  \omega_{l} (s)  \omega_{l'}(s) ds = 1$ if $l = l'$ and 0 otherwise, and  define $\boldsymbol {\Omega}_{z}= \boldsymbol {I}_{M}$ similarly in terms of the $\tau_m(z)$'s. General bases will not have $\boldsymbol {\Omega}_{s}$  and $\boldsymbol {\Omega}_{z}$ equal to the identity.  Let $\omega_l'' (\cdot)$ and $\tau_m'' (\cdot)$ be the second derivatives of $\omega_l(\cdot)$ and $\tau_m(\cdot),$ respectively; let $\boldsymbol{\omega}''(s)$ and $\boldsymbol{ \tau}''(z)$ be the vectors of these $L$ and $M$ functions. Define ${\boldsymbol \Omega}_{ss}=\int \boldsymbol{\omega}''(s)\boldsymbol{\omega}''(s)^T ds$ and ${\boldsymbol \Omega}_{zz}=\int \boldsymbol{\tau}''(z)\boldsymbol{\tau}''(z)^T dz.$

It follows that $$||\gamma_{k}||^{2} = \sum_{l,l'} \sum_{m,m'} \big\{ \int_{\mathcal{S}}  \omega_{l} (s)  \omega_{l'}(s) ds \big\} \big\{\int_{\mathcal{Z}}  \tau_{m} (z)  \tau_{m'}(z) dz \big\} \beta_{klm}^{2}=\boldsymbol{\beta}_{k}^T \boldsymbol {\beta}_{k}\ ,\ $$ where the last equality follows from the orthogonality of the bases. Similarly, we obtain  $|| \gamma''_{k,s} ||^{2} = \boldsymbol {\beta}_{k}^{T} \left(\boldsymbol {\Omega}_{ss} \otimes  \boldsymbol {I}_{M}\right) \boldsymbol {\beta}_{k}$ and $|| \gamma''_{k,z} ||^{2} = \boldsymbol {\beta}_{k}^{T}  \left( \boldsymbol {I}_{L} \otimes  \boldsymbol {\Omega}_{zz} \right) \boldsymbol {\beta}_{k}.$
The penalty $P_{\phi} (\gamma_{k})$ may now be expressed as a sparse-smooth regularization \citep{meier2008group,meier2009grplasso}
$$P_{\phi} (\gamma_{k}) =  \left( \boldsymbol {\beta}_{k}^{T} \boldsymbol {Q}_{\phi} \boldsymbol {\beta}_{k} \right) ^{1/2},$$
where $\boldsymbol {Q}_{\phi} =  \boldsymbol {I}_{LM} + \phi_{1}(\boldsymbol {\Omega}_{ss} \otimes  \boldsymbol {I}_{M}) +  \phi_{2}(\boldsymbol {I}_{L} \otimes  \boldsymbol {\Omega}_{zz})$ is a $LM \times LM$  symmetric positive-definite matrix. Cholesky decomposition on $\boldsymbol {Q}_{\phi}$ gives $\boldsymbol {Q}_{\phi} = \boldsymbol {R}_{\phi} \boldsymbol {R}^{T}_{\phi},$ where $\boldsymbol {R}_{\phi}$ is a lower triangular non-singular matrix. Using \eqref{approx_FLM} and $\boldsymbol {R}_{\phi},$ reparametrize the model coefficients $\boldsymbol {\beta}_{k}$ as $\boldsymbol {\widetilde \beta}_{k} = \boldsymbol {R}^{T}_{\phi} \boldsymbol {\beta}_{k}$ and transform each $\widetilde{\boldsymbol {X}}_{ki}$ to $\boldsymbol {W}_{ki} = \boldsymbol {R}^{-1}_{\phi} \widetilde{\boldsymbol {X}}_{ki}.$ It follows that \eqref{penalized_estimation} can be written equivalently as 
\begin{equation}
\label{final_penazation}
\sum^{N}_{i=1} ( y_{i} - \sum_{k=1}^{K} \boldsymbol {W}^{T}_{ki} \boldsymbol {\widetilde \beta}_{k} )^{2} + \lambda \sum^{K}_{k=1} || \boldsymbol {\widetilde \beta}_{k} ||,  
\end{equation}

which is similar to the group LASSO penalized criterion described in \cite{yuan2006model,yang2013gglasso,yang2015fast}.

We use the \textit{groupwise-majorization-descent} (GMD) algorithm described by \cite{yang2013gglasso} to solve~\eqref{final_penazation}, which is implemented by the \texttt{R} package {\tt gglasso} \citep{yang2013gglasso}.  A benefit of the GMD algorithm is that it does not require group-wise orthonormality and is applicable for a general design matrix. In addition, for given values of the tuning parameters $\lambda$ and $\phi,$ the minimizer of \eqref{final_penazation} with respect to $\boldsymbol{{\widetilde \beta}}_{k} $ has a close-form solution; denote the solution by $\boldsymbol{ \widehat  \beta}_{k}$ by omitting its dependence on $\lambda$ and $\phi$.  

It is straightforward to restructure the estimated basis coefficients in $\boldsymbol{ \widehat  \beta}_{k}$ to give $\boldsymbol{ \widehat {B}}_{k}$, and we have $$\widehat \gamma_{k}(s,z) \approx \boldsymbol {\omega}(s)^T \boldsymbol {\widehat B_{k}} \boldsymbol {\tau}(z). $$  Let $\mathcal{K}_{\lambda,\phi} $ be the estimated index set after solving $\eqref{final_penazation}$ under $\lambda$ and $\phi;$  and denote the set of their corresponding estimates by $\mathcal{B}_{\lambda,\phi} = \{\widehat \gamma_{k}(\cdot,\cdot); k \in \mathcal{K}_{\lambda,\phi} \}$. In reality, the sparseness and smoothing parameters, $\lambda$ and $\phi$ are unknown and need to be chosen empirically; for instance by cross-validation. We discuss this aspect in more detail in Section  \ref{tuning}.

The penalized criterion \eqref{penalized_estimation} does not allow for different shrinkage and smoothness for different functional predictors, and assumes equal weight. This may inflate the number of false positives in variable selection and to mitigate this problem, adaptive estimation  is recommended for variable selection in high-dimensional data \citep{meier2009grplasso,tutz2010feature,gertheiss2013variable}. We too adopt this approach and discuss it next.

\subsection{Adaptive penalized criterion} \label{model_strategy2.1}
We generalize the criterion \eqref{penalized_estimation} by adaptive penalized criterion. Specifically, we use initial weights $w = \{f_{k}, g_{k}, h_{k}\}^{K}_{k=1}$ to introduce prior information on the relative importance and smoothness of functional predictors, where $f_k$ is related to the sparse penalty factor and $g_k$ and $h_k$ are related to the smooth penalty factors in the $s$ and $z$ dimension, respectively. Define the adaptive penalty function by 
\begin{equation}
\label{penalized_estimation_ADAPT}
P_{\phi} (\gamma_{k}) = \left[ \boldsymbol {\beta}_{k}^{T} \left\{ f_{k} \boldsymbol {I}_{LM} +   g_{k} \phi_{1} (\boldsymbol {\Omega}_{ss} \otimes  \boldsymbol {I}_{M}) +    h_{k} \phi_{2}(\boldsymbol {I}_{L} \otimes  \boldsymbol {\Omega}_{zz}) \right  \} \boldsymbol {\beta}_{k} \right]^{1/2}
\end{equation}

where  the weights $w$ are strictly positive and calculated based on the parameter estimates $\{\widetilde \gamma_{k0}(\cdot, \cdot)\}^{K}_{k=1}$ associated with a functional additive model \citep{mclean2014functional, scheipl2015functional} fit without a sparseness penalty. The subscript 0 is used to distinguish the initial estimates from the model estimates of \eqref{final_penazation}. Define the weights by $f_{k} =  1 / ||\widetilde \gamma_{k0}||^{d}, g_{k} =  1 / ||\widetilde \gamma''_{k,s0}||^{d},$ and $h_{k} =  1 / ||\widetilde \gamma''_{k,z0}||^{d},$ $d > 0;$ for details we refer to \cite{zou2006adaptive, gertheiss2013variable,guo2015improved,ciuperca2016adaptive,ivanoff2016adaptive}. Alternatively, one can adopt a semi-adaptive approach using only $f_k$ while keeping equal weights for smooth penalty factors such as $g_k = 1$ and $h_k = 1$ for all $k.$ Like \cite{yang2015fast}, the weights based on the number of parameters within each group can also be adopted alternatively. Here, as $||\widetilde \gamma_{k0}||$ decreases to $0$, $f_k$ increases to $\infty$ which yields sparser solution for $\gamma_{k}$ for a given $\lambda.$ Similarly, as  $g_k$ and $h_k$ increase to $\infty$, the regression surface yields linear pattern in both directions for a given $\phi.$ We denote the minimizer of the criteria with penalty \eqref{penalized_estimation_ADAPT} by  $\mathcal{B}_{\lambda, \phi}$ associated with $\mathcal{K}_{\lambda, \phi}$ for a given $\lambda$ and $\phi.$

\subsection{Selection of the tuning parameters, $\lambda$ and $\phi$} \label{tuning}

A widely used method to select tuning parameters is $K$-fold cross-validation (CV). We use 5-fold block CV which has been found appealing for data with temporal correlations (see \cite{roberts2017cross}).

Specifically, we partition the data into 5 equally-sized sections; let $n = 1, \dots, 5$ index the folds and denote by $y_{i_{n,m}}$ be the $m$th ordered response of the $n$th fold corresponding to the instance $i_{n,m}$, and $N_{n}$ be the total number of observations in the $n$th subsample. We aim for the $N_n$ to be as equal as possible across $n$.  The test set is formed by one of the five folds; the remaining four folds form the training set. 

For fixed values of $\lambda,$ $\phi,$ and $n,$ the model parameters are estimated based on the training set (e.g., all data with the $n$th fold removed), and the prediction accuracy is evaluated based on the performance of the test set (e.g., the $n$th fold); i.e., $PE_{out,n} = \sqrt{ \sum_{m=1}^{N_{n}}   (y_{i_{n,m}}  - \widehat y_{i_{n,m}} )^2 / N_{n} },$ where $\widehat y_{i_{n,m}}$ is the predicted value for $y_{i_{n,m}}.$  

For each value of the tuning parameters, we compute the average $PE_{out,n}$'s and select the values corresponding to the minimum prediction error. Furthermore, CV with the one standard error rule can be also adopted; the main idea is to select the simplest model whose numerical performance is comparable with the optimal model that is chosen by minimum prediction error \citep{friedman2001elements, krstajic2014cross,yang2015fast}.

\subsection{Sequential Adaptive Functional Empirical (SAFE) Selection}
\label{model_strategy3}

Most variable selection algorithms can not guarantee the exclusion of noise variables \citep{meinshausen2009p}. In particular, the group LASSO approach has been shown to select a larger number of groups than necessary \citep{meier2008group,gertheiss2013variable}; this may happen due to the uncertainty in selecting the optimum combination of tuning parameters. As a remedy, we propose to perform the selection step twice where the second stage considers only the covariates $k \in \mathcal{K}^{(1)}_{\lambda^{\dagger}, \phi^{\dagger}}$; where the superscript (1) denotes the first stage, and $\lambda^{\dagger}$ and $\phi^{\dagger}$ are the optimal tuning parameters determined by CV in the first stage. The adaptive weights for the second stage are then calculated based on $\mathcal{B}^{(1)}_{\lambda^{\dagger}, \phi^{\dagger}}.$

The proposed idea is motivated by the relaxed LASSO \citep{meinshausen2007relaxed} approach, however, the fitting procedure is different in terms of calculating the adaptive weights for the second stage. In contrast to \cite{meinshausen2007relaxed}, our solution path of the second stage is driven by both $\mathcal{B}^{(1)}_{\lambda^{\dagger}, \phi^{\dagger}}$ and the selection of second-stage tuning parameters  $\lambda$ and $\phi.$ In \cite{meinshausen2007relaxed}, the estimates of the second stage are not driven by weights but rather based on the exhaustive search of the sparse tuning parameter. This idea was also promoted by \cite{wei2010consistent} and \cite{guo2015improved} who adopted a two-stage variable selection strategy for the scalar-on-scalar regression.

With weights re-calculated, we minimize the objective function  \eqref{penalized_estimation_ADAPT} using only those $\mathcal{K}^{(1)}_{\lambda^{\dagger}, \phi^{\dagger}}$ with respect to $\lambda$ and $\phi.$ Let $\mathcal{K}^{(2)}_{\lambda^{\ast}, \phi^{\ast}}$ be the estimated index set resulting from the second stage of the regularization fit under the optimal combination of tuning parameters $\lambda^{\ast}$ and $\phi^{\ast}$ found by CV.

\subsection{Post-selection inference and prediction} \label{model_strategy4}

Shrinkage penalties cause the estimates of the non-zero coefficients to be biased towards zero \citep{zhao2017defense,friedman2001elements}. In the similar spirit to \cite{leeb2015various, zhao2017defense}, once $\mathcal{K}^{(2)}_{\lambda^{\ast}, \phi^{\ast}}$ is determined, we refit the model using the second order smooth regularization to reduce the prediction bias. Specifically, we solve the following penalized criterion  

\begin{equation}
\label{approx_FLM2}
\sum^{N}_{i=1} \left\{ y_{i} -  \sum_{k \in \mathcal{K}^{(2)}_{\lambda^{\ast},\phi^{\ast}}} \int_{\mathcal{S}} X_{k,i}(s) \gamma_{k}(s, z_{i}) ds \right\}^{2} + \sum_{ k \in \mathcal{K}^{(2)}_{\lambda^{\ast}, \phi^{\ast}}}  P_{\phi}(\gamma_{k});
\end{equation}
where $\phi = \{\phi_{1}, \phi_{2} \}$ are the smoothing parameters that control the curvature of the fit and $P_{\phi}(\gamma_{k}) =  \phi_{1} || \gamma''_{k,s} ||^2 + \phi_{2} || \gamma''_{k,z} ||^2$ is the smoothing penalty function as described in \cite{wood2006low} and \cite{eilers2003multivariate}. We choose the optimal tuning parameters using CV as discussed in Section \ref{tuning}.  

The model \eqref{approx_FLM2} is conditional on the event that the same data is being used twice: once in the variable selection stage and again in the post-selection fit on the selected subset. \cite{wu2009genome,  zhao2017defense} argued that the use of standard prediction (confidence) intervals or \textit{p}-values without adjustment are invalid as they neglect the complex selection procedure to define the reduced model in the first place. Thus an adjustment is required to construct valid prediction intervals or obtain valid \textit{p}-values \citep{tibshirani2014significance, fithian2014optimal, lee2016exact}. To the best of our knowledge, post-selection inference for the group LASSO in the context of functional data is still an open problem. Alternatively, inference after selection based on data splitting  is commonly adopted for high-dimensional data; see \cite{wasserman2009high, meinshausen2009p}. We extend these ideas to the case of functional covariate and focus on post-selection predictive inference based on data splitting and construct split conformal prediction bands following \cite{Lei2017}; for details of the algorithm we refer to the Supplementary Material, Section \ref{algorithm}.

\section{Data Application: EMG selection for finger/wrist movement} 
\label{Data analysis}

In this section, we present the variable selection and prediction results for our data collected from an AB subject across multiple postures and movement patterns as described in Section~\ref{data}.  Recall there are 16 EMG signals: 14 coming from different forearm muscles that could potentially contribute to finger or wrist movements and 2 that are randomly generated noise.  The ideal selection scheme would pick one EMG signal for extension and another EMG for flexion for both finger and wrist movements, but we do not enforce this restriction in the estimation.  This selection should be consistent across postures and movement patterns.  In addition to identifying these EMG signals, we are also interested in a model with clear interpretabililty of the regression surfaces and high predictive ability.

Here we consider fitting the procedure described in Section \ref{Proposed} to our data application. Let $i \in \{1,\dots,N\}$ denote the current time point and $y_i$ be the velocity at time point $i$. Let $z_i$ be the finger or wrist position at time point $i$. Define  $\mathcal{S}=[-\delta,0]$ for some integer $\delta >0$ as the time window for the recent past EMG signal relative to the current time $i,$ so the $\delta+1$ $s_r$'s make up the sequence $\{-\delta,-\delta+1,\ldots,0\}$. For $i\geq\delta+1$, let $\left\{ X_{k,-\delta}, X_{k,-\delta+1}, \ldots, X_{k,0} \right\}$ be the realizations of the $k$th EMG signal at the original time points $\{i-\delta, i-\delta+1, \ldots, i\}.$ Recall from Section \ref{sec:process} that we proposed and justified a recent past time window that spanned roughly 1/3 second, making $\delta=40$ since data were collected at 120Hz.  The following results were found to be relatively insensitive to other $\delta$ within a reasonable neighborhood of 40.

\begin{figure}[H]
	\centering
	\caption{Reference forearm muscles for finger (top) and wrist (bottom) movements.}     
	\label{Reference}
	\noindent\makebox[\textwidth]{
		\includegraphics[width=0.70\textwidth]{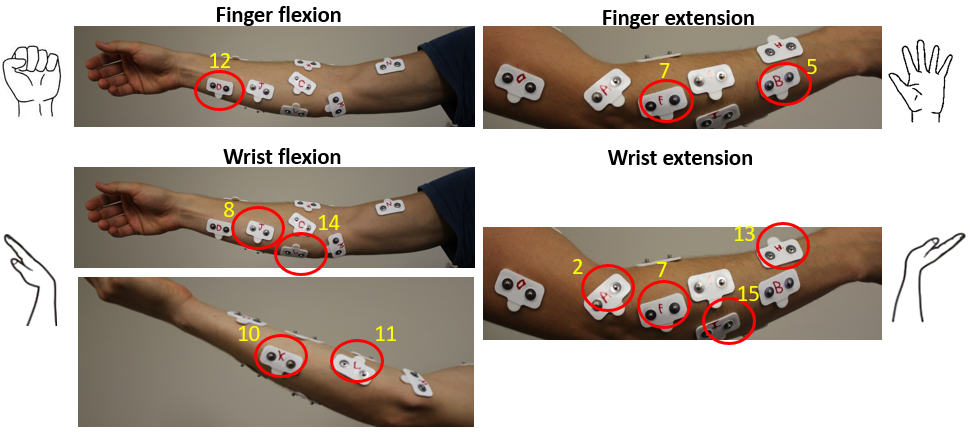}
	}
\end{figure}

The underlying muscle-movement mechanism for an AB person is known to clinicians; this prior information about the underlying true signals allows us to define the correct model size in the data application. To evaluate the model selection performance, we partition $\mathcal{K}$ according to important movement classes based on expert knowledge. Let $\mathcal{K}_F$ and $\mathcal{K}_E$ denote the partitioning subsets contributing to flexion and extension movement, respectively, which depend on finger or wrist movements. Figure \ref{Reference} shows in circled labels the clinically relevant muscles for different movement classes for an AB person. Specifically for finger movements, $\mathcal{K}_F = \{12\}$ and $\mathcal{K}_E = \{5, 7\};$ these muscles are known as \textit{flexor digitorum} and \textit{extensor digitorum}, respectively. Similarly for wrist movements, $\mathcal{K}_F = \{8, 10, 11, 14\}$ which are known as \textit{flexor carpi ulnaris}, \textit{flexor digitorum superficialis}, \textit{flexor carpi ulnaris}, and \textit{flexor carpi radials,} respectively, and $\mathcal{K}_E = \{2, 7, 13, 15\}$ which are coined as \textit{extensor carpi radialis longus, extensor digitorum, extensor carpi radialis brevis}, and \textit{extensor carpi ulnaris,} respectively. All muscles associated with a specific movement class produce highly correlated signals. The expert consensus is that one muscle in $\mathcal{K}_F$ and another in $\mathcal{K}_E$ are sufficient to define the relationship between the movements and EMG signals.  The ideal index set identifies one muscle in each movement class or equivalently $|\mathcal{K} \cap \mathcal{K}_F| = 1$ and $|\mathcal{K} \cap \mathcal{K}_E| = 1$, where $|\cdot|$ denotes the cardinality of a set. 

We use the method described in Section \ref{Proposed} to study which EMG signals $X_{k,i}$'s are related to the velocity and determine representative members of the sets $\mathcal{K}_F$ and $\mathcal{K}_E$. We also consider three alternative approaches; the first competitor is the method proposed by \cite{gertheiss2013variable} in which the authors propose the functional variable selection technique using adaptive group LASSO (agLASSO) type penalized criterion inducing both sparsity and smoothness. The second competitor is the method proposed by \cite{pannu2017robust} extending the idea of \cite{gertheiss2013variable}; here the authors use the objective function based on the least absolute deviation and select the functional variables using group lasso penalty. This approach is explicitly designed to account for potential outliers in the functional predictors, reducing overfitting of the regression surfaces. We refer this approach to LAD-gLASSO. The third competitor \citep{fan2015functional} uses functional additive regression (FAR) with a groupwise smoothly clipped absolute deviation (gSCAD) penalty, denoted by FAR-gSCAD.  Unlike other approaches, FAR-gSCAD imposes penalty on the integral $\int_{\mathcal{S}} X_{k,i}(s) \gamma_{k}(s) ds$ assuming the integral is well-defined. 	Note that all these methods are designed to assess the relationship between the scalar responses and functional predictors. In stark contrast to SAFE-gLASSO, all three competitors assume non-varying smooth coefficients, $\gamma_{k}(\cdot)$'s defined on $\mathcal{S}$, which accounts for the passive forces but assumes the relationship is not position dependent.

The performance of all the methods is evaluated in terms of how close they yield an ideal selection of the EMG signals as well as their prediction ability. Specifically if $\mathcal{\widehat  K}$ denotes an estimator produced by one of the methods, we evaluate its performance by
\begin{itemize}
	\item[$\bullet$] Size $= |\mathcal{\widehat  K}|,$ ideally size = 2 with $|\mathcal{\widehat  K} \cap \mathcal{K}_F | = 1$ and $|\mathcal{\widehat  K} \cap \mathcal{K}_E | = 1;$    
	
	\item[$\bullet$] Sparsity (SP) $= 1 - |\mathcal{\widehat  K}|/ K.$ In our application, $K$ is 16 and the optimal sparsity  is 0.88 which is associated with the ideal model size 2. Define the relative sparsity (RSP) with respect to the optimal sparsity by RSP = SP/0.88;     
	
	\item[$\bullet$] False positive rate (FPR)$=FP\, /\,|\mathcal{K}^c|;$ where we define the number of false positives (\textit{FP}) or falsely identified EMG signals in $\mathcal{\widehat  K}$ by $FP=|\mathcal{\widehat  K}  \cap \mathcal{K}^c|$, and $\mathcal{K}^c$ is the complement of $\mathcal{K};$ 
	
	\item[$\bullet$] True positive rate (TPR)$=TP/2;$  where the number of true positives (\textit{TP})  is defined by $TP=\mathbbm{1}(|\mathcal{\widehat  K}  \cap \mathcal{K}_F| \geq 1)+\mathbbm{1}(|\mathcal{\widehat  K}  \cap \mathcal{K}_E| \geq 1)$ and focuses only on whether we capture the two index sets $\mathcal{K}_F$ and $\mathcal{K}_E;$

	\item[$\bullet$] Mean squared error (MSE) $ = \sum_{i=1}^{N} (y_{i} - \widehat y_{i})^{2} / N.$

\end{itemize}
Note that in general one can not calculate FPRs and TPRs in a data application as the underlying truth is unknown. Our definitions for the above matrices are somewhat  unconventional focusing on group identification of $\mathcal{K}_F$ and $\mathcal{K}_E$ rather than identification of all $\mathcal{K}.$

\subsection{Computational Details}

We briefly describe the computational details of our implementation as performed in \texttt{R}, starting with the necessary steps after data processing including construction of the recent past EMG curves using $\delta=40$. We approximate the $\gamma_{k}(\cdot, \cdot)$ using a tensor product of two orthogonal B-splines with $L=12$ basis functions in the $s$ direction and $M=22$ basis functions in the $z$ direction.  Knots are placed uniformly throughout $\mathcal{S}$ and $\mathcal{Z}$.  The tuning parameters to produce the final estimates at each stage are chosen using 5-fold CV following the procedure described in \cite{roberts2017cross}.  

For the competitors, the smooth coefficient functions $\{\gamma_{k}(\cdot)\}^{16}_{k=1}$ are modeled  using B-splines with 12 basis functions and the sparsity-smoothness tuning parameters are estimated by 5-fold CV \citep{roberts2017cross} for all the competing methods. We used the {\tt R} package  {\tt grplasso} \citep{meier2009grplasso} to fit the agLASSO \citep{gertheiss2013variable} and {\tt rqPen} \citep{sherwood2017package} to fit the robust LAD-gLASSO \citep{pannu2017robust}. We used the code provided by the corresponding author \cite{fan2015functional} for FAR-gSCAD.

\subsection{Variable selection, prediction, and implication} \label{results_data}

Table \ref{HAND2} shows the results of the model selection performance for the finger movements. In most cases all the competing methods select one EMG signal that contributes for finger extension and another for finger flexion. All methods perform fairly well in terms of TPR.  In particular, the proposed SAFE-gLASSO attains desirable values of TPRs, FPRs, and RSPs across all settings.  We found that the first stage of SAFE-gLASSO tended to select more EMG signals than needed but the second stage corrected the surplus. In contrast, agLASSO exhibits good numerical performance reflected by the optimal levels of FPRs and RSPs but did have poor TPR in the last two settings.  LAD-gLASSO exhibited large FPRs and RSPs, indicating a high Type I error rate.  FAR-gSCAD demonstrates good model selection performance but selected more EMG signals than necessary, as indicated by its RSP values.  The numerical performances are similar for the wrist movements and due to the interest of the space, we include the results in Table \ref{WRIST2}  of the Supplementary Materials, Section \ref{Wrist analysis}.

\begin{figure}[h]
	\caption{Comparison of MSEs based on alternative approaches. Results correspond to finger (left) and wrist (right) movements with different patterns and postures.} 
	\quad\quad\quad
	\includegraphics[width=0.55\textwidth]{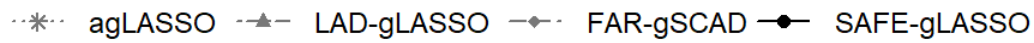} 
	\quad\quad\quad\\  
	\includegraphics[width=0.5\textwidth]{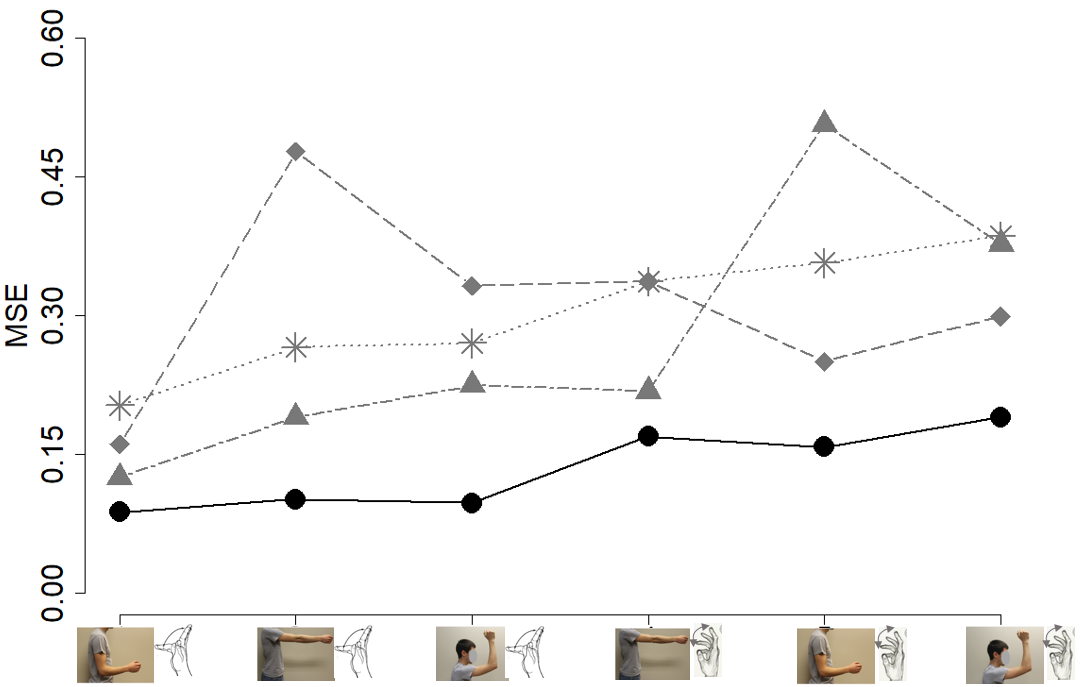}
	\hspace{-0.25cm} 
	\includegraphics[width=0.5\textwidth]{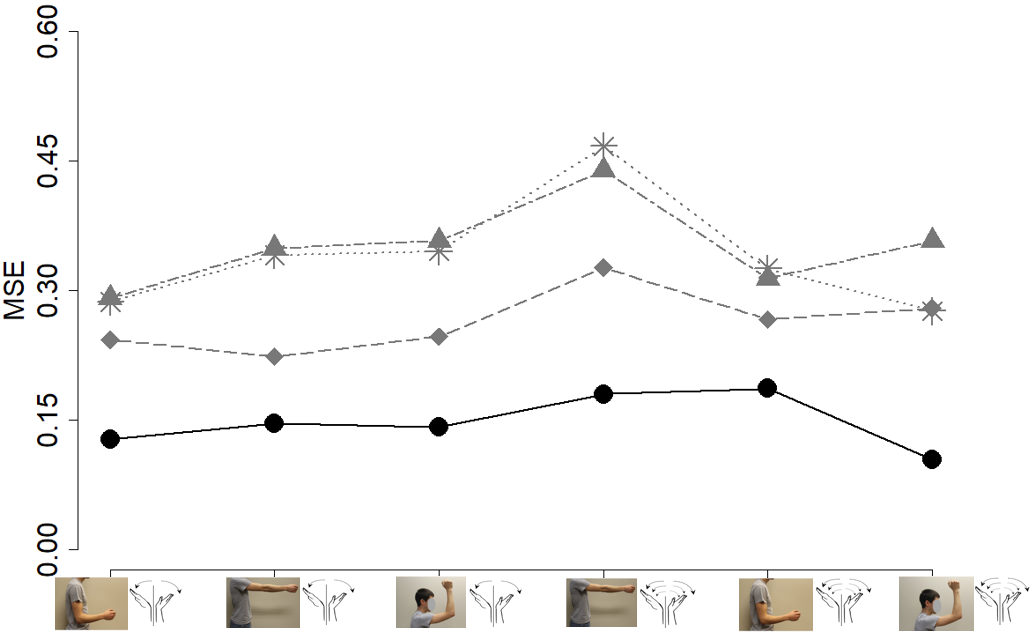} 
	\label{MSE_EMG}
\end{figure}

Figure \ref{MSE_EMG} illustrates graphically the MSEs of the competitive methods. For both finger and wrist movements, SAFE-gLASSO outperforms the competitors across all patterns and postures. The proposed approach was able to consistently fit the data better with a smaller model size, specifically relative to FAR-gSCAD and LAD-gLASSO.

\begin{figure}[h]
	\caption{Predicted velocity corresponding to consistent finger movements based on SAFE-gLASSO (red dashed line) and agLASSO (blue dashed line). Vertical reference lines are drawn at 1.45 and 1.88 second at which movements due to passive forces occur in the absence of muscle contractions. Shaded regions correspond to pointwise 95\%  split conformal prediction bands for SAFE-gLASSO.} 

	\centering
	\includegraphics[width=0.5\textwidth]{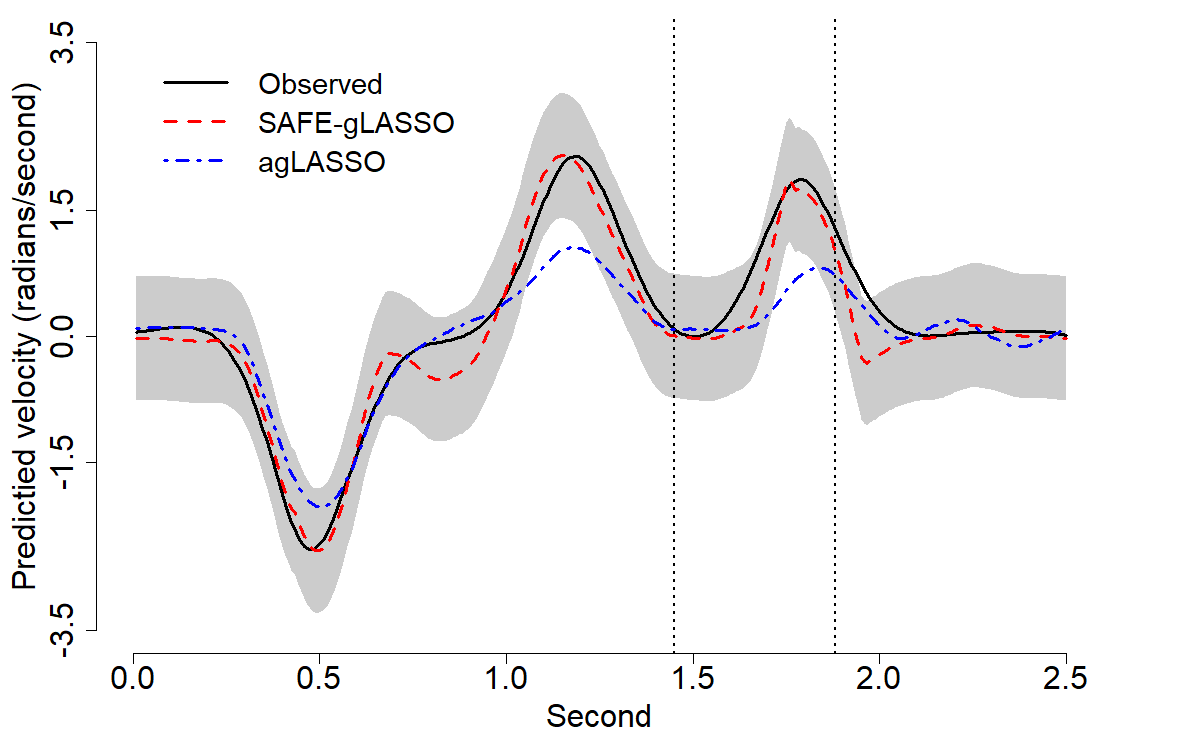}
	\label{SAFE-gLASSO_agLASSO}
\end{figure}

Figure \ref{SAFE-gLASSO_agLASSO} shows a segment of the fitted velocities based on SAFE-gLASSO and agLASSO where the shaded region corresponds to the split conformal prediction bands following \cite{Lei2017}.  SAFE-gLASSO yields superior prediction trajectory than agLASSO in the events with a larger absolute velocity.  In the highlighted event of Figure \ref{SAFE-gLASSO_agLASSO}, fingers return to the neutral position around 1.45 to 1.88 seconds due to passive forces and while both methods predict movement, agLASSO greatly underestimates the observed velocity. The improvement is most directly attributable to the inclusion of position information.  The agLASSO relies on the assumption that the effect of the EMG signals on finger/wrist movements is invariant across $z_i,$ which appears to limit the fit across a variety of movement events.

Since our methodology offers low-dimensional modeling with negligible false positives, biomedical engineers can use our method as a screening approach and focus primarily on the selected subsets of forearm muscles in collecting data from TRAs. This approach significantly reduces the burden of data collection with respect to time and cost. Furthermore, since our model and that by \cite{crouch2016lumped} account for similar biomechanical phenomenon, our variable selection results may be directly applied to their method.

\begin{table}[h]
	\tiny
	\caption{Finger movements EMG signal selection for consistent (top three rows) and random (bottom three row) patterns at different postures and results with superscript $\dagger$ correspond to the first stage of SAFE-gLASSO. RSPs, number of variables (square brackets), and the percentages of TPRs and FPRs are presented.}
	\label{HAND2}
	\centering
	\noindent\makebox[\textwidth]{
		\begin{tabular}{ccccccccccccc}
			& \multicolumn{3}{c}{agLASSO} & \multicolumn{3}{c}{LAD-gLASSO} & \multicolumn{3}{c}{FAR-gSCAD} & \multicolumn{3}{c}{SAFE-gLASSO} \\ \cline{2-13} 
			Pattern & RSP & TPR & FPR & RSP & TPR & FPR & RSP & TPR & FPR & RSP & TPR & FPR \\ \cline{2-13} 
			\begin{minipage}{.05\textheight}
				\includegraphics[width=10mm,height=5mm]{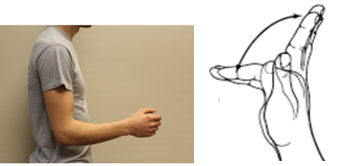}
			\end{minipage} & \begin{tabular}[c]{@{}c@{}}1.00\\ {[}2{]}\end{tabular} & \begin{tabular}[c]{@{}c@{}}100\\ {[}2{]}\end{tabular} & \begin{tabular}[c]{@{}c@{}}0\\ {[}0{]}\end{tabular} & \begin{tabular}[c]{@{}c@{}}0.57\\ {[}8{]}\end{tabular} & \begin{tabular}[c]{@{}c@{}}100\\ {[}3{]}\end{tabular} & \begin{tabular}[c]{@{}c@{}}38\\ {[}5{]}\end{tabular} & \begin{tabular}[c]{@{}c@{}}0.92\\ {[}3{]}\end{tabular} & \begin{tabular}[c]{@{}c@{}}100\\ {[}3{]}\end{tabular} & \begin{tabular}[c]{@{}c@{}}0\\ {[}0{]}\end{tabular} & \begin{tabular}[c]{@{}c@{}}1.00\\ {[}2{]}{[}2{]}$^{\dagger}$\end{tabular} & \begin{tabular}[c]{@{}c@{}}100\\ {[}2{]}{[}2{]}$^{\dagger}$\end{tabular} & \begin{tabular}[c]{@{}c@{}}0\\ {[}0{]}{[}0{]}$^{\dagger}$\end{tabular} \\
			\multicolumn{1}{l}{} & \multicolumn{1}{l}{} & \multicolumn{1}{l}{} & \multicolumn{1}{l}{} & \multicolumn{1}{l}{} & \multicolumn{1}{l}{} & \multicolumn{1}{l}{} & \multicolumn{1}{l}{} & \multicolumn{1}{l}{} & \multicolumn{1}{l}{} & \multicolumn{1}{l}{} & \multicolumn{1}{l}{} & \multicolumn{1}{l}{} \\
			\begin{minipage}{.05\textheight}
				\includegraphics[width=10mm,height=5mm]{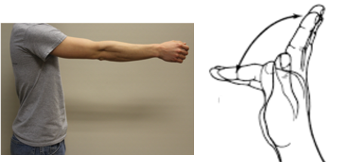}
			\end{minipage} & \begin{tabular}[c]{@{}c@{}}1.00\\ {[}2{]}\end{tabular} & \begin{tabular}[c]{@{}c@{}}100\\ {[}2{]}\end{tabular} & \begin{tabular}[c]{@{}c@{}}0\\ {[}0{]}\end{tabular} & \begin{tabular}[c]{@{}c@{}}0.72\\ {[}6{]}\end{tabular} & \begin{tabular}[c]{@{}c@{}}100\\ {[}3{]}\end{tabular} & \begin{tabular}[c]{@{}c@{}}23\\ {[}3{]}\end{tabular} & \begin{tabular}[c]{@{}c@{}}0.92\\ {[}3{]}\end{tabular} & \begin{tabular}[c]{@{}c@{}}100\\ {[}3{]}\end{tabular} & \begin{tabular}[c]{@{}c@{}}0\\ {[}0{]}\end{tabular} & \begin{tabular}[c]{@{}c@{}}1.00\\ {[}2{]}{[}2{]}$^{\dagger}$\end{tabular} & \begin{tabular}[c]{@{}c@{}}100\\ {[}2{]}{[}2{]}$^{\dagger}$\end{tabular} & \begin{tabular}[c]{@{}c@{}}0\\ {[}0{]}{[}0{]}$^{\dagger}$\end{tabular} \\
			\multicolumn{1}{l}{} & \multicolumn{1}{l}{} & \multicolumn{1}{l}{} & \multicolumn{1}{l}{} & \multicolumn{1}{l}{} & \multicolumn{1}{l}{} & \multicolumn{1}{l}{} & \multicolumn{1}{l}{} & \multicolumn{1}{l}{} & \multicolumn{1}{l}{} & \multicolumn{1}{l}{} & \multicolumn{1}{l}{} & \multicolumn{1}{l}{} \\
			\begin{minipage}{.05\textheight}
				\includegraphics[width=10mm,height=5mm]{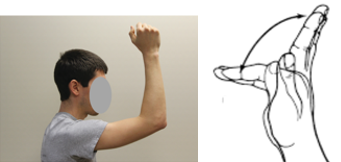}
			\end{minipage} & \begin{tabular}[c]{@{}c@{}}1.00\\ {[}2{]}\end{tabular} & \begin{tabular}[c]{@{}c@{}}100\\ {[}2{]}\end{tabular} & \begin{tabular}[c]{@{}c@{}}0\\ {[}0{]}\end{tabular} & \begin{tabular}[c]{@{}c@{}}0.57\\ {[}8{]}\end{tabular} & \begin{tabular}[c]{@{}c@{}}100\\ {[}3{]}\end{tabular} & \begin{tabular}[c]{@{}c@{}}38\\ {[}5{]}\end{tabular} & \begin{tabular}[c]{@{}c@{}}0.92\\ {[}3{]}\end{tabular} & \begin{tabular}[c]{@{}c@{}}100\\ {[}3{]}\end{tabular} & \begin{tabular}[c]{@{}c@{}}0\\ {[}0{]}\end{tabular} & \begin{tabular}[c]{@{}c@{}}0.92\\ {[}3{]}{[}3{]}$^{\dagger}$\end{tabular} & \begin{tabular}[c]{@{}c@{}}100\\ {[}3{]}{[}3{]}$^{\dagger}$\end{tabular} & \begin{tabular}[c]{@{}c@{}}0\\ {[}0{]}{[}0{]}$^{\dagger}$\end{tabular} \\
			\multicolumn{1}{l}{} & \multicolumn{1}{l}{} & \multicolumn{1}{l}{} & \multicolumn{1}{l}{} & \multicolumn{1}{l}{} & \multicolumn{1}{l}{} & \multicolumn{1}{l}{} & \multicolumn{1}{l}{} & \multicolumn{1}{l}{} & \multicolumn{1}{l}{} & \multicolumn{1}{l}{} & \multicolumn{1}{l}{} & \multicolumn{1}{l}{} \\
			\begin{minipage}{.05\textheight}
				\includegraphics[width=10mm,height=5mm]{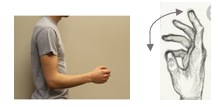}
			\end{minipage} & \begin{tabular}[c]{@{}c@{}}1.00\\ {[}2{]}\end{tabular} & \begin{tabular}[c]{@{}c@{}}100\\ {[}2{]}\end{tabular} & \begin{tabular}[c]{@{}c@{}}0\\ {[}0{]}\end{tabular} & \begin{tabular}[c]{@{}c@{}}0.78\\ {[}5{]}\end{tabular} & \begin{tabular}[c]{@{}c@{}}100\\ {[}2{]}\end{tabular} & \begin{tabular}[c]{@{}c@{}}23\\ {[}3{]}\end{tabular} & \begin{tabular}[c]{@{}c@{}}1.00\\ {[}2{]}\end{tabular} & \begin{tabular}[c]{@{}c@{}}100\\ {[}2{]}\end{tabular} & \begin{tabular}[c]{@{}c@{}}0\\ {[}0{]}\end{tabular} & \begin{tabular}[c]{@{}c@{}}1.00\\ {[}2{]}{[}3{]}$^{\dagger}$\end{tabular} & \begin{tabular}[c]{@{}c@{}}100\\ {[}2{]}{[}2{]}$^{\dagger}$\end{tabular} & \begin{tabular}[c]{@{}c@{}}0\\ {[}0{]}{[}1{]}$^{\dagger}$\end{tabular} \\
			\multicolumn{1}{l}{} & \multicolumn{1}{l}{} & \multicolumn{1}{l}{} & \multicolumn{1}{l}{} & \multicolumn{1}{l}{} & \multicolumn{1}{l}{} & \multicolumn{1}{l}{} & \multicolumn{1}{l}{} & \multicolumn{1}{l}{} & \multicolumn{1}{l}{} & \multicolumn{1}{l}{} & \multicolumn{1}{l}{} & \multicolumn{1}{l}{} \\
			\begin{minipage}{.05\textheight}
				\includegraphics[width=10mm,height=5mm]{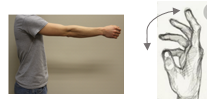}
			\end{minipage} & \begin{tabular}[c]{@{}c@{}}1.06\\ {[}1{]}\end{tabular} & \begin{tabular}[c]{@{}c@{}}50\\ {[}1{]}\end{tabular} & \begin{tabular}[c]{@{}c@{}}0\\ {[}0{]}\end{tabular} & \begin{tabular}[c]{@{}c@{}}0.85\\ {[}4{]}\end{tabular} & \begin{tabular}[c]{@{}c@{}}100\\ {[}2{]}\end{tabular} & \begin{tabular}[c]{@{}c@{}}15\\ {[}2{]}\end{tabular} & \begin{tabular}[c]{@{}c@{}}0.92\\ {[}3{]}\end{tabular} & \begin{tabular}[c]{@{}c@{}}100\\ {[}3{]}\end{tabular} & \begin{tabular}[c]{@{}c@{}}0\\ {[}0{]}\end{tabular} & \begin{tabular}[c]{@{}c@{}}1.00\\ {[}2{]}{[}3{]}$^{\dagger}$\end{tabular} & \begin{tabular}[c]{@{}c@{}}100\\ {[}2{]}{[}3{]}$^{\dagger}$\end{tabular} & \begin{tabular}[c]{@{}c@{}}0\\ {[}0{]}{[}0{]}$^{\dagger}$\end{tabular} \\
			\multicolumn{1}{l}{} & \multicolumn{1}{l}{} & \multicolumn{1}{l}{} & \multicolumn{1}{l}{} & \multicolumn{1}{l}{} & \multicolumn{1}{l}{} & \multicolumn{1}{l}{} & \multicolumn{1}{l}{} & \multicolumn{1}{l}{} & \multicolumn{1}{l}{} & \multicolumn{1}{l}{} & \multicolumn{1}{l}{} & \multicolumn{1}{l}{} \\
			\begin{minipage}{.05\textheight}
				\includegraphics[width=10mm,height=5mm]{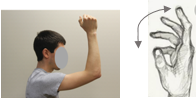}
			\end{minipage} & \begin{tabular}[c]{@{}c@{}}1.06\\ {[}1{]}\end{tabular} & \begin{tabular}[c]{@{}c@{}}50\\ {[}1{]}\end{tabular} & \begin{tabular}[c]{@{}c@{}}0\\ {[}0{]}\end{tabular} & \begin{tabular}[c]{@{}c@{}}0.85\\ {[}4{]}\end{tabular} & \begin{tabular}[c]{@{}c@{}}100\\ {[}3{]}\end{tabular} & \begin{tabular}[c]{@{}c@{}}8\\ {[}1{]}\end{tabular} & \begin{tabular}[c]{@{}c@{}}0.92\\ {[}3{]}\end{tabular} & \begin{tabular}[c]{@{}c@{}}100\\ {[}3{]}\end{tabular} & \begin{tabular}[c]{@{}c@{}}0\\ {[}0{]}\end{tabular} & \begin{tabular}[c]{@{}c@{}}0.92\\ {[}3{]}{[}3{]}$^{\dagger}$\end{tabular} & \begin{tabular}[c]{@{}c@{}}100\\ {[}2{]}{[}2{]}$^{\dagger}$\end{tabular} & \begin{tabular}[c]{@{}c@{}}0\\ {[}1{]}{[}1{]}$^{\dagger}$\end{tabular} \\ \cline{2-13} 
		\end{tabular}
	}
\end{table}

\subsection{Estimation of smooth coefficients} \label{Interpretation} 

\begin{figure}[h]
	\centering
	\caption{Regression surfaces for finger flexion (top-left), finger extension (top-right), wrist flexion (bottom-left), and wrist extension (bottom-right). The selected forearm muscles are pointed out by red circles.}     
	\label{EFFECTS}
	\includegraphics[width=1\textwidth]{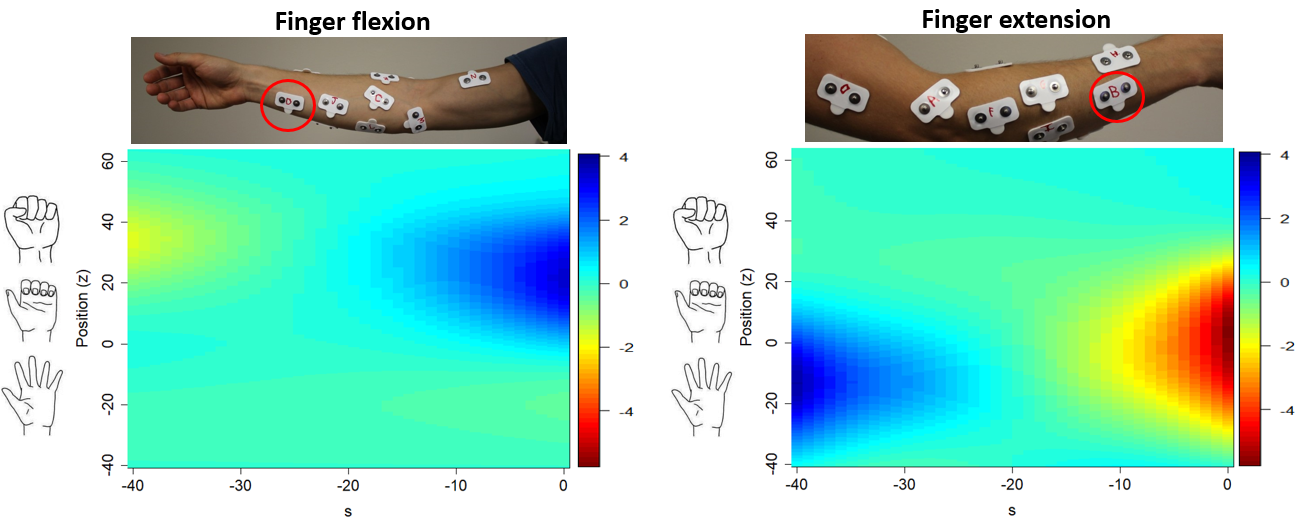}\\
	\includegraphics[width=1\textwidth]{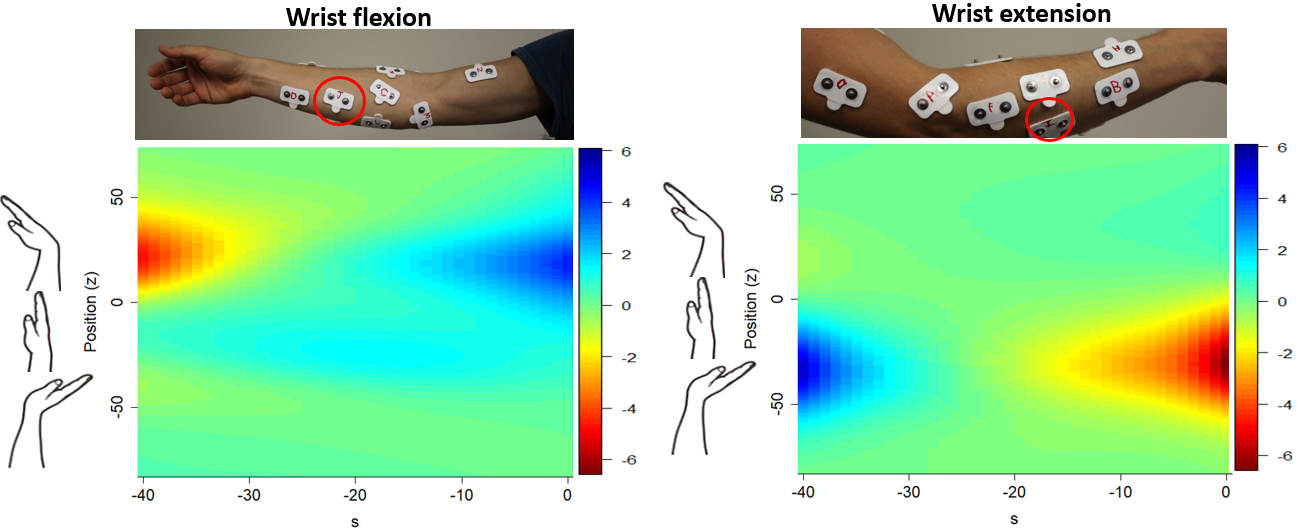}
\end{figure}

Figure \ref{EFFECTS} illustrates the estimated coefficient functions for the  finger (top) and wrist (bottom) movements. For the purpose of illustration, we focus on interpreting the second posture shown in Table \ref{HAND2}. In particular the top and bottom panels of Figure \ref{EFFECTS} correspond to consistent finger and wrist movement, respectively, for the two identified EMG signals from SAFE-gLASSO.  For finger movement, there is a clear distinction in the movement contributions: concurrent contraction of \textit{extensor digitorum} is responsible for finger extension (positive velocity) while \textit{flexor digitorum} leads to flexion (negative velocity).  This matches with the intuition of the biomechanical system. As expected, these associations also depend on the positions.  Indeed, the impact of the left surface on the velocity is most important for angles between $20$ and $55$ radians which correspond to finger flexion.  Similarly, the impact of the right surface is most relevant between angles $-40$ and $10$ radians which are primarily associated with finger extension.  The \textit{flexor digitorum} only leads to finger flexion when the hand is in a neutral position (0 to 15 radians). 

Finger flexion occurs at other positions occur due to past behavior of the \textit{extensor digitorum}. The observed concurrent relationship described above has the opposite historical relationship for the positions when the signals are presently active. This implies that past activation of one of these EMG signals can lead to the opposite type of concurrent movement they produce. In particular, this corresponds to two difficult cases where the model tries to establish the systematic relationship between: (1) passive force movements in the absence of muscle contraction, and (2) lack of movements in the presence of consistent muscle contraction due to physical constraints. Our model borrows information from the past, where the muscles were active, to predict such passive movements. Failing to appropriately account for passive forces results in poor prediction performance in those cases as shown in Figure \ref{SAFE-gLASSO_agLASSO}. 

The bottom panels of Figure \ref{EFFECTS} plot the estimated regression coefficients for the selected EMG channels for wrist movements. The interpretation of the regression surfaces for the wrist flexion/extension follows the same intuition as of hand movements. Concurrently, \textit{flexor carpi ulnaris} leads to wrist flexion around 0 to 50 radians while \textit{extensor carpi ulnaris}  yields wrist extension around -60 to -15 radians.  When the wrist is in a neutral position, \textit{flexor carpi ulnaris} has to be flexed to keep the wrist upright around -5 to -15 radians. As before, the observed concurrent relationship exhibits the opposite historical association for wrist flexion and extension.

The regression surfaces for other patterns of finger/wrist movements can be interpreted in the similar manner; we report the regression surfaces corresponding to other patterns in Section \ref{regression_surface_finger} of the Supplementary Material.

\section{Simulation study}
\subsection{Simulation mimicking the data application} \label{Simulation} 
In this section, we consider a simulation study that mimicks the  EMG data from Section \ref{Data analysis}. Specifically, we focus on finger movements with consistent movement. Let the observed data be $\left[y_i, z_i, \{X_{k,i}(s_r); r = 1, \ldots, 41\},  k = 1, \cdots, 16, i = 1,\dots, N \right];$ where $y_i$ is now the simulated velocity at time $i,$ and all other data match the descriptions from section~\ref{Data analysis}. In particular, consider the generating model $y_{i} =  \alpha + \sum_{k=1}^{16} \int_{\mathcal{S}} X_{k,i}(s) \gamma_{k}(s, z_{i}) ds + \epsilon_{i}$ where $\gamma_{k}(\cdot, \cdot) = 0$ for all $k$ except $k = 5$ and $k = 12,$  and  $\gamma_{5}(\cdot, \cdot)$ and $\gamma_{12}(\cdot, \cdot)$  are the estimated effects from one of the scenarios. In addition,  $\epsilon_{i}$ is a zero-mean error process with isotropic covariance function described by $$ \text{cov}(y_i,y_{i'}) = \sigma_{h}^{2} \left[ \mathbbm{1}(i = i') + \theta  \exp \big \{ - (|i - i'|/\eta)^2 \big\} \right].$$  Here $\theta$ is related to the dominant sources of dependence; $\theta = 0$ means that the responses are uncorrelated while large $\theta$ reflects higher degree of autocorrelation between the responses. In addition, $\eta$ controls the strength of correlation between any two measurements.  The choice of $\eta$ is driven by the parameter $\kappa$, the correlation between any two consecutive measurements, using the relationship $\eta = |i - i'| / \sqrt {-log (\kappa)}$.

The simulation study varies over three factors.  The first factor we examine is the dominant  sources of dependence determined by $\theta.$
\begin{itemize}
	\item[\textit{A1.}] \textit{Dominant white noise.} $\theta << 1.$
	\item[\textit{A2.}] \textit{Equal process.} $\theta = 1.$
	\item[\textit{A3.}] \textit{Dominant dependent process.} $\theta >> 1.$
\end{itemize}
The second factor is the strength of the correlation between successive measurements. 
\begin{itemize}
	\item[\textit{B1.}] \textit{Low correlation.} $\kappa = 0.2.$
	\item[\textit{B2.}] \textit{High correlation.} $\kappa = 0.9.$
\end{itemize}
The third factor is the magnitude of noise variance $\sigma^{2}_{h}.$ 
\begin{itemize}
	\item[\textit{C1.}] \textit{Small noise.} $\sigma_{h}^{2} = 0.01\widehat \sigma^{2}$ corresponds to the situation of having smaller variance than that of the original data.
	\item[\textit{C2.}] \textit{Large noise.} $\sigma_{h}^{2} =  \widehat \sigma^{2}$ corresponds to the case of equal variance.
\end{itemize}
Calculate signal-to-noise ratio (SNR) as SNR = $\text{var}(\mu_i)/\text{var}(\epsilon_{i}),$ where $\mu_i = \alpha + \sum_{k=1}^{16} \int_{\mathcal{S}} X_{k,i}(s) \gamma_{k}(s, z_{i}) ds.$  In particular, we consider SNR = \{875, 445, 80, 8, 4, 0.8\}. The results are based on 100 independent samples for each combination of the simulation
settings.

Table \ref{high_CORR} illustrates the numerical performance corresponding to the simulated kinematic data for different SNRs. The results are consistent to the  findings of finger movements in Section \ref{Data analysis}. As before, we observe that the mean model size of the second stage of the proposed is nearer to the truth (i.e. 2) than that of the first stage; see the ``SP" column under SAFE-gLASSO in Table \ref{high_CORR}. This is expected due to the fact that by adopting the two-stage variable selection scheme, we shrink the surplus variables in the second stage of the procedure. The proposed approach exhibits lower FPRs than that of the agLASSO, LAD-gLASSO, and FAR-gSCAD across all simulation settings. Similar to the other approaches, SAFE-gLASSO also attains high TPRs.

\begin{figure}[h]
	\caption{Comparison of MSEs based on competing approaches for high (left) and low (right) SNRs with high  correlation coefficient (\textit{B2}). Simulated data corresponds to the consistent finger movements. Reported are the results based on 100 simulations.} 
	\includegraphics[width=0.45\textwidth]{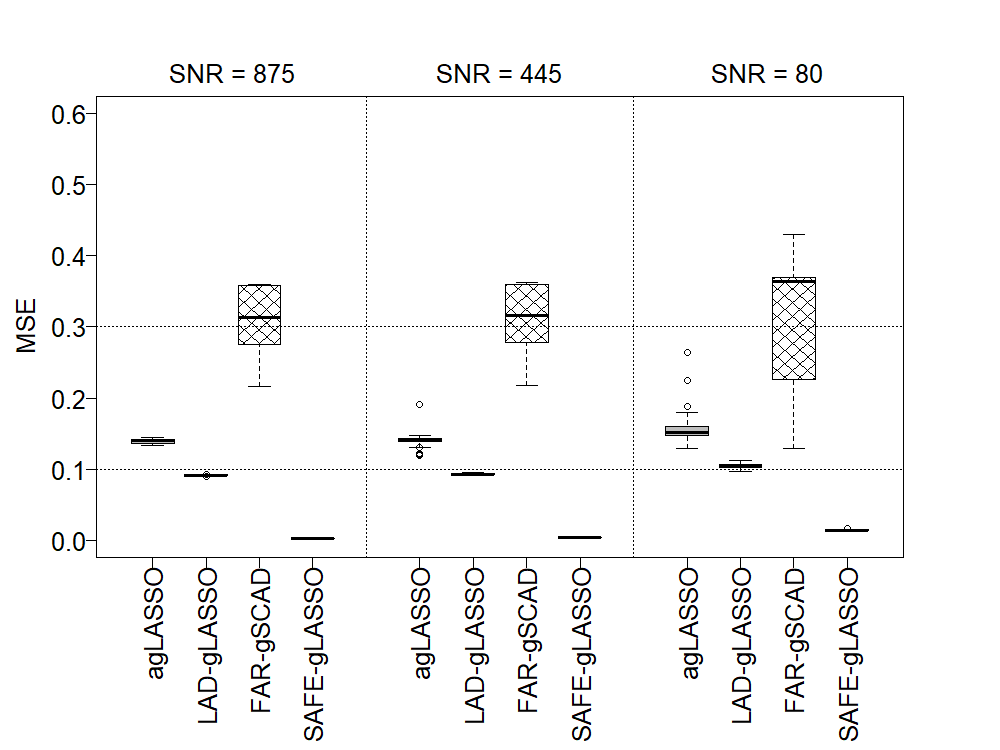}
	\includegraphics[width=0.45\textwidth]{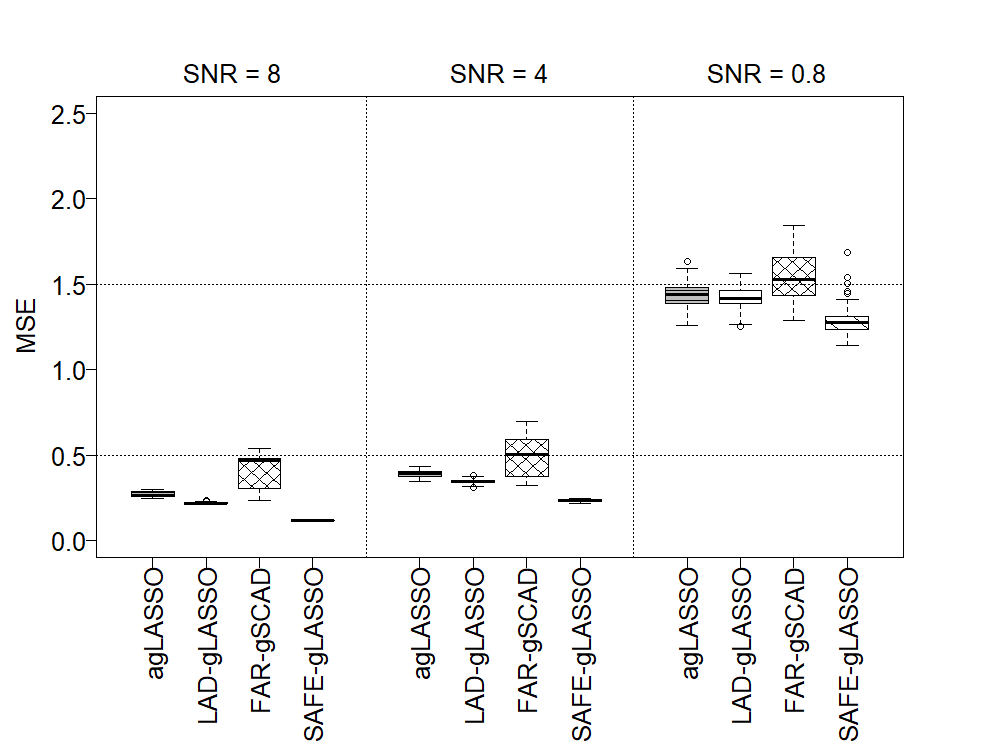}
	\label{high_CORR_MSE}
\end{figure}

SAFE-gLASSO exhibited superior prediction performance relative to the competing methods across all simulation settings; see Figure \ref{high_CORR_MSE}. These findings are in agreement with the results of Section \ref{Data analysis}. As expected the prediction accuracy improves with the SNR; compare the box-plots corresponding to the SAFE-gLASSO for different SNRs in Figure \ref{high_CORR_MSE}. Unlike the proposed approach, the numerical performance of the alternative methods suffers due to not considering positing-varying smooth coefficients as the methods fail to quantify the systematic relationship between the velocity and muscle activities of a  biomechanical model. We report the additional simulation results in the Supplementary Materials, Section \ref{validation_EMG}. 
\begin{table}[H]
	\tiny
	\centering
	\caption{Analysis of finger movements with fixed motion. Data is generated assuming noise variance \textit{C1} and \textit{C2} with different dominant processes (\textit{A1}-\textit{A3}) for high correlation coefficient (\textit{B2}). Reported are the SPs (\%), model size (in square brackets), TPRs (\%), and FPRs (\%) averaged over 100 simulations. Results with superscript $\dagger$ correspond to the first stage of SAFE-gLASSO.}
	\label{high_CORR}
	\noindent\makebox[\textwidth]{
		\begin{tabular}{cccccccccccccc}
			&  & \multicolumn{3}{c}{agLASSO} & \multicolumn{3}{c}{LAD-gLASSO} & \multicolumn{3}{c}{FAR-gSCAD} & \multicolumn{3}{c}{SAFE-gLASSO} \\ \cline{3-14} 
			Setting & SNR & SP & TPR & FPR & SP & TPR & FPR & SP & TPR & FPR & SP & TPR & FPR \\ \cline{3-14} 
			$\textit{A1} + \textit{C1}$ & 875 & \begin{tabular}[c]{@{}c@{}}87\\ {[}2.09{]}\end{tabular} & 100 & 1 & \begin{tabular}[c]{@{}c@{}}66\\ {[}5.41{]}\end{tabular} & 100 & 24 & \begin{tabular}[c]{@{}c@{}}81\\ {[}3.00{]}\end{tabular} & 100 & 7 & \begin{tabular}[c]{@{}c@{}}88\\ {[}2.00{]}{[}2.00{]}$^{\dagger}$\end{tabular} & \begin{tabular}[c]{@{}c@{}}100\\ {[}100{]}$^{\dagger}$\end{tabular} & \begin{tabular}[c]{@{}c@{}}0\\ $[0]^{\dagger}$\end{tabular} \\ \cline{3-14} 
			$\textit{A2} + \textit{C1}$ & 445 & \begin{tabular}[c]{@{}c@{}}86\\ {[}2.18{]}\end{tabular} & 100 & 1 & \begin{tabular}[c]{@{}c@{}}66\\ {[}5.42{]}\end{tabular} & 100 & 24 & \begin{tabular}[c]{@{}c@{}}81\\ {[}3.00{]}\end{tabular} & 100 & 7 & \begin{tabular}[c]{@{}c@{}}88\\ {[}2.00{]}{[}2.00{]}$^{\dagger}$\end{tabular} & \begin{tabular}[c]{@{}c@{}}100\\ {[}100{]}$^{\dagger}$\end{tabular} & \begin{tabular}[c]{@{}c@{}}0\\ {[}0{]}$^{\dagger}$\end{tabular} \\ \cline{3-14} 
			$\textit{A3} + \textit{C1}$ & 80 & \begin{tabular}[c]{@{}c@{}}87\\ {[}2.08{]}\end{tabular} & 99 & 1 & \begin{tabular}[c]{@{}c@{}}68\\ {[}5.09{]}\end{tabular} & 100 & 22 & \begin{tabular}[c]{@{}c@{}}81\\ {[}2.98{]}\end{tabular} & 100 & 7 & \begin{tabular}[c]{@{}c@{}}88\\ {[}2.00{]}{[}2.00{]}$^{\dagger}$\end{tabular} & \begin{tabular}[c]{@{}c@{}}100\\ {[}100{]}$^{\dagger}$\end{tabular} & \begin{tabular}[c]{@{}c@{}}0\\ {[}0{]}$^{\dagger}$\end{tabular} \\ \cline{3-14} 
			$\textit{A1} + \textit{C2}$ & 8 & \begin{tabular}[c]{@{}c@{}}88\\ {[}2.00{]}\end{tabular} & 100 & 0 & \begin{tabular}[c]{@{}c@{}}74\\ {[}4.22{]}\end{tabular} & 100 & 15 & \begin{tabular}[c]{@{}c@{}}81\\ {[}2.97{]}\end{tabular} & 100 & 7 & \begin{tabular}[c]{@{}c@{}}88\\ {[}2.00{]}{[}2.00{]}$^{\dagger}$\end{tabular} & \begin{tabular}[c]{@{}c@{}}100\\ {[}100{]}$^{\dagger}$\end{tabular} & \begin{tabular}[c]{@{}c@{}}0\\ {[}0{]}$^{\dagger}$\end{tabular} \\ \cline{3-14} 
			$\textit{A2} + \textit{C2}$ & 4 & \begin{tabular}[c]{@{}c@{}}87\\ {[}2.03{]}\end{tabular} & 100 & 0 & \begin{tabular}[c]{@{}c@{}}71\\ {[}4.58{]}\end{tabular} & 100 & 18 & \begin{tabular}[c]{@{}c@{}}81\\ {[}2.91{]}\end{tabular} & 100 & 7 & \begin{tabular}[c]{@{}c@{}}88\\ {[}2.00{]}{[}2.00{]}$^{\dagger}$\end{tabular} & \begin{tabular}[c]{@{}c@{}}100\\ {[}100{]}$^{\dagger}$\end{tabular} & \begin{tabular}[c]{@{}c@{}}0\\ {[}0{]}$^{\dagger}$\end{tabular} \\ \cline{3-14} 
			$\textit{A3} + \textit{C2}$ & 0.8 & \begin{tabular}[c]{@{}c@{}}87\\ {[}2.16{]}\end{tabular} & 98 & 2 & \begin{tabular}[c]{@{}c@{}}69\\ {[}5.00{]}\end{tabular} & 100 & 21 & \begin{tabular}[c]{@{}c@{}}82\\ {[}2.96{]}\end{tabular} & 100 & 7 & \begin{tabular}[c]{@{}c@{}}87\\ {[}2.02{]}{[}2.27{]}$^{\dagger}$\end{tabular} & \begin{tabular}[c]{@{}c@{}}95\\ {[}99{]}$^{\dagger}$\end{tabular} & \begin{tabular}[c]{@{}c@{}}1\\ {[}2{]}$^{\dagger}$\end{tabular} \\ \cline{3-14} 
		\end{tabular}
	}
\end{table}
\subsection{Numerical experiment} \label{Mathematical_simul}
Next we consider another simulation study where data are generated from a purely mathematical perspective. The simulated data is $\left[y_i, z_i, \{X_{k,i}(s_r); r = 1, \ldots, 100\}, k = 1, \cdots, 10, i=1,\dots, 500 \right],$ where $y_i$ is a scalar response at an instance $i,$ $z_i$ is a scalar covariate observed at $i$,  $X_{k,i}(s_r)$'s are the realizations of the $k$th functional predictor such that $s_r \in [0,1]$.  The 10 functional predictors are generated similar to \cite{gertheiss2013variable} and \cite{pannu2017robust}.  The nonzero functional coefficients $\{\gamma_{k}(\cdot, \cdot); k = 1, 2\}$'s are varying over $z,$ $z_i \in [-1, 1],$ and defined as $\gamma_{1}(s, z) = 1 + \sqrt{2} C z k  +  \sqrt{2} k \text{cos} (\pi s)$ and $\gamma_{2}(s, z) = 1 + C \text{exp} (-0.5 z)  +  s + 0.5 s^2$. Note that $C$ controls the strength of functional coefficients over $z.$  The error variance $\sigma^{2}$ is calculated based on different SNRs; i.e., SNR = $\{0.5, 1, 5\}.$  More details about the functional predictors, functional coefficients, and simulation results are summarized in the Supplementary Materials, Section \ref{toy}.

As expected, when $C = 0,$ meaning the functional coefficient should have only one dimension, $s$, all competitors showed comparable performance across all metrics, having high TPR and similar MSE.  All methods except LAD-gLASSO had low FPR, which consistently selected 5 of the 10 possible variables. As the difficulty of the model increases, say for $C = 5$  the competitors perform much worse than SAFE-gLASSO.  FAR-gSCAD and agLASSO have low FPR, but tend to select only 1 of the 2 important variables, while LAD-gLASSO has high TPR but also high FPR, similar to the $C=0$ scenario.  As expected, the MSE performance also deteriorates for the competitors as $C$ increases.

We also investigated the coverage probabilities and length of the prediction intervals for the SAFE-gLASSO models. Denote the lower and upper bound of the in-sample prediction interval by $C_{roo}(X_{i}) = (C^{l}_{roo}(X_{i}), C^{u}_{roo}(X_{i}))$, which is constructed  following \textit{rank-one-out} (ROO) split conformal prediction inference \citep{Lei2017}. Define the pointwise average coverage probability by $CP_{roo}  = \sum^{N}_{i = 1} \mathbbm{1} \{ y_{i} \in C_{roo}(X_{i}) \} / N$ where $N$ is the total number of in-sample observations.   Calculate the expected length of the interval by $\sum^{N}_{i = 1} \{ C^{u}_{roo}(X_{i}) -  C^{l}_{roo}(X_{i}) \} / N.$ Let $i^{*}$ index the instance at which new information of the predictor $X_{i^{*}}$ is recorded; where $i^{*} = 1, \ldots, M,$ and M is the total number of future observations.  Define the pointwise average coverage probability for the future observations by $CP_{split}  = \sum^{M}_{i^{*}=1} \mathbbm{1} \{y_{i^{*}} \in C_{split}(X_{i^{*}}) \} / M;$  where $C_{split}(X_{i^{*}})$ is the prediction intervals computed using the split conformal prediction inference \citep{Lei2017}. In our simulation, we consider $M = 250.$

Table \ref{prediction_interval} demonstrates the predictive inference of the proposed approach in terms of the actual coverage and expected length of the interval for both the in-sample ($y_{i}$) and future ($y_{i^{*}}$) observations. In general, the average coverage stays around the nominal levels irrespective of the complexity of the model as defined by $C;$ see the results for $C = 0$ and $C = 10.$ As expected, the width of the interval increases as $C$ departs from 0 but reduces for larger SNR; see the average length of the interval for $C = 10$ at miscoverage level $\alpha = 0.20$ for $y_{i}.$ We also observe the similar phenomenon with the prediction intervals for the future observations.

\begin{table}[H]
	\tiny
	\centering
	\caption{\small Average pointwise coverage of 80\%, 90\%, 95\%, and 99\% prediction bands for the in-sample ($y_{i}$) and future ($y_{i^{*}}$) observations using ROO and split conformal prediction inference, respectively; standard errors (in parenthesis) and average lengths (in square bracket) of prediction bands are reported. Results are based on 100 simulations.}
	\label{prediction_interval}
	\noindent\makebox[\textwidth]{
		\begin{tabular}{ccccccccc}
			&  & \multicolumn{3}{c}{SNR = 0.5} &  & \multicolumn{3}{c}{SNR = 5} \\ \cline{3-5} \cline{7-9} 
			& $\alpha$ & C = 0 & C = 5 & C = 10 &  & C = 0 & C = 5 & C = 10 \\ \cline{3-5} \cline{7-9} 
			{$y_{i}$} & 0.01 & 0.990 (0.010) {[}1.169{]} & 0.990 (0.010) {[}5.202{]} & 0.991 (0.010) {[}9.572{]} &  & 0.990 (0.010) {[}0.367{]} & 0.990 (0.010) {[}2.626{]} & 0.990 (0.010) {[}5.120{]} \\ \cline{3-5} \cline{7-9} 
			& 0.05 & 0.949 (0.022) {[}0.875{]} & 0.950 (0.022) {[}3.870{]} & 0.950 (0.022) {[}7.123{]} &  & 0.949 (0.022) {[}0.277{]} & 0.950 (0.022) {[}1.848{]} & 0.950 (0.022) {[}3.586{]} \\ \cline{3-5} \cline{7-9} 
			& 0.10 & 0.900 (0.030) {[}0.735{]} & 0.900 (0.030) {[}3.234{]} & 0.900 (0.030) {[}5.971{]} &  & 0.900 (0.030) {[}0.232{]} & 0.900 (0.030) {[}1.489{]} & 0.900 (0.030) {[}2.874{]} \\ \cline{3-5} \cline{7-9} 
			& 0.20 & 0.800 (0.040) {[}0.572{]} & 0.799 (0.040) {[}2.504{]} & 0.800 (0.040) {[}4.653{]} &  & 0.800 (0.040) {[}0.181{]} & 0.800 (0.040) {[}1.115{]} & 0.799 (0.040) {[}2.136{]} \\ \cline{3-5} \cline{7-9} 
			&  &  &  &  &  &  &  &  \\ \cline{3-5} \cline{7-9} 
			{$y_{i^{*}}$} & 0.01 & 0.992 (0.009) {[}1.168{]} & 0.990 (0.010) {[}5.209{]} & 0.992 (0.009) {[}9.657{]} &  & 0.990 (0.010) {[}0.366{]} & 0.989 (0.011) {[}2.598{]} & 0.989 (0.011) {[}5.062{]} \\ \cline{3-5} \cline{7-9} 
			& 0.05 & 0.949 (0.022) {[}0.871{]} & 0.948 (0.022) {[}3.860{]} & 0.951 (0.022) {[}7.162{]} &  & 0.948 (0.022) {[}0.277{]} & 0.951 (0.022) {[}1.841{]} & 0.951 (0.022) {[}3.568{]} \\ \cline{3-5} \cline{7-9} 
			& 0.10 & 0.898 (0.030) {[}0.732{]} & 0.900 (0.030) {[}3.228{]} & 0.904 (0.029) {[}5.992{]} &  & 0.898 (0.030) {[}0.232{]} & 0.902 (0.030) {[}1.484{]} & 0.901 (0.030) {[}2.860{]} \\ \cline{3-5} \cline{7-9} 
			& 0.20 & 0.802 (0.040) {[}0.571{]} & 0.802 (0.040) {[}2.817{]} & 0.806 (0.040) {[}4.667{]} &  & 0.798 (0.040) {[}0.182{]} & 0.803 (0.040) {[}1.115{]} & 0.803 (0.040) {[}2.136{]} \\ \cline{3-5} \cline{7-9} 
		\end{tabular}
	}
\end{table}

\section{Discussion}
\label{Discuss}

In this paper, we proposed a covariate-dependent, scalar-on-function regression model that appropriately accounts for the biomechanical processes involved in hand movement, such as passive forces and physical constraints.  The functional predictors were the recent past behavior of EMG signals measured across multiple muscles in the subject's forearm and the responses were finger and wrist velocity.  The functional coefficients for each EMG signal were allowed to vary based on the current finger or wrist position.  The bivariate coefficients were then approximated using a tensor product of rich basis expansions that were then estimated with a combined multi-dimensional smoothing and sparseness penalty, which is an extension of \cite{gertheiss2013variable}. We developed a two-step variable selection procedure, called Sequential Adaptive Functional Empirical group LASSO (SAFE-gLASSO), that was shown through numerical investigations to have superior performance over standard selection approaches \citep{gertheiss2013variable,pannu2017robust,fan2015functional} by reducing the number of false positives irrespective of model complexity.

The results of the data application showed SAFE-gLASSO was able to identify the important EMG signals for finger and wrist movement for an AB subject.  Furthermore, the estimated varying functional coefficients were relatively sparse, easy to interpret, and had exceptional predictive performance compared to standard selection approaches.  Our model and fitting algorithm have great potential to outperform current state-of-the-art data driven methods for prosthesis control such as pattern recognition because they ignore biomechanical constraints, do not perform variable selection, and are prone to overfitting.  The variable selection results from SAFE-gLASSO could also be incorporated in the method by \cite{crouch2016lumped}, which currently does not perform variable selection and uses a planar link-segment dynamic model.  Although our model mathematically differs from theirs, both models account for the biomechanical system with enough similarity that our variable selection results still apply.  Our model does have the advantage in that, after training our model, there is minimal data processing required to produce predictions in the prosthetic limb, reducing the burden of real-time data collection.

There are many opportunities for extensions and new applications of the approaches taken in this paper.  The two-step fitting approach used in SAFE-gLASSO easily applies to the more common univariate functional coefficients.  We also discussed an approach to assess the predictive ability based on data splitting, which as far as the authors are aware has not been applied to functional regression.  Additionally, our model, which uses a tensor product basis, can also be applied in a more general setting with many functional predictors having functional coefficients varying over multiple covariates.  This would be of interest for this data application if the coefficients were considered to vary significantly across postures.

In our current developments, the functional covariates were assumed to be observed without error on a fine grid of points. Extensions to a situation where the functional measurements are perturbed by error or when the grid points $s_r$'s are sparse would require a preliminary smoothing of the functional covariates using existing approaches  \cite{yao2005functional,xiao2016fast} and then SAFE-gLASSO may be employed on the smoothed functional covariates.  In a preliminary investigation, not reported here, we observed that the variable selection performance of SAFE-gLASSO is unaffected due to the use of estimated smooth profiles in the place of true functional predictors. However prediction error of the responses is affected by large noise variance of the functional predictors, as would any method.  

As in non-parametric regressions, our method relies on the tensor products of basis functions; as a result, the number of parameters to estimate can explode very quickly. Even though our method can tackle the dimensionality issue, it becomes computationally intensive with large number of basis functions. Note that we adopt CV based approach in selecting the tuning parameters for the data application. We also acknowledge that there are other methods to select optimum tuning parameters and CV based approach may not be theoretically the best approach \citep{gertheiss2013variable}. While the method works well for our data application, future research is needed to investigate the optimality in selecting the tuning parameters. In addition, we focused on the development of a subject-specific modeling procedure and were not concerned with estimating subject-specific variability.  By accounting for this source of variability, we can develop highly functional, user-specific robotic prosthetic limb that perform well across multiple subjects through estimation of population parameters.


\section*{Acknowledgments}

The project described was supported by Division of Information and Intelligent System, National Science Foundation through award number 1527202. The content is solely the responsibility of the authors and does not necessarily represent the official views of the NSF. A.-M. Staicu's research was supported by NSF Grant number DMS1454942.

\bibliographystyle{natbib}
\small
\bibliography{ciTe_NSF}

\newpage
\appendix
\section*{Supplementary Material}

This Supplementary Material consists of five sections. Section \ref{GAM} provides details for the fit without sparse penalty. Section \ref{additional_SMG_results} presents additional analysis results for the kinematic data. Section \ref{validation_EMG} and Section \ref{toy} provide additional simulation results. Section \ref{algorithm} details the algorithm of the post-selection predictive inference.

\section{Generalized additive model}\label{GAM}

Let the observed data, in general, be $\left[y_i, z_i, \{X_{k,i}(s_r); r = 1, \ldots, R\},  k = 1, \cdots, K, i = 1, \dots, N\right],$ where all terms bear the usual meaning as before and are described in the manuscript. Write the penalized criterion for the model (4.3) with the smoothing penalty only in matrix form as  
\begin{equation*}
\label{approx_FLM_MATRIX}
(\boldsymbol {Y} - \boldsymbol {\widetilde X}\boldsymbol {\beta} )^{T} (\boldsymbol {Y} - \boldsymbol {\widetilde X} \boldsymbol {\beta} ) + \phi_{1} \boldsymbol {\beta}^{T} \boldsymbol {\Omega}_{s} \boldsymbol {\beta} + \phi_{2} \boldsymbol {\beta}^{T} \boldsymbol {\Omega}_{z} \boldsymbol {\beta};
\end{equation*}    
where $\boldsymbol {Y}$ is a $N \times 1$ vector with elements $y_{i}$'s, $\boldsymbol { \widetilde X} = ( \boldsymbol {\widetilde X}_{1}| \cdots | \boldsymbol {\widetilde X}_{K}  ),$  $\boldsymbol {\widetilde X}_{k}$ is an $N \times LM$ matrix having the $i$th row $\boldsymbol {\widetilde X}^{T}_{k,i}$, $\boldsymbol {\beta} = (\boldsymbol {\beta}^{T}_{1}, \cdots, \boldsymbol {\beta}^{T}_{K})^{T}$ is a vector of $LM$ parameters, 
$\boldsymbol {\Omega}_{s} = \text{diag} \{\boldsymbol {\Omega}_{s1} \otimes \boldsymbol {I}_{M}, \cdots, \boldsymbol {\Omega}_{sK} \otimes \boldsymbol {I}_{M}\},$ and $\boldsymbol {\Omega}_{z} = \text{diag} \{\boldsymbol {I}_{L} \otimes \boldsymbol {\Omega}_{z1}, \cdots, \boldsymbol {I}_{L} \otimes \boldsymbol {\Omega}_{zK}\}.$ Following \cite{ruppert2003semiparametric, wood2017generalized}, taking the derivative with respect to $\boldsymbol {\beta}$ yields $\boldsymbol {\widetilde \beta} = (\boldsymbol { \widetilde X}^{T} \boldsymbol { \widetilde X} + \phi_{1} \boldsymbol {\Omega}_{s} + \phi_{2} \boldsymbol {\Omega}_{z})^{-1} \boldsymbol { \widetilde X}^{T} \boldsymbol {Y};$ we obtain the initial estimates of the regression coefficients as $\widetilde \beta_{k0}(\cdot, \cdot)$'s for a given $\phi_{sz}$. The corresponding Bayesian posterior covariance matrix is $ \text{Var}(\boldsymbol {\widetilde \beta}) = (\boldsymbol { \widetilde X}^{T} \boldsymbol { \widetilde X} + \phi_{1} \boldsymbol {\Omega}_{s} + \phi_{2} \boldsymbol {\Omega}_{z})^{-1}  \widehat \sigma^{2},$ where $\widehat \sigma^{2}$ is estimated from the residual sum of squares. Predict the responses as $\widehat Y = \boldsymbol {\widetilde X} \widetilde{\boldsymbol {\beta}}.$ 

The $\phi_{1}$ and $\phi_2$ are unknown in practice; one approach to select the optimal tuning parameters is block CV. The post-selection model (4.7) is fitted using the above intuition but with the reduced subset $\mathcal{K}^{(2)}_{\lambda^{\ast},\phi^{\ast}}.$


\section{Additional results of EMG selection for movement data}\label{additional_SMG_results}

In this section, we present more results of the analysis of the EMG and movement data.

\subsection{Data restructuring of EMG signals}\label{restructure}

\begin{figure}[H]
	\centering
	\caption{Visualization of data restructuring of EMG signals and joint finger angles at 0.9 second and 2.3 second.  Time at current position (blue dot on black line) used to extract concurrent and previous $\delta$ values of the two EMG signals, shown in red and green in (a).  The time domain for each set of past $\delta+1$ EMG signal measurements is rescaled to $[-\delta,0].$ Reconstructed EMG curves are plotted in gray lines on the top-right and bottom-right panels, and two curves corresponding to 0.9 second and 2.3 second are highlighted in (b) and (c) for EMG-7 (green solid line) and EMG-12 (red solid line) respectively.}
	\includegraphics[width=0.85\textwidth]{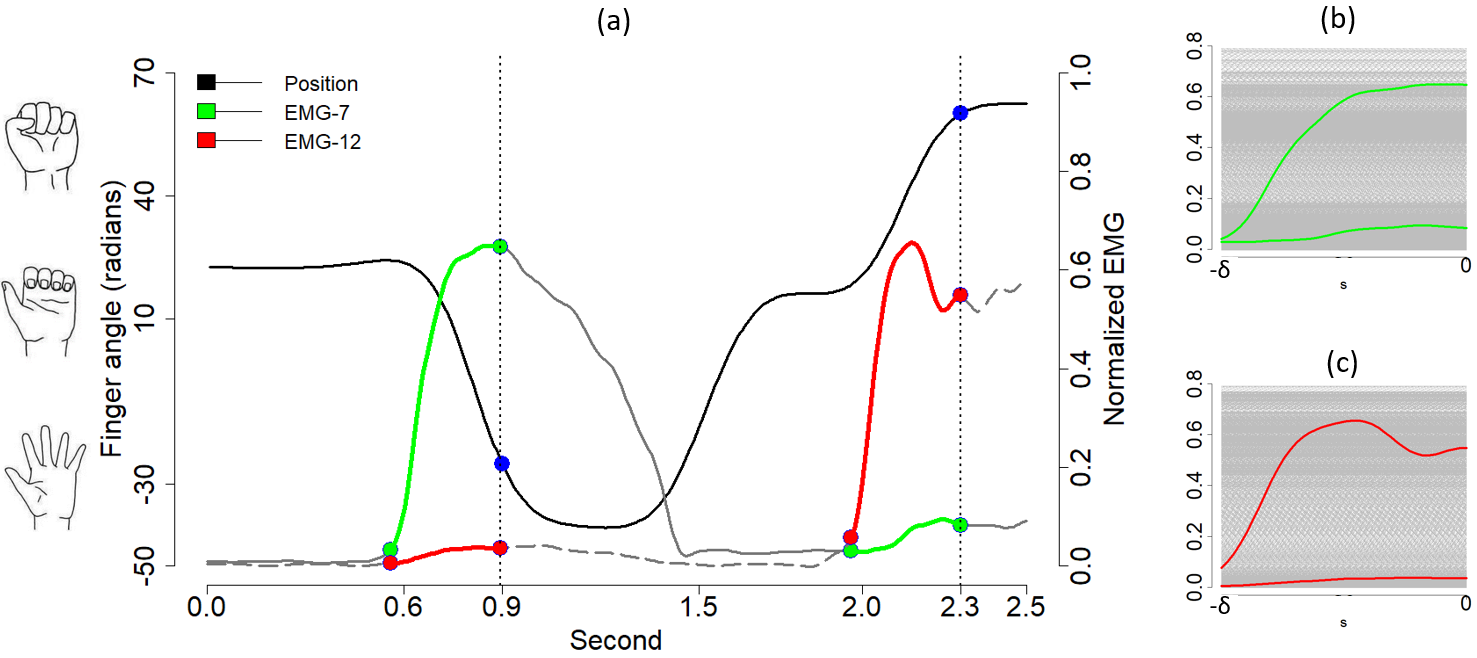}
	\label{Historical} 
\end{figure}

Our idea is to characterize the velocity at an instance $i$ as a function of the recent past EMG signals. Figure \ref{Historical} displays visually the data reconstruction of functional predictors using the recent past EMG signals. 

\subsection{Variation in EMG curves}
\begin{figure}[H]
	\centering
	\caption{\small Restructured EMG curves corresponding to the \textit{extensor digitorum} muscle and the first three eigenfunctions from left to right.}     
	\label{emgHistorical}
	\hspace*{\fill}%
	\includegraphics[width=0.26\textwidth]{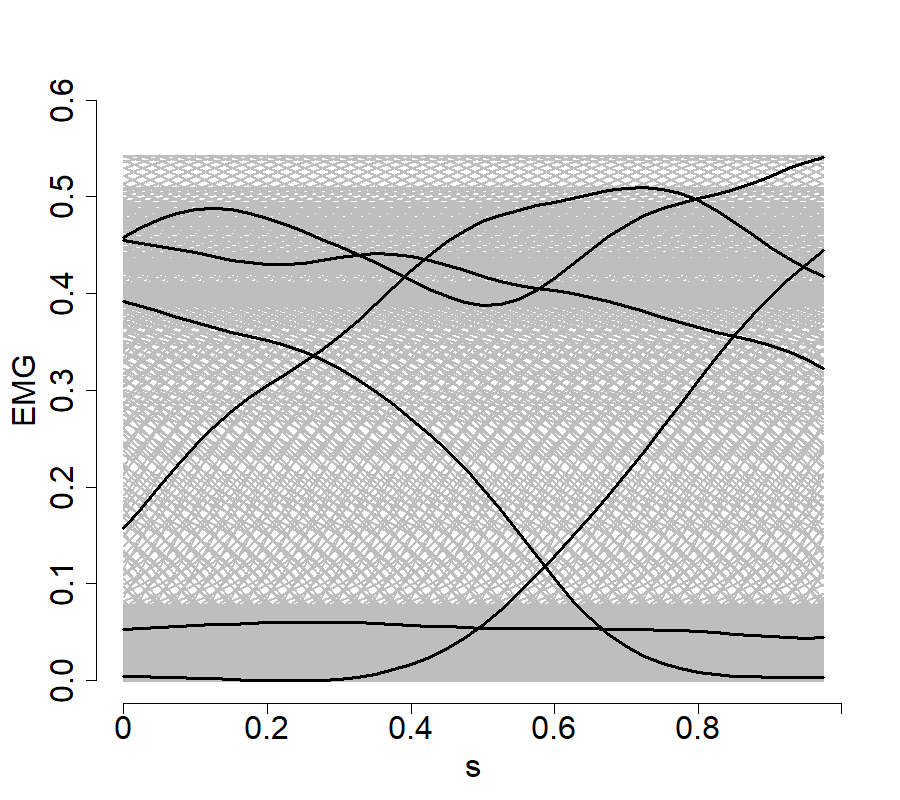}\hfill 
	\includegraphics[width=0.24\textwidth]{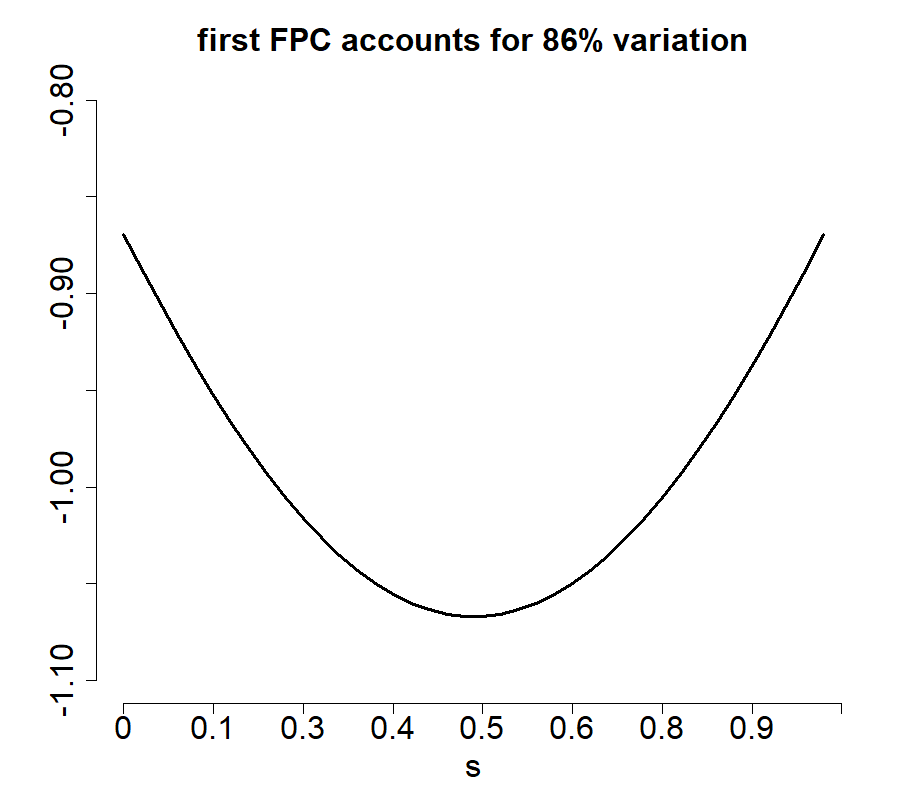}\hfill 
	\includegraphics[width=0.24\textwidth]{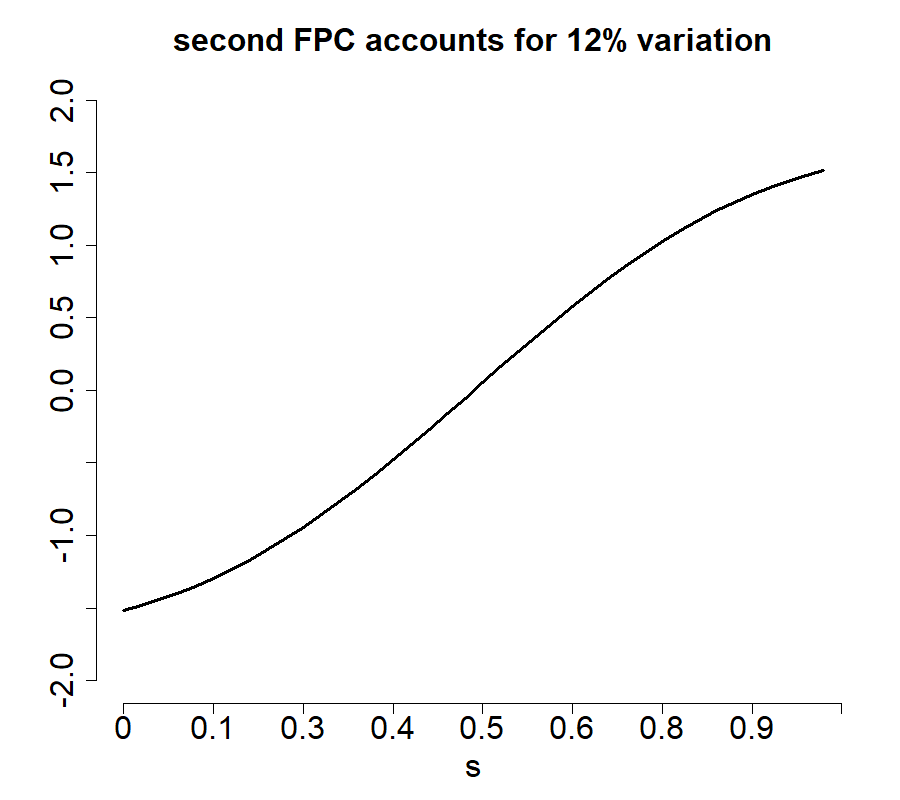}\hfill
	\includegraphics[width=0.24\textwidth]{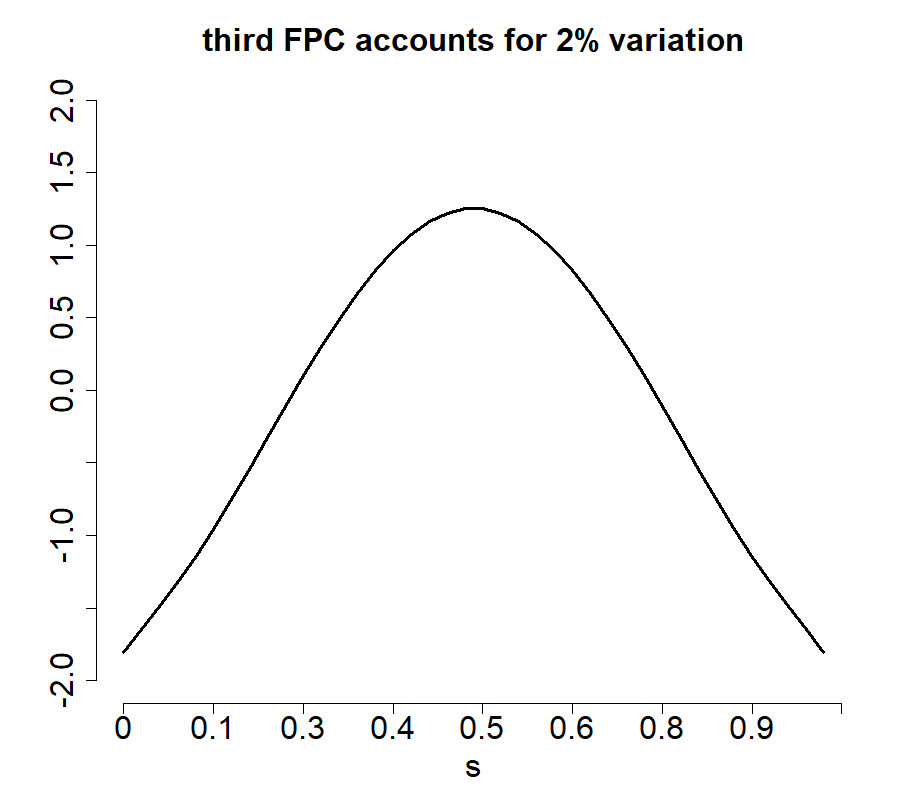}\hfill
\end{figure}

We use functional principal component analysis technique to examine the main sources of variability in the curves.  Figure \ref{emgHistorical} illustrates the EMG curves $X_{k,i}(\cdot)$'s associated with a muscle (\textit{flexor digitorum}) in the forearm that contributes to finger flexion. The first three functional principal components (FPCs) of the estimated marginal covariance of $X_{k,i}(\cdot)$'s are also plotted. We observe one key feature that solely explains the majority of variation in the curves associated with the EMG signal; see the first FPC in Figure \ref{emgHistorical}.

\subsection{Variable selection for wrist movements}\label{Wrist analysis}

Table \ref{WRIST2} shows the results of the model selection performance for the wrist movements with consistent and varying patterns. In most cases, the competing methods select one muscle for wrist extension and another for flexion. As described in the manuscript, there are many potential muscles that contribute to similar wrist movements. Therefore selection between the alike muscles is desired to reduce the redundancy of EMG information.  Notice both agLASSO and SAFE-gLASSO attain optimal RSPs. In contrast FAR-gSCAD shows suboptimal RSPs and LAD-gLASSO performs poorly in selecting desired variables; follow the column ``RSP" for competing methods in Table \ref{WRIST2}. While both these methods have tendency to select noise variables in addition to the true positives, the problem in the selection by LAD-gLASSO is severe; follow the column ``FPR" in Table \ref{WRIST2}.

\begin{table}[H]
	\tiny
	\centering
	\caption{Wrist movements EMG signal selection. RSPs, model size (square brackets), and the percentages of TPRs and FPRs are presented for consistent (top three rows) and random (bottom three row) patterns at different postures. Results with superscript $\dagger$ correspond to the first stage of SAFE-gLASSO.}
	\label{WRIST2}
	\noindent\makebox[\textwidth]{
		\begin{tabular}{ccccccccccccc}
			& \multicolumn{3}{c}{agLASSO} & \multicolumn{3}{c}{LAD-gLASSO} & \multicolumn{3}{c}{FAR-gSCAD} & \multicolumn{3}{c}{SAFE-gLASSO} \\ \cline{2-13} 
			Pattern & RSP & TPR & FPR & RSP & TPR & FPR & RSP & TPR & FPR & RSP & TPR & FPR \\ \cline{2-13} 
			\begin{minipage}{.05\textheight}
				\includegraphics[width=10mm,height=5mm]{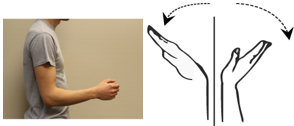}
			\end{minipage}
			& \begin{tabular}[c]{@{}c@{}}1.00\\ {[}2{]}\end{tabular} & \begin{tabular}[c]{@{}c@{}}100\\ {[}2{]}\end{tabular} & \begin{tabular}[c]{@{}c@{}}0\\ {[}0{]}\end{tabular} & \begin{tabular}[c]{@{}c@{}}0.57\\ {[}8{]}\end{tabular} & \begin{tabular}[c]{@{}c@{}}100\\ {[}6{]}\end{tabular} & \begin{tabular}[c]{@{}c@{}}20\\ {[}2{]}\end{tabular} & \begin{tabular}[c]{@{}c@{}}0.92\\ {[}3{]}\end{tabular} & \begin{tabular}[c]{@{}c@{}}100\\ {[}3{]}\end{tabular} & \begin{tabular}[c]{@{}c@{}}0\\ {[}0{]}\end{tabular} & \begin{tabular}[c]{@{}c@{}}0.92\\ {[}3{]}{[}4{]}$^{\dagger}$\end{tabular} & \begin{tabular}[c]{@{}c@{}}100\\ {[}3{]}{[}4{]}$^{\dagger}$\end{tabular} & \begin{tabular}[c]{@{}c@{}}0\\ {[}0{]}{[}0{]}$^{\dagger}$\end{tabular} \\
			\multicolumn{1}{l}{} & \multicolumn{1}{l}{} & \multicolumn{1}{l}{} & \multicolumn{1}{l}{} & \multicolumn{1}{l}{} & \multicolumn{1}{l}{} & \multicolumn{1}{l}{} & \multicolumn{1}{l}{} & \multicolumn{1}{l}{} & \multicolumn{1}{l}{} & \multicolumn{1}{l}{} & \multicolumn{1}{l}{} & \multicolumn{1}{l}{} \\
			\begin{minipage}{.05\textheight}
				\includegraphics[width=10mm,height=5mm]{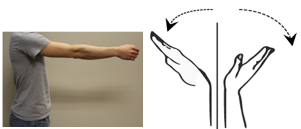}
			\end{minipage}
			& \begin{tabular}[c]{@{}c@{}}1.00\\ {[}2{]}\end{tabular} & \begin{tabular}[c]{@{}c@{}}100\\ {[}2{]}\end{tabular} & \begin{tabular}[c]{@{}c@{}}0\\ {[}0{]}\end{tabular} & \begin{tabular}[c]{@{}c@{}}0.50\\ {[}9{]}\end{tabular} & \begin{tabular}[c]{@{}c@{}}100\\ {[}5{]}\end{tabular} & \begin{tabular}[c]{@{}c@{}}40\\ {[}4{]}\end{tabular} & \begin{tabular}[c]{@{}c@{}}0.85\\ {[}4{]}\end{tabular} & \begin{tabular}[c]{@{}c@{}}100\\ {[}4{]}\end{tabular} & \begin{tabular}[c]{@{}c@{}}0\\ {[}0{]}\end{tabular} & \begin{tabular}[c]{@{}c@{}}1.00\\ {[}2{]}{[}2{]}$^{\dagger}$\end{tabular} & \begin{tabular}[c]{@{}c@{}}100\\ {[}2{]}{[}2{]}$^{\dagger}$\end{tabular} & \begin{tabular}[c]{@{}c@{}}0\\ {[}0{]}{[}0{]}$^{\dagger}$\end{tabular} \\
			\multicolumn{1}{l}{} & \multicolumn{1}{l}{} & \multicolumn{1}{l}{} & \multicolumn{1}{l}{} & \multicolumn{1}{l}{} & \multicolumn{1}{l}{} & \multicolumn{1}{l}{} & \multicolumn{1}{l}{} & \multicolumn{1}{l}{} & \multicolumn{1}{l}{} & \multicolumn{1}{l}{} & \multicolumn{1}{l}{} & \multicolumn{1}{l}{} \\
			\begin{minipage}{.05\textheight}
				\includegraphics[width=10mm,height=5mm]{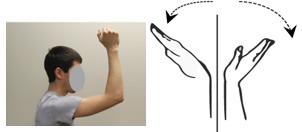}
			\end{minipage}
			& \begin{tabular}[c]{@{}c@{}}1.00\\ {[}2{]}\end{tabular} & \begin{tabular}[c]{@{}c@{}}100\\ {[}2{]}\end{tabular} & \begin{tabular}[c]{@{}c@{}}0\\ {[}0{]}\end{tabular} & \begin{tabular}[c]{@{}c@{}}0.57\\ {[}8{]}\end{tabular} & \begin{tabular}[c]{@{}c@{}}100\\ {[}3{]}\end{tabular} & \begin{tabular}[c]{@{}c@{}}50\\ {[}5{]}\end{tabular} & \begin{tabular}[c]{@{}c@{}}0.92\\ {[}3{]}\end{tabular} & \begin{tabular}[c]{@{}c@{}}100\\ {[}3{]}\end{tabular} & \begin{tabular}[c]{@{}c@{}}0\\ {[}0{]}\end{tabular} & \begin{tabular}[c]{@{}c@{}}1.00\\ {[}2{]}{[}2{]}$^{\dagger}$\end{tabular} & \begin{tabular}[c]{@{}c@{}}100\\ {[}2{]}{[}2{]}$^{\dagger}$\end{tabular} & \begin{tabular}[c]{@{}c@{}}0\\ {[}0{]}{[}0{]}$^{\dagger}$\end{tabular} \\
			\multicolumn{1}{l}{} & \multicolumn{1}{l}{} & \multicolumn{1}{l}{} & \multicolumn{1}{l}{} & \multicolumn{1}{l}{} & \multicolumn{1}{l}{} & \multicolumn{1}{l}{} & \multicolumn{1}{l}{} & \multicolumn{1}{l}{} & \multicolumn{1}{l}{} & \multicolumn{1}{l}{} & \multicolumn{1}{l}{} & \multicolumn{1}{l}{} \\
			\begin{minipage}{.05\textheight}
				\includegraphics[width=10mm,height=5mm]{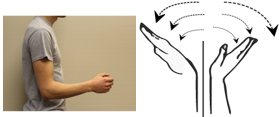}
			\end{minipage}
			& \begin{tabular}[c]{@{}c@{}}1.00\\ {[}2{]}\end{tabular} & \begin{tabular}[c]{@{}c@{}}100\\ {[}2{]}\end{tabular} & \begin{tabular}[c]{@{}c@{}}0\\ {[}0{]}\end{tabular} & \begin{tabular}[c]{@{}c@{}}0.57\\ {[}8{]}\end{tabular} & \begin{tabular}[c]{@{}c@{}}100\\ {[}4{]}\end{tabular} & \begin{tabular}[c]{@{}c@{}}40\\ {[}4{]}\end{tabular} & \begin{tabular}[c]{@{}c@{}}0.92\\ {[}3{]}\end{tabular} & \begin{tabular}[c]{@{}c@{}}100\\ {[}2{]}\end{tabular} & \begin{tabular}[c]{@{}c@{}}10\\ {[}1{]}\end{tabular} & \begin{tabular}[c]{@{}c@{}}1.00\\ {[}2{]}{[}2{]}$^{\dagger}$\end{tabular} & \begin{tabular}[c]{@{}c@{}}100\\ {[}2{]}{[}2{]}$^{\dagger}$\end{tabular} & \begin{tabular}[c]{@{}c@{}}0\\ {[}0{]}{[}0{]}$^{\dagger}$\end{tabular} \\
			\multicolumn{1}{l}{} & \multicolumn{1}{l}{} & \multicolumn{1}{l}{} & \multicolumn{1}{l}{} & \multicolumn{1}{l}{} & \multicolumn{1}{l}{} & \multicolumn{1}{l}{} & \multicolumn{1}{l}{} & \multicolumn{1}{l}{} & \multicolumn{1}{l}{} & \multicolumn{1}{l}{} & \multicolumn{1}{l}{} & \multicolumn{1}{l}{} \\
			\begin{minipage}{.05\textheight}
				\includegraphics[width=10mm,height=5mm]{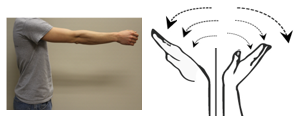}
			\end{minipage}
			& \begin{tabular}[c]{@{}c@{}}1.00\\ {[}2{]}\end{tabular} & \begin{tabular}[c]{@{}c@{}}100\\ {[}2{]}\end{tabular} & \begin{tabular}[c]{@{}c@{}}0\\ {[}0{]}\end{tabular} & \begin{tabular}[c]{@{}c@{}}0.72\\ {[}6{]}\end{tabular} & \begin{tabular}[c]{@{}c@{}}100\\ {[}4{]}\end{tabular} & \begin{tabular}[c]{@{}c@{}}20\\ {[}2{]}\end{tabular} & \begin{tabular}[c]{@{}c@{}}0.92\\ {[}3{]}\end{tabular} & \begin{tabular}[c]{@{}c@{}}100\\ {[}3{]}\end{tabular} & \begin{tabular}[c]{@{}c@{}}0\\ {[}0{]}\end{tabular} & \begin{tabular}[c]{@{}c@{}}1.00\\ {[}2{]}{[}3{]}$^{\dagger}$\end{tabular} & \begin{tabular}[c]{@{}c@{}}100\\ {[}2{]}{[}2{]}$^{\dagger}$\end{tabular} & \begin{tabular}[c]{@{}c@{}}0\\ {[}0{]}{[}1{]}$^{\dagger}$\end{tabular} \\
			\multicolumn{1}{l}{} & \multicolumn{1}{l}{} & \multicolumn{1}{l}{} & \multicolumn{1}{l}{} & \multicolumn{1}{l}{} & \multicolumn{1}{l}{} & \multicolumn{1}{l}{} & \multicolumn{1}{l}{} & \multicolumn{1}{l}{} & \multicolumn{1}{l}{} & \multicolumn{1}{l}{} & \multicolumn{1}{l}{} & \multicolumn{1}{l}{} \\
			\begin{minipage}{.05\textheight}
				\includegraphics[width=10mm,height=5mm]{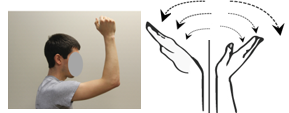}
			\end{minipage}
			& \begin{tabular}[c]{@{}c@{}}1.00\\ {[}2{]}\end{tabular} & \begin{tabular}[c]{@{}c@{}}100\\ {[}2{]}\end{tabular} & \begin{tabular}[c]{@{}c@{}}0\\ {[}0{]}\end{tabular} & \begin{tabular}[c]{@{}c@{}}0.50\\ {[}9{]}\end{tabular} & \begin{tabular}[c]{@{}c@{}}100\\ {[}6{]}\end{tabular} & \begin{tabular}[c]{@{}c@{}}30\\ {[}3{]}\end{tabular} & \begin{tabular}[c]{@{}c@{}}0.92\\ {[}3{]}\end{tabular} & \begin{tabular}[c]{@{}c@{}}100\\ {[}3{]}\end{tabular} & \begin{tabular}[c]{@{}c@{}}0\\ {[}0{]}\end{tabular} & \begin{tabular}[c]{@{}c@{}}0.92\\ {[}3{]}{[}4{]}$^{\dagger}$\end{tabular} & \begin{tabular}[c]{@{}c@{}}100\\ {[}3{]}{[}3{]}$^{\dagger}$\end{tabular} & \begin{tabular}[c]{@{}c@{}}0\\ {[}0{]}{[}1{]}$^{\dagger}$\end{tabular} \\ \cline{2-13} 
		\end{tabular}
	}
\end{table}

\subsection{Residual diagnostics}\label{residual_diagnostic}
Reported are the results based on a varying finger movement data. We also carry out the residual diagnostics to check the statistical assumptions for other data sets in a similar fashion, and hence we omit it here.

\begin{figure}[H]
	\centering
	\caption{Diagnostics of the residual assumption of the statistical model for kinematic data; scatter plot of the residuals against predicted values (left), histogram of the residuals (middle), and normal Q-Q plot (right).}     
	\label{residual}
	\includegraphics[width=0.35\textwidth]{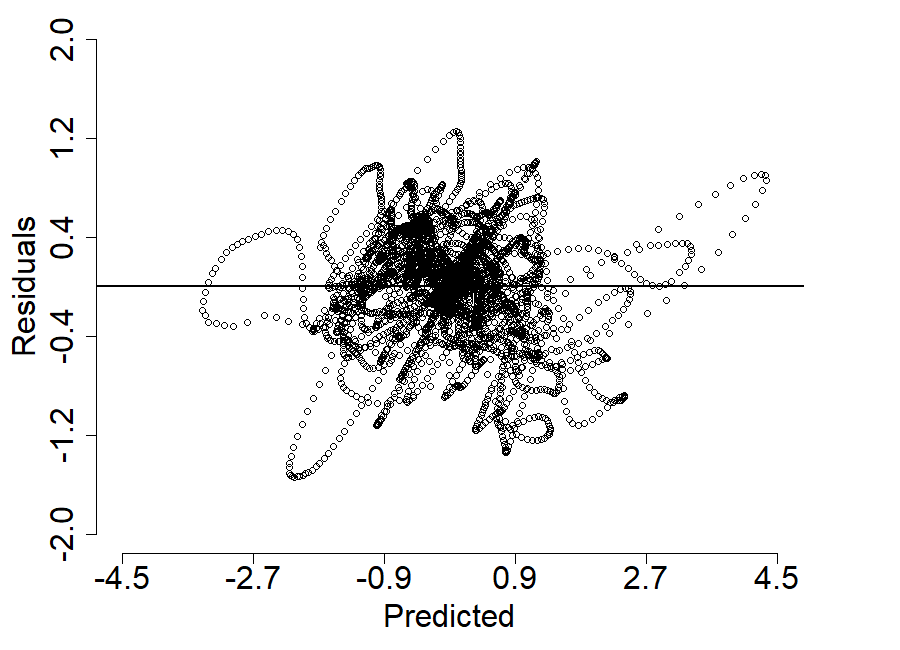}
	\includegraphics[width=0.35\textwidth]{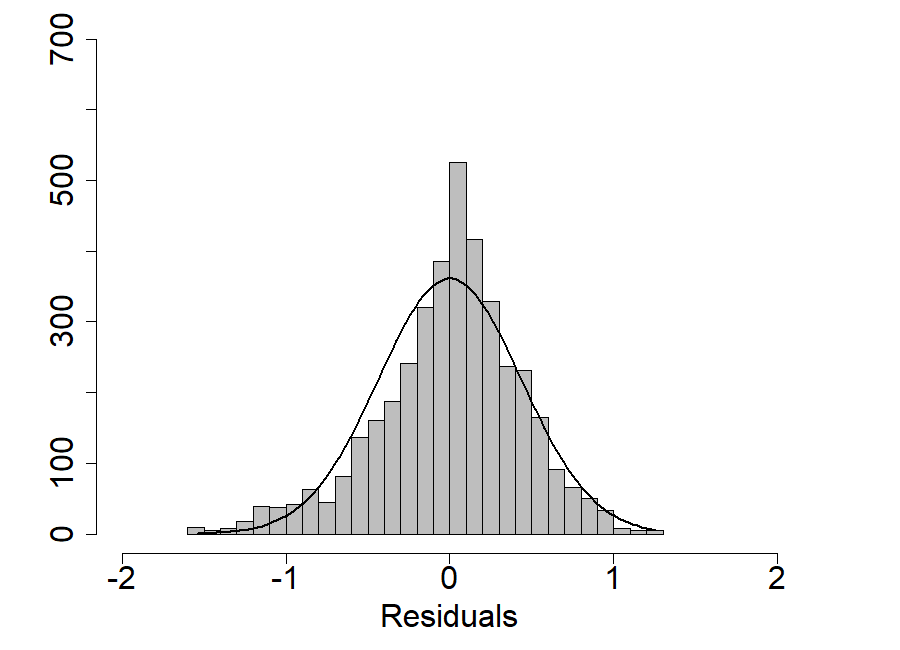}
	\includegraphics[width=0.35\textwidth]{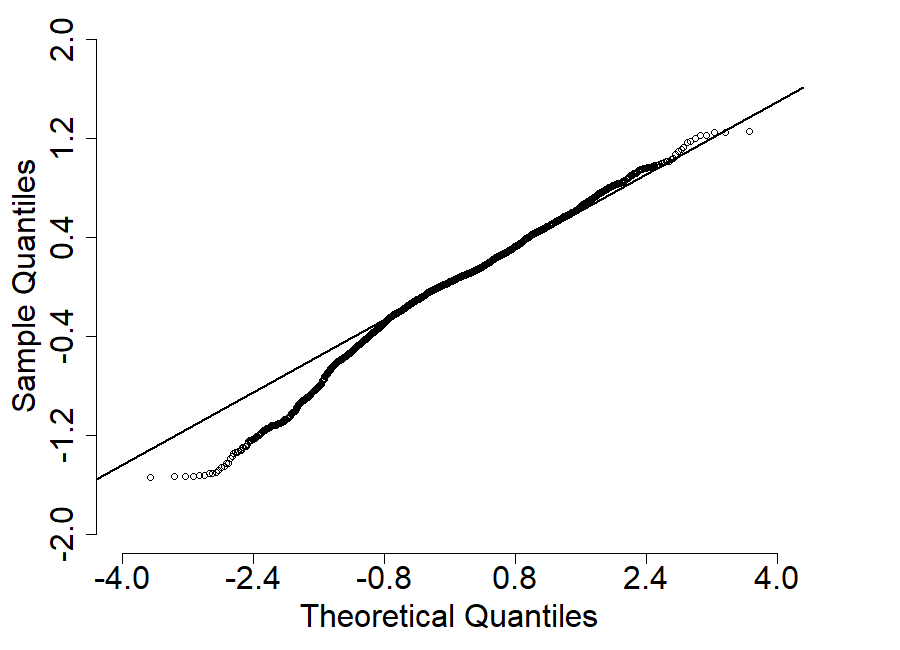}\\ 
\end{figure}

The left panel of Figure \ref{residual} checks the constant error variance assumption for the model (4.1). Let the residual at an instance $i$ be $r_i = y_i - \widehat y_i;$ where $y_i$ and $\widehat y_i$ are the observed and predicted responses, respectively. Notice
the residuals ``bounce randomly" around zero and there is no systematic structure. In addition, no one residual ``stands out" from the underlying random pattern implying the absence of outliers.  The middle and right
panels of Figure \ref{residual} assess the normality assumption of the model via the density curve and \textit{Q-Q} plot based on the residuals. 

\begin{figure}[H]
	\centering
	\caption{Autocorrelation (left) and partial autocorrelation (right) of the residuals corresponding to the varying finger movements.}     
	\label{LAG}
	\includegraphics[width=0.5\textwidth]{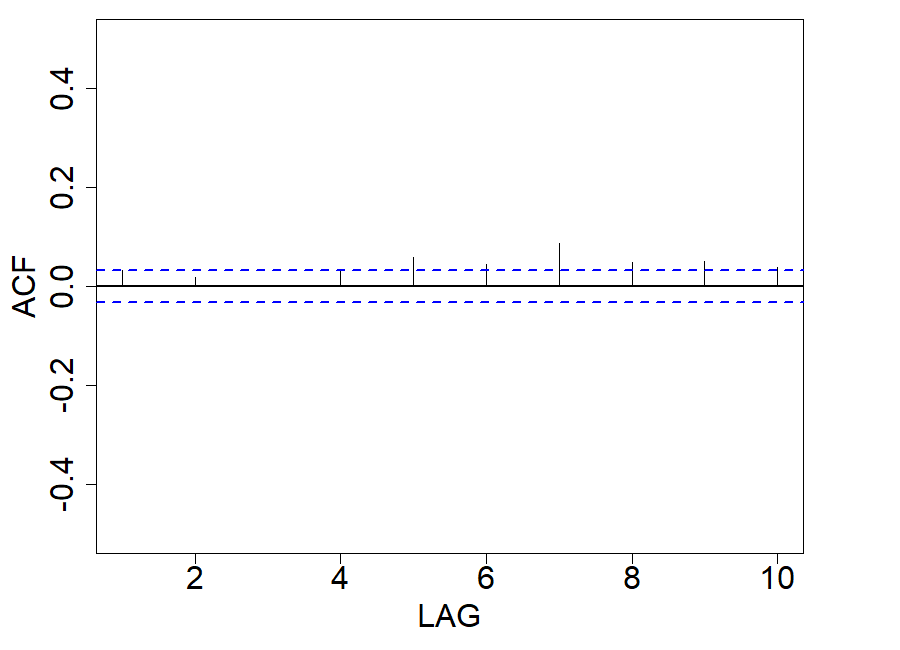}\hfill 
	\includegraphics[width=0.5\textwidth]{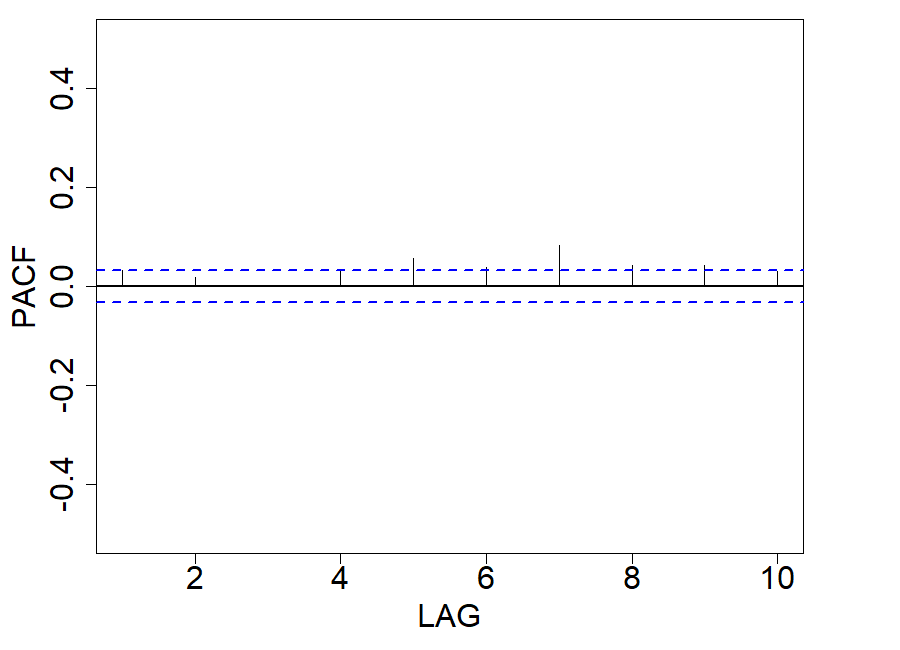}
\end{figure}

Figure \ref{LAG} illustrates the underlying dependence structure in the data. We use the auto-correlation function (ACF) and the partial ACF (PACF) to measure the strength of correlation between the successive measurements. We consider the mean model (4.7) and examine the auto-regressive patterns of the residuals. We do not observe any significant autocorrelation in Figure \ref{LAG}. 


%


\subsection{Regression surfaces for finger movements}\label{regression_surface_finger}

\begin{figure}[H]
	\centering
	\caption{Postures for finger/wrist extension and flexion.}     
	\label{postures}
	\includegraphics[width=0.5\textwidth]{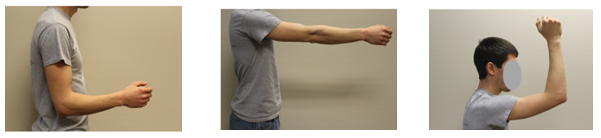}
\end{figure}

In this section, we interpret the regression surfaces corresponding to the finger movements with consistent and varying patterns; Figure \ref{postures} illustrates the corresponding postures. Reported are the estimates based on the second stage of SAFE-gLASSO.

\begin{figure}[H]
	\centering
	\caption{Regression surface for finger flexion (left) and extension (right). Reported are the estimates corresponding to consistent finger movements with neutral posture depicted in the left panel of Figure \ref{postures}.}     
	\label{hand1}
	\noindent\makebox[\textwidth]{
		\includegraphics[width=1\textwidth]{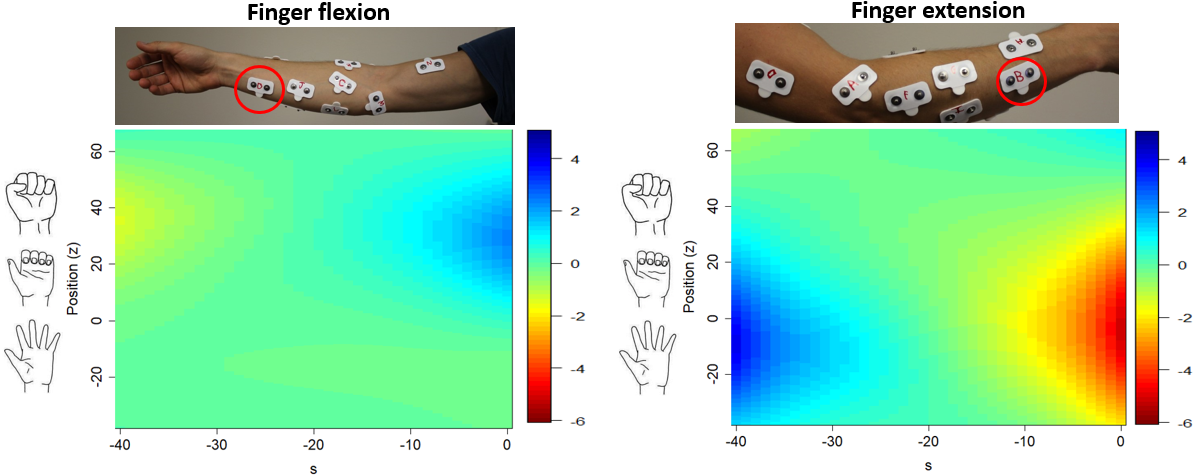}
	}
\end{figure}

\begin{figure}[H]
	\centering
	\caption{Regression surfaces for finger flexion (left) and extension (right). Reported are the estimates corresponding to consistent finger movements with posture depicted in the right panel of Figure \ref{postures}.}     
	\label{hand2}
	\noindent\makebox[\textwidth]{
		\includegraphics[width=1\textwidth]{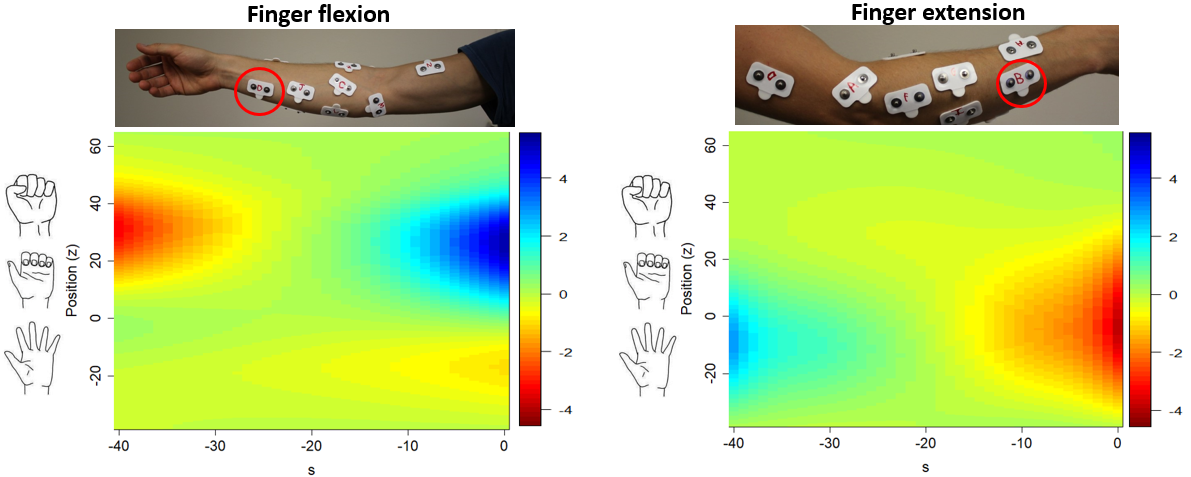}
	}
\end{figure}

\begin{figure}[H]
	\centering
	\caption{Regression surfaces for finger flexion (left) and extension (right). Reported are the estimates corresponding to varying finger movements with neutral posture depicted in the left panel of Figure \ref{postures}.}     
	\label{hand3}
	\noindent\makebox[\textwidth]{
		\includegraphics[width=1\textwidth]{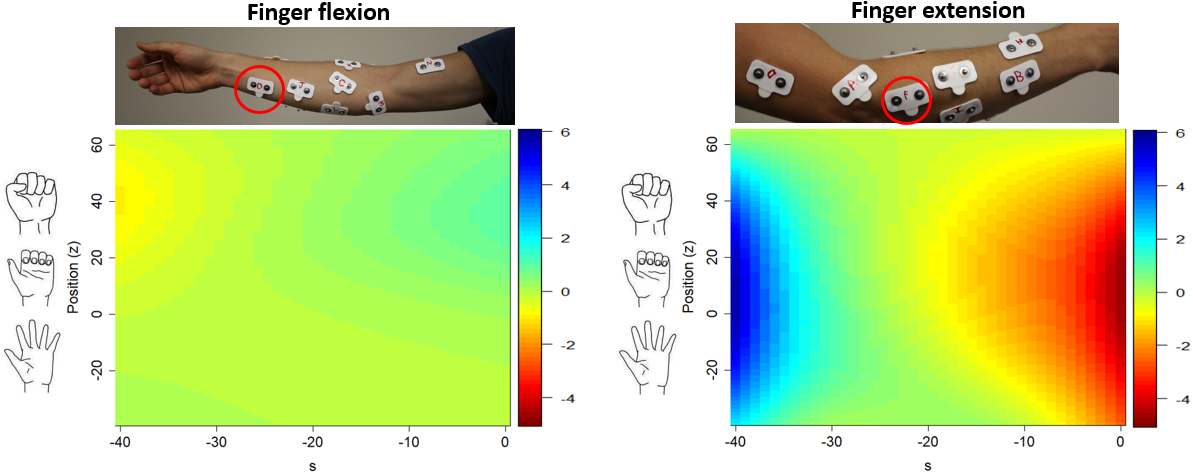}
	}
\end{figure}

\begin{figure}[H]
	\centering
	\caption{Regression surface for finger flexion (left) and extension (right). Reported are the estimates corresponding to varying finger movements with posture depicted in the middle panel of Figure \ref{postures}.}     
	\label{hand5}
	\noindent\makebox[\textwidth]{
		\includegraphics[width=1\textwidth]{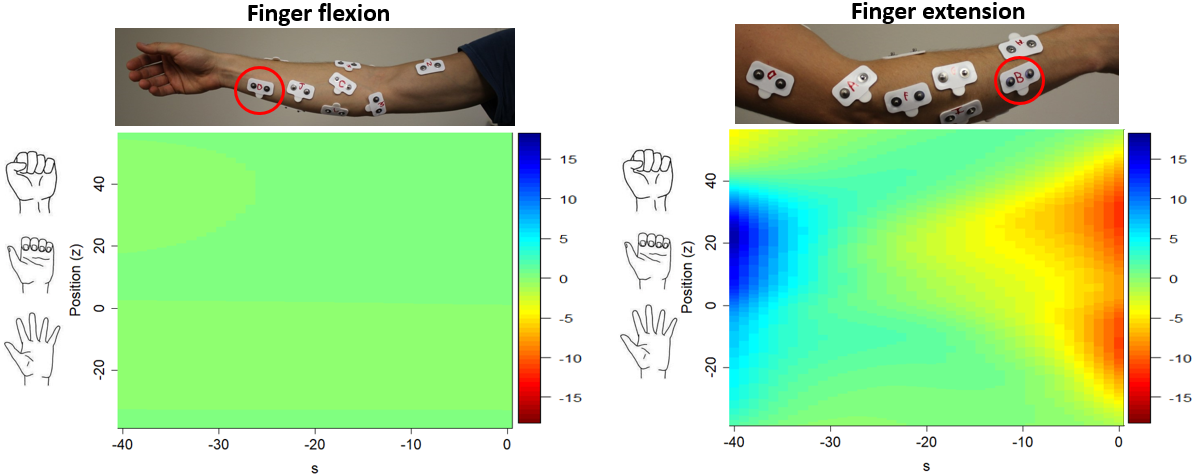}
	}
\end{figure}

\begin{figure}[H]
	\centering
	\caption{Regression surfaces for finger flexion (left) and extension (right). Reported are the estimates corresponding to varying finger movements with posture depicted in the right panel of Figure \ref{postures}.}     
	\label{hand6}
	\noindent\makebox[\textwidth]{
		\includegraphics[width=1\textwidth]{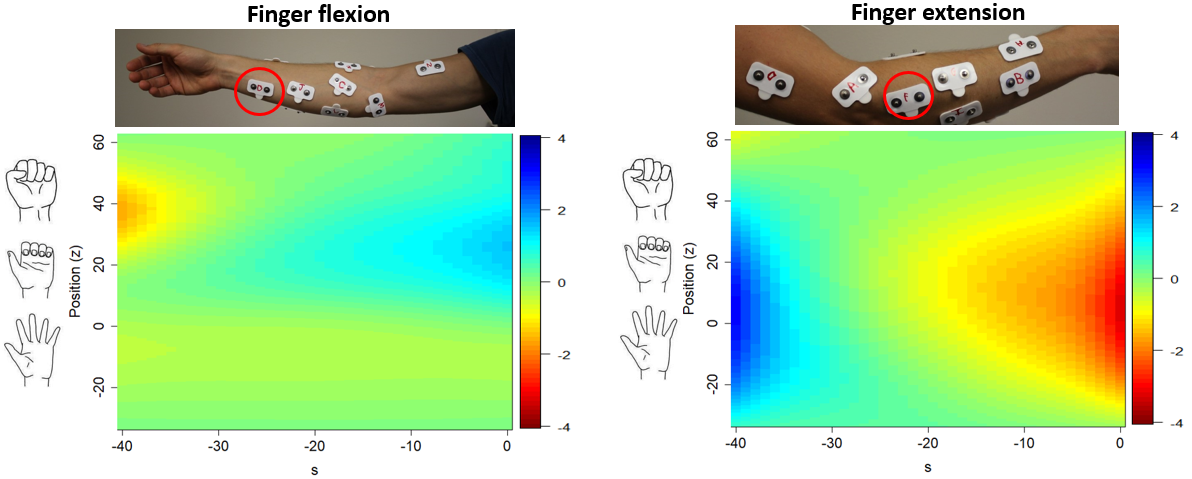}
	}
\end{figure}

\begin{itemize}
	\item[$\bullet$]  Figure \ref{hand1} corresponds to the neutral posture demonstrated in the left panel of Figure \ref{postures}. SAFE-gLASSO selects two muscles: \textit{extensor digitorum} (muscle 5) and \textit{flexor digitorum} (muscle 12). The interpretation of the \textit{extensor digitorum}'s (right panel) activity on finger extension is as same as before;  the impact is most relevant between  $-30$ and $20$ radians which are primarily associated with finger opening. On the other hand, the \textit{flexsor digitorum} is active for the positions $20$ and $40$ radians which correspond to making a fist. As before, the muscle mechanism at the past time window depicts the opposite features  to that of the concurrent relationship and specifically describes the involuntary movements due to the passive forces when finger is fully flexed (left) and extended (right).

	\item[$\bullet$] Figure \ref{hand2}  corresponds to the posture demonstrated in the right panel of Figure \ref{postures}. Both stages of the SAFE-gLASSO select two muscles: one is \textit{extensor digitorum} (muscle 5) and the other is \textit{flexor digitorum} (muscle 12). The impact of the left surface on velocity is most important for positions between $10$ and $60$ radians which correspond to finger closing.  Similarly, the impact of the right surface is most relevant between $-30$ and $20$ radians which are primarily associated with finger opening.  The corresponding historical relationship follows the same intuition as before. 
	
	\item[$\bullet$] Figure \ref{hand3}, Figure \ref{hand5}, and Figure \ref{hand6} correspond to the postures demonstrated in the left, middle, and right panel of Figure \ref{postures}, respectively. The interpretation of the corresponding regression surfaces follows the same idea as before. Notice that the smooth estimates in the left surfaces for Figure \ref{hand3} and Figure \ref{hand5} are not zero.  Here one important specific is that the magnitude of the effect of \textit{flexor digitorum} is considerably smaller than than that of the \textit{extensor digitorum}. This happens due to  
	the fact that user focuses primarily on opening the finger at varying pattern and does not make the fist properly. This is not entirely surprising since the pattern at these postures is not directive by the experimenter but solely determined by the user.

	
\end{itemize} 

\subsection{Regression surfaces for wrist movements}\label{regression_surface_wrist}

\begin{figure}[H]
	\centering
	\caption{Regression surfaces for wrist flexion (left) and extension (right). Reported are the estimates corresponding to consistent wrist movements with neutral posture depicted in the left panel of Figure \ref{postures}.}     
	\label{wrist1}
	\noindent\makebox[\textwidth]{
		\includegraphics[width=1\textwidth]{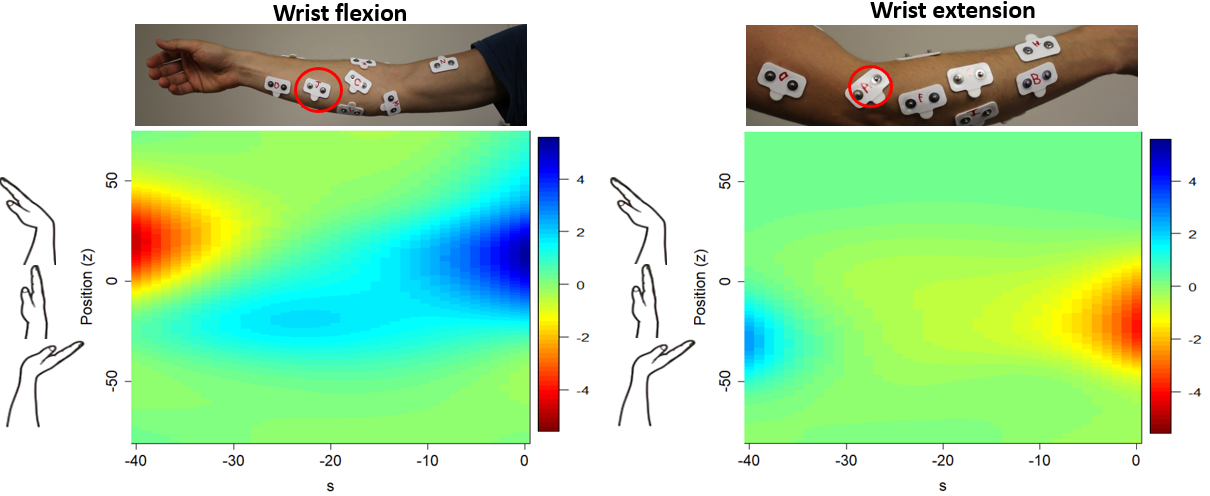}
	}
\end{figure}

\begin{figure}[H]
	\centering
	\caption{Regression surfaces for wrist flexion (left) and extension (right). Reported are the estimates corresponding to consistent wrist movements with posture depicted in the right panel of Figure \ref{postures}.}     
	\label{wrist3}
	\noindent\makebox[\textwidth]{
		\includegraphics[width=1\textwidth]{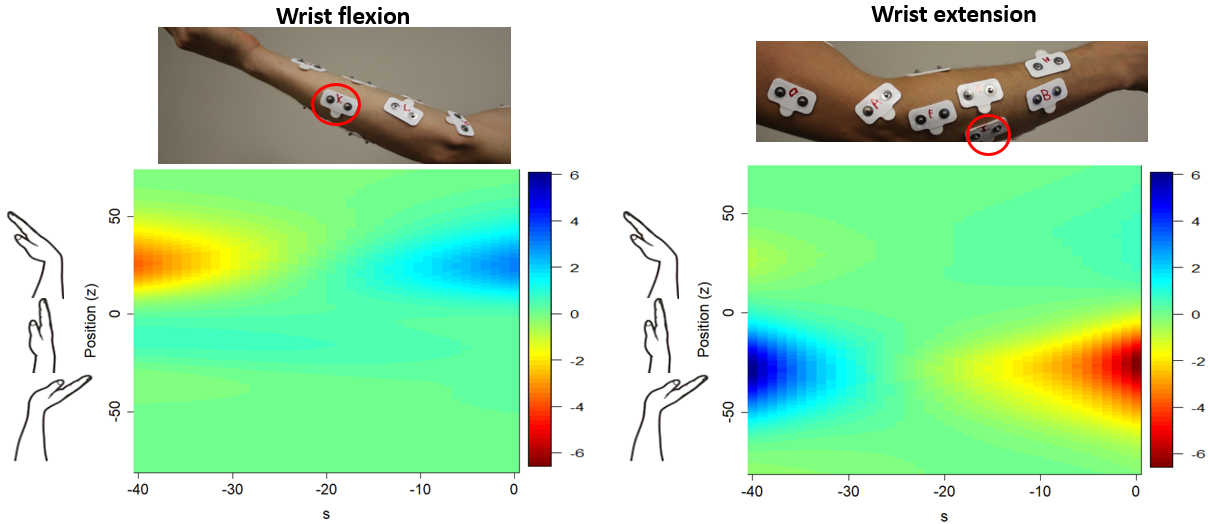}
	}
\end{figure}

\begin{figure}[H]
	\centering
	\caption{Regression surfaces for wrist flexion (left) and extension (right). Reported are the estimates corresponding to varying wrist movements with neutral posture depicted in the left panel of Figure \ref{postures}.}     
	\label{wrist4}
	\noindent\makebox[\textwidth]{
		\includegraphics[width=1\textwidth]{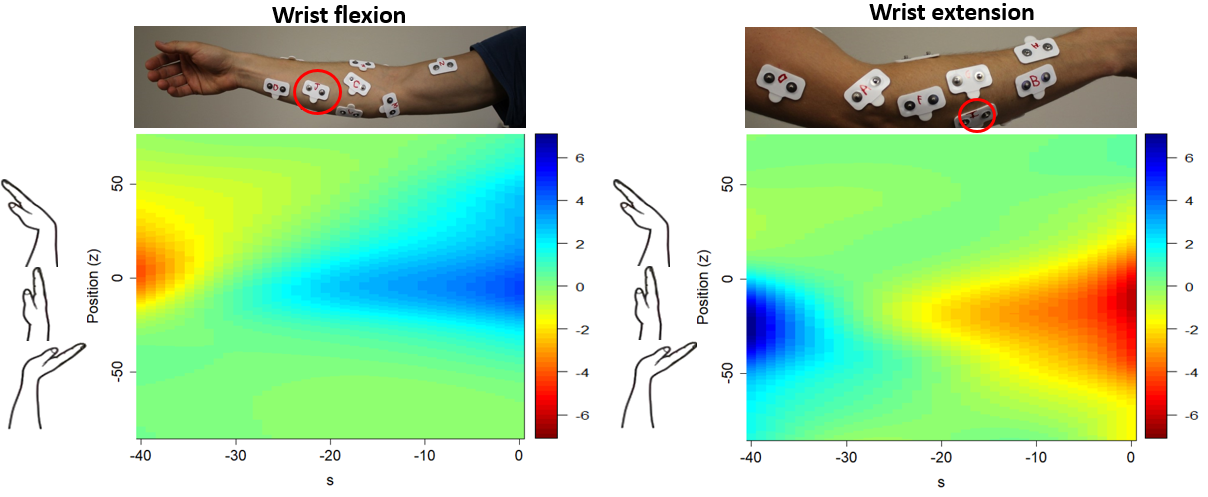}
	}
\end{figure}

\begin{figure}[H]
	\centering
	\caption{Regression surfaces for wrist flexion (left) and extension (right). Reported are the estimates corresponding to varying wrist movements with posture depicted in the middle panel of Figure \ref{postures}.}     
	\label{wrist5}
	\noindent\makebox[\textwidth]{
		\includegraphics[width=1\textwidth]{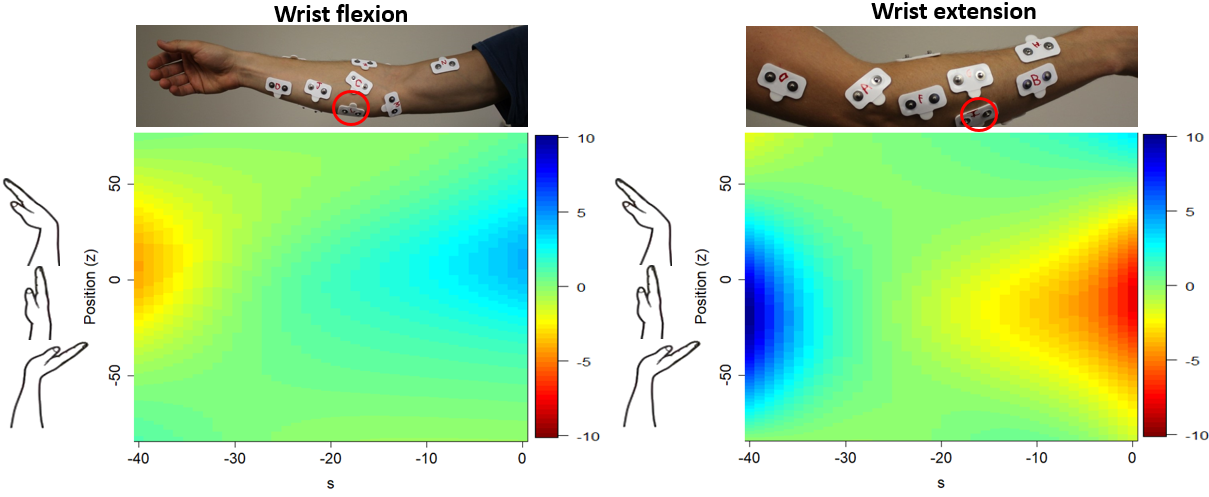}
	}
\end{figure}

\begin{figure}[H]
	\centering
	\caption{Regression surfaces for wrist flexion (left) and extension (right). Reported are the estimates corresponding to varying wrist movements with posture depicted in the right panel of Figure \ref{postures}.}     
	\label{wrist6}
	\noindent\makebox[\textwidth]{
		\includegraphics[width=1\textwidth]{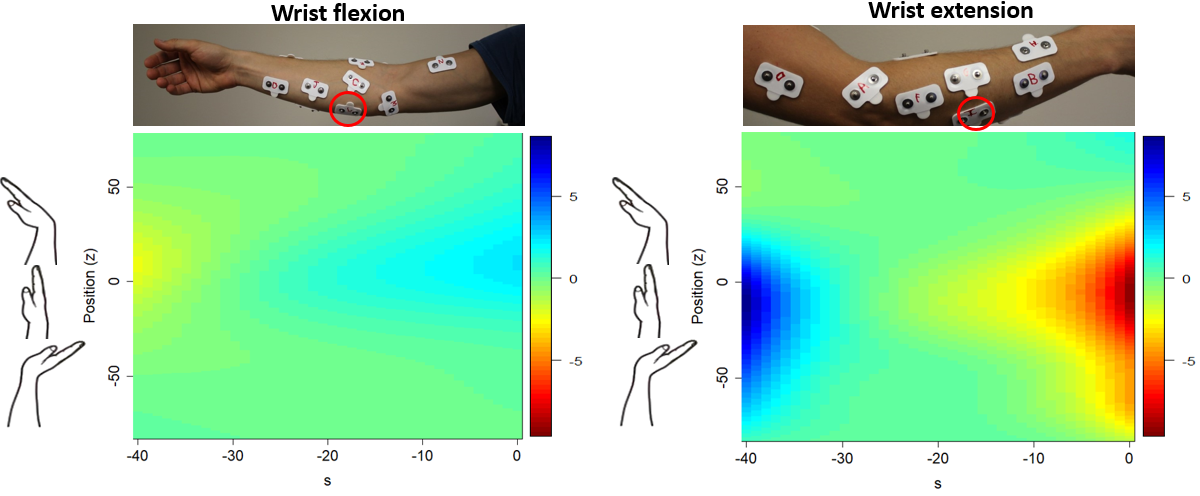}
	}
\end{figure}

In this section, we plot the regression surfaces corresponding to wrist movements based on the stage of SAFE-gLASSO.
\begin{itemize}
	\item[$\bullet$]  Figure \ref{wrist1} corresponds to the neutral posture demonstrated in the left panel of Figure \ref{postures}. The second stage of SAFE-gLASSO select \textit{flexor carpi ulnaris}, \textit{flexor digitorum superficialis}, and \textit{extensor carpi radialis longus}. While the first two muscles contribute for wrist flexion, the last one leads to extension. Since the regression surfaces associated with both \textit{flexor carpi ulnaris} and \textit{flexor digitorum superficialis} are similar, we report and interpret the effect of  \textit{flexor carpi ulnaris} only. The impact of this muscle on velocity is most important for positions between $5$ and $50$ radians. We also notice some activity of this muscle when the wrist is in the neutral state, say around $-20$ and $0$ radians. This says that the \textit{flexor carpi ulnaris} needs to be flexed to hold the wrist is in the upright position. In contrast, the impact of the \textit{extensor carpi radialis longus} is most relevant between positions $-50$ and $-10$ radians which are primarily associated with wrist extension. The corresponding muscles have opposite historical relationship which in particular quantifies the movements due to the physical constraints and passive forces.   
	
	\item[$\bullet$]  Figure \ref{wrist3} corresponds to the posture demonstrated in the right panel of Figure \ref{postures}. Unlike previous case, the impact of the \textit{flexor carpi ulnaris} on velocity is most important for angles between $10$ and $50$ radians which correspond to wrist flexion. In contrast, the impact of the \textit{extensor carpi ulnaris
	} (right surface) is most relevant between positions $-50$ and $-20$ radians which are primarily associated with wrist extension.

	\item[$\bullet$]  Figure \ref{wrist4}, Figure \ref{wrist5}, and Figure \ref{wrist6} correspond to the postures demonstrated in the left, middle, and right panel of Figure \ref{postures}. The interpretation of the muscle mechanism follows the same intuition as before and hence we omit it here.
	
\end{itemize}

\subsection{Regression surfaces based on competing approaches}\label{regression_surface_finger_agLASSO}

\begin{figure}[H]
	\centering
	\caption{\small Estimates of smooth coefficients for the selected muscles:  \textit{flexor digitorum} (left) and \textit{extensor digitorum} (right) corresponding to agLASSO (top panels) and LAD-gLASSO (bottom panels).}     
	\label{surface}
	\includegraphics[width=0.6\textwidth]{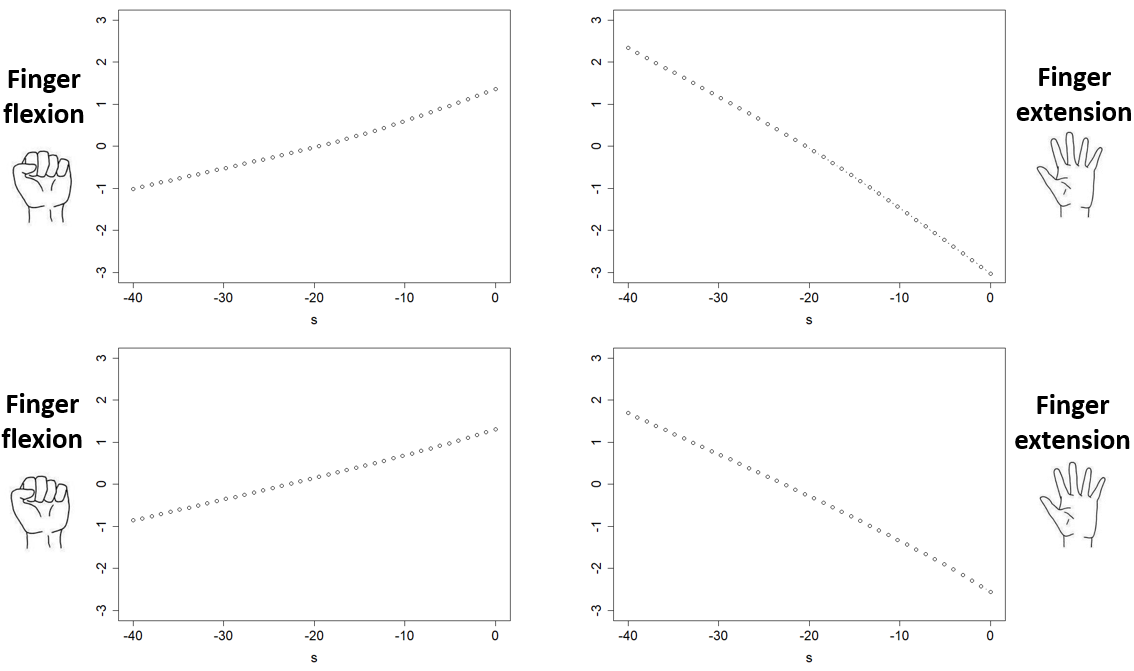}
\end{figure}

Figure \ref{surface} plots the regression surfaces based on the competing approaches for the finger movements with fixed pattern (i.e., second row in Table 1). We notice that the effect of the selected EMG signals on velocity is linear in both agLASSO and LAD-gLASSO. Concurrently, say at $s = 0$, the effect of the \textit{flexor digitorum} is negative to velocity while the effect of the \textit{extensor digitorum} is positive. Historically, say at $s = -40,$ they imply the opposite effects which matches with the surface interpreted by SAFE-gLASSO. While agLASSO and SAFE-gLASSO account for passive forces, they fail to acknowledge the changes in the regression surfaces across different finger positions. In particular, they implicitly assume that the muscle mechanism is same irrespective of the finger opening/closing positions which is indeed unrealistic. As pointed out in \cite{james2009functional}, the smooth estimates based on FAR-gSCAD is difficult to interpret due to the issue of identifiability; and hence we omit it here. 

\section{Additional simulation results for the data mimicking application}\label{validation_EMG}

\begin{figure}[]
	\centering
	\caption{\small Simulated velocity across different simulation settings at SNR = $\{875, 80, 8, 4, 0.8\}.$ Data correspond to non-varying finger movements.}     
	\label{EMG 7&12}  
	\hspace*{\fill}%
	\includegraphics[width=0.33\textwidth]{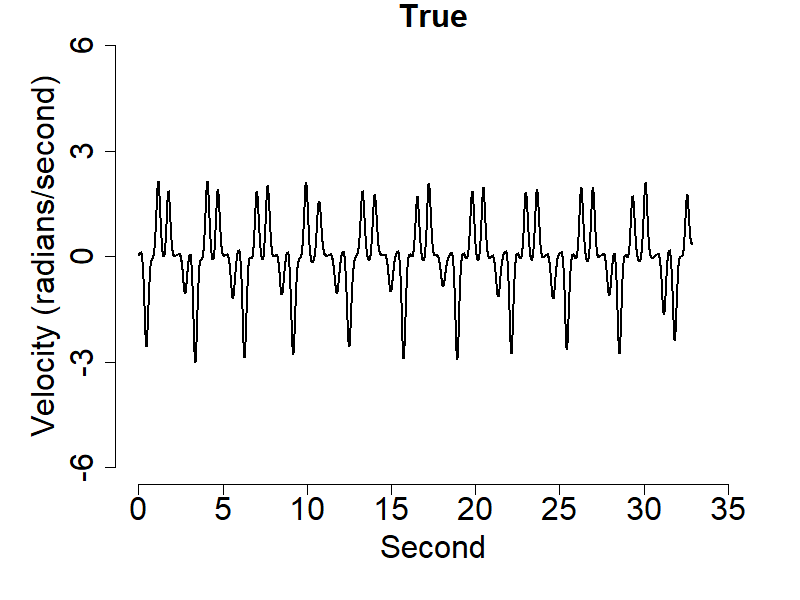}\hfill 
	\includegraphics[width=0.33\textwidth]{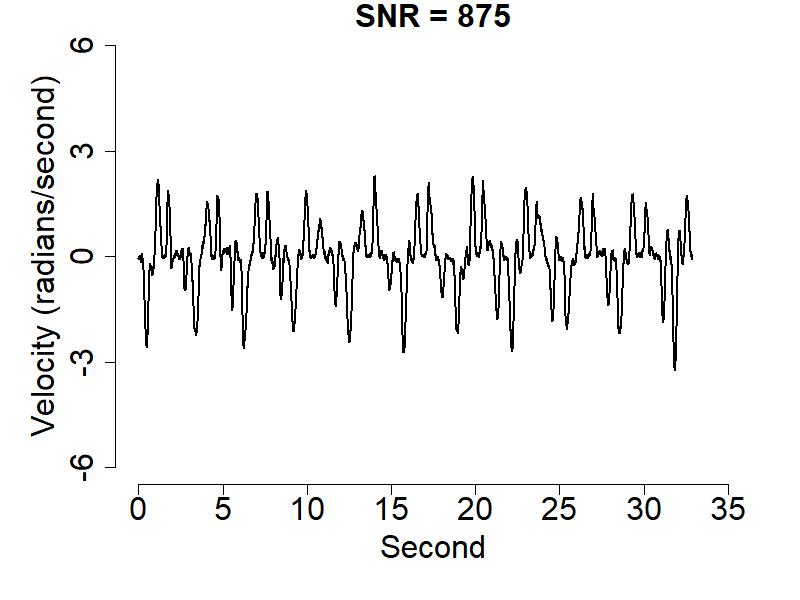}\hfill 
	\includegraphics[width=0.33\textwidth]{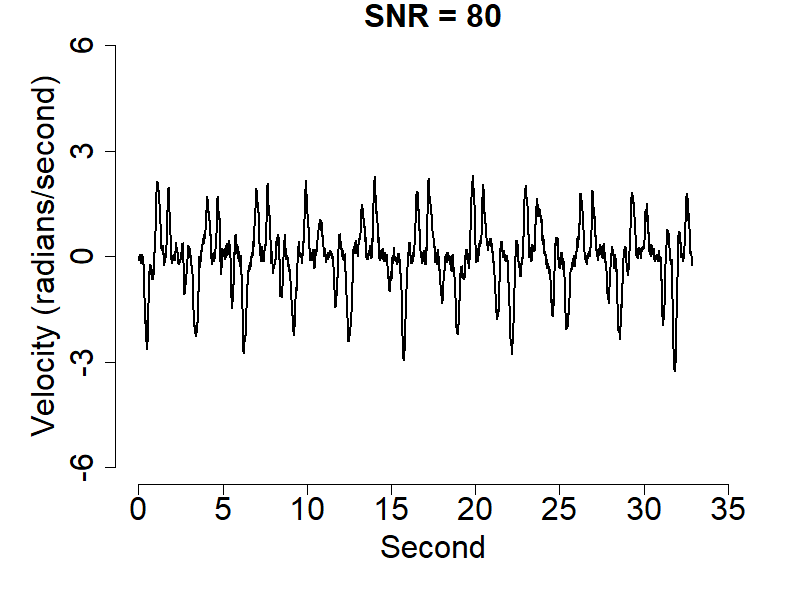}\hfill
	\hspace*{\fill}\\
	\includegraphics[width=0.33\textwidth]{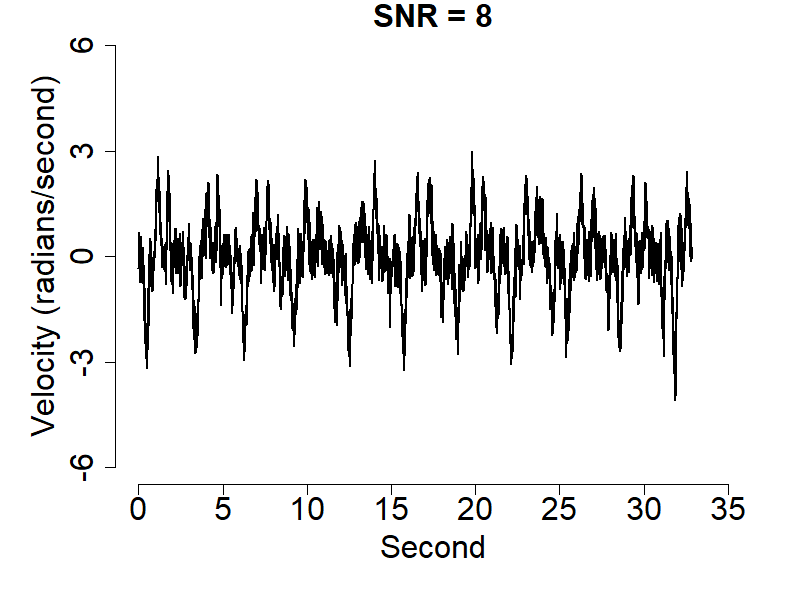}\hfill
	\includegraphics[width=0.33\textwidth]{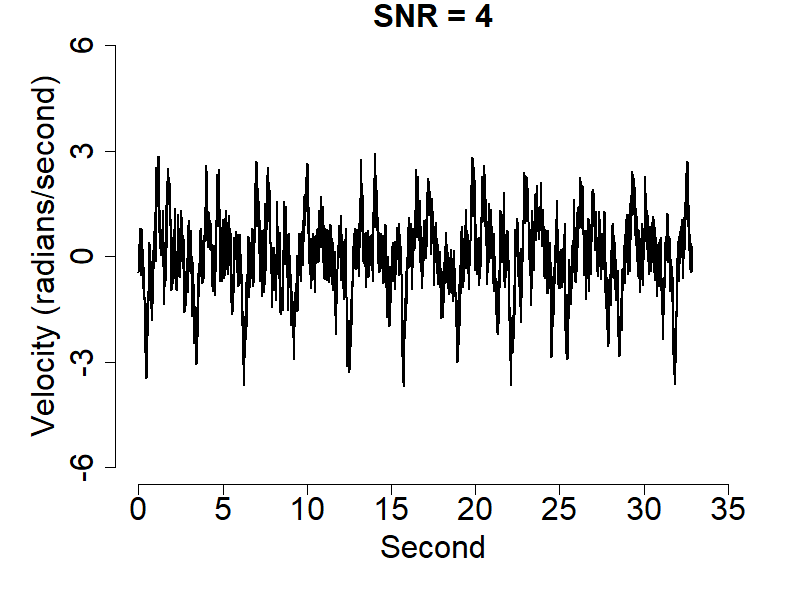}\hfill
	\includegraphics[width=0.33\textwidth]{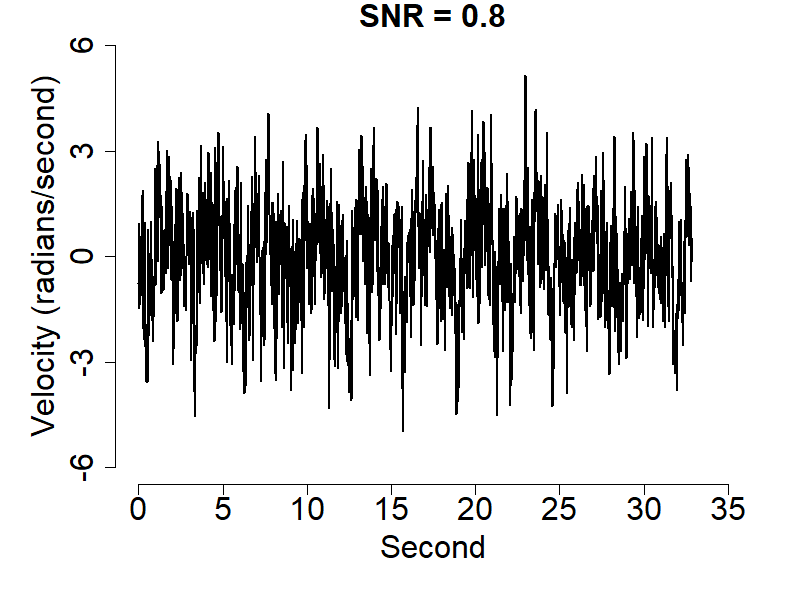}\hfill
	\hspace*{\fill}
\end{figure}

Figure \ref{EMG 7&12} depicts the simulated velocity for finger movements at different scenario. As we increase the level of noise, we depart more from the truth; compare the top-left and bottom-right panels.

\begin{figure}[H]
	\caption{Comparison of MSEs based on competing approaches for high (left) and low (right) SNRs with high  correlation coefficient (\textit{B2}). Simulated data corresponds to the non-varying finger movements. Reported are the results based on 100 simulations.} 
	\includegraphics[width=0.49\textwidth]{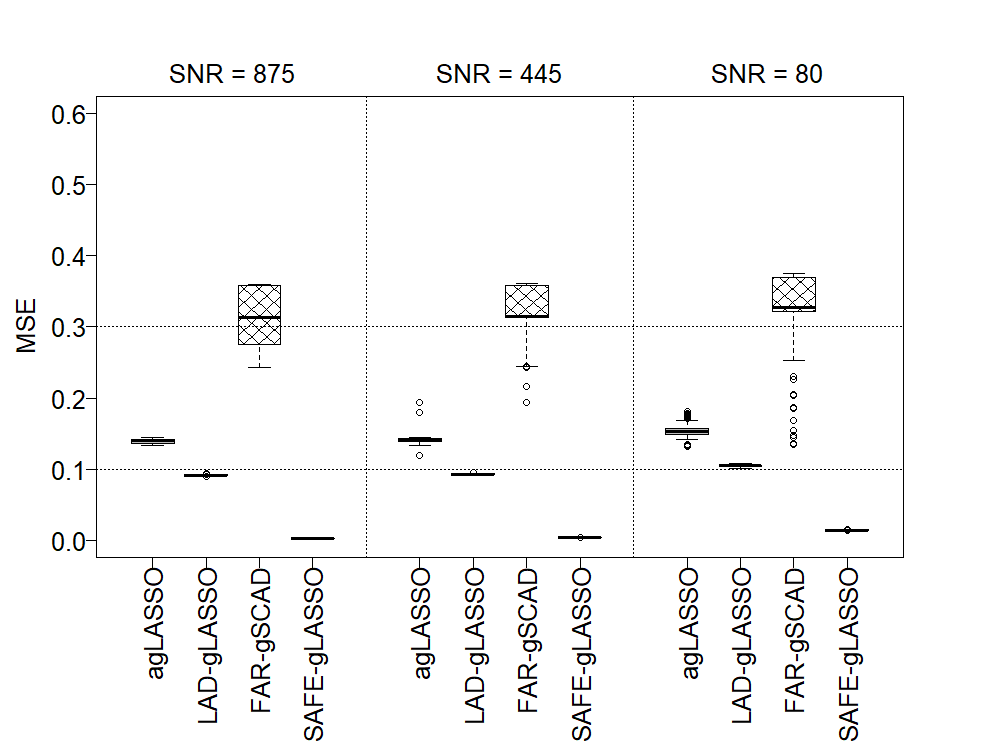}
	\includegraphics[width=0.49\textwidth]{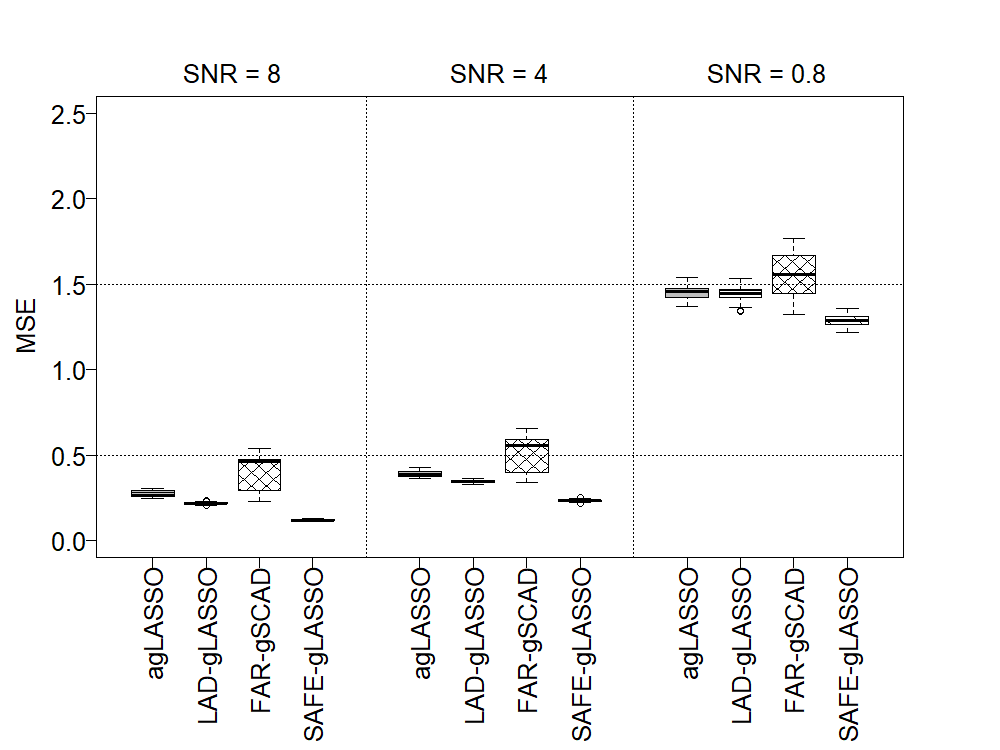}
	\label{low_CORR_MSE}
\end{figure}

Table \ref{low_CORR} illustrates the numerical performance corresponding to the simulated kinematic data for different SNRs at low correlation coefficient. The results are consistent to the  findings of finger movements in Section 5 and Section 6. 
The proposed approach has lower or equal FPRs than that of the competitors across different SNRs. However, at SNR = 0.8, the numerical performance of the competitors deteriorates. As before, the model size for the second stage of SAFE-gLASSO is smaller than that of the first stage; see the column ``SP" for the proposed at SNR = 0.8. In addition, the proposed method outperforms the competitors in terms of prediction accuracy at all SNRs; see Figure \ref{low_CORR_MSE}.  

\begin{table}[H]
	\tiny
	\centering
	\caption{Analysis of finger movements with fixed motion. Data is generated assuming noise variance \textit{C1} and \textit{C2} with different dominant processes (\textit{A1}-\textit{A3}) for low correlation coefficient (\textit{B1}). Reported are the SPs (\%), model size (in square brackets), TPRs (\%), and FPRs (\%) averaged over 100 simulations. Results with superscript $\dagger$ correspond to the first stage of SAFE-gLASSO.}
	\label{low_CORR}
	\noindent\makebox[\textwidth]{
		\begin{tabular}{lccccccccccccc}
			\multicolumn{1}{c}{} &  & \multicolumn{3}{c}{agLASSO} & \multicolumn{3}{c}{LAD-gLASSO} & \multicolumn{3}{c}{FAR-gSCAD} & \multicolumn{3}{c}{SAFE-gLASSO} \\ \cline{3-14} 
			\multicolumn{1}{c}{Setting} & SNR & SP & TPR & FPR & SP & TPR & FPR & SP & TPR & FPR & SP & TPR & FPR \\ \cline{3-14} 
			$\textit{A1} + \textit{C1}$ & 875 & \begin{tabular}[c]{@{}c@{}}87\\ {[}2.13{]}\end{tabular} & 100 & 1 & \begin{tabular}[c]{@{}c@{}}67\\ {[}5.32{]}\end{tabular} & 100 & 24 & \begin{tabular}[c]{@{}c@{}}81\\ {[}3.00{]}\end{tabular} & 100 & 7 & \begin{tabular}[c]{@{}c@{}}88\\ {[}2.00{]}{[}2.00{]}$^{\dagger}$\end{tabular} & \begin{tabular}[c]{@{}c@{}}100\\ {[}100{]}$^{\dagger}$\end{tabular} & \begin{tabular}[c]{@{}c@{}}0\\ $[0]^{\dagger}$\end{tabular} \\ \cline{3-14} 
			$\textit{A2} + \textit{C1}$ & 445 & \begin{tabular}[c]{@{}c@{}}86\\ {[}2.13{]}\end{tabular} & 100 & 1 & \begin{tabular}[c]{@{}c@{}}67\\ {[}5.35{]}\end{tabular} & 100 & 24 & \begin{tabular}[c]{@{}c@{}}81\\ {[}3.00{]}\end{tabular} & 100 & 7 & \begin{tabular}[c]{@{}c@{}}88\\ {[}2.00{]}{[}2.00{]}$^{\dagger}$\end{tabular} & \begin{tabular}[c]{@{}c@{}}100\\ {[}100{]}$^{\dagger}$\end{tabular} & \begin{tabular}[c]{@{}c@{}}0\\ {[}0{]}$^{\dagger}$\end{tabular} \\ \cline{3-14} 
			$\textit{A3} + \textit{C1}$ & 80 & \begin{tabular}[c]{@{}c@{}}87\\ {[}2.05{]}\end{tabular} & 100 & 0 & \begin{tabular}[c]{@{}c@{}}68\\ {[}5.01{]}\end{tabular} & 100 & 22 & \begin{tabular}[c]{@{}c@{}}81\\ {[}3.00{]}\end{tabular} & 100 & 7 & \begin{tabular}[c]{@{}c@{}}88\\ {[}2.00{]}{[}2.00{]}$^{\dagger}$\end{tabular} & \begin{tabular}[c]{@{}c@{}}100\\ {[}100{]}$^{\dagger}$\end{tabular} & \begin{tabular}[c]{@{}c@{}}0\\ {[}0{]}$^{\dagger}$\end{tabular} \\ \cline{3-14} 
			$\textit{A1} + \textit{C2}$ & 8 & \begin{tabular}[c]{@{}c@{}}88\\ {[}2.00{]}\end{tabular} & 100 & 0 & \begin{tabular}[c]{@{}c@{}}74\\ {[}4.15{]}\end{tabular} & 100 & 15 & \begin{tabular}[c]{@{}c@{}}81\\ {[}2.99{]}\end{tabular} & 100 & 7 & \begin{tabular}[c]{@{}c@{}}88\\ {[}2.00{]}{[}2.00{]}$^{\dagger}$\end{tabular} & \begin{tabular}[c]{@{}c@{}}100\\ {[}100{]}$^{\dagger}$\end{tabular} & \begin{tabular}[c]{@{}c@{}}0\\ {[}0{]}$^{\dagger}$\end{tabular} \\ \cline{3-14} 
			$\textit{A2} + \textit{C2}$ & 4 & \begin{tabular}[c]{@{}c@{}}87\\ {[}2.00{]}\end{tabular} & 100 & 0 & \begin{tabular}[c]{@{}c@{}}72\\ {[}4.56{]}\end{tabular} & 100 & 18 & \begin{tabular}[c]{@{}c@{}}81\\ {[}2.90{]}\end{tabular} & 100 & 6 & \begin{tabular}[c]{@{}c@{}}88\\ {[}2.00{]}{[}2.00{]}$^{\dagger}$\end{tabular} & \begin{tabular}[c]{@{}c@{}}100\\ {[}100{]}$^{\dagger}$\end{tabular} & \begin{tabular}[c]{@{}c@{}}0\\ {[}0{]}$^{\dagger}$\end{tabular} \\ \cline{3-14} 
			$\textit{A3} + \textit{C2}$ & 0.8 & \begin{tabular}[c]{@{}c@{}}87\\ {[}2.01{]}\end{tabular} & 100 & 2 & \begin{tabular}[c]{@{}c@{}}69\\ {[}4.91{]}\end{tabular} & 100 & 21 & \begin{tabular}[c]{@{}c@{}}82\\ {[}2.86{]}\end{tabular} & 100 & 6 & \begin{tabular}[c]{@{}c@{}}88\\ {[}2.00{]}{[}2.03{]}$^{\dagger}$\end{tabular} & \begin{tabular}[c]{@{}c@{}}100\\ {[}100{]}$^{\dagger}$\end{tabular} & \begin{tabular}[c]{@{}c@{}}0\\ {[}0{]}$^{\dagger}$\end{tabular} \\ \cline{3-14} 
		\end{tabular}
	}
\end{table}

\section{Numerical experiment}\label{toy}

\begin{figure}[h]
	\centering
	\caption{Functional measurements corresponding to the first two predictors; three curves are chosen randomly and displayed as black solid lines.}   
	\includegraphics[width=0.45\textwidth]{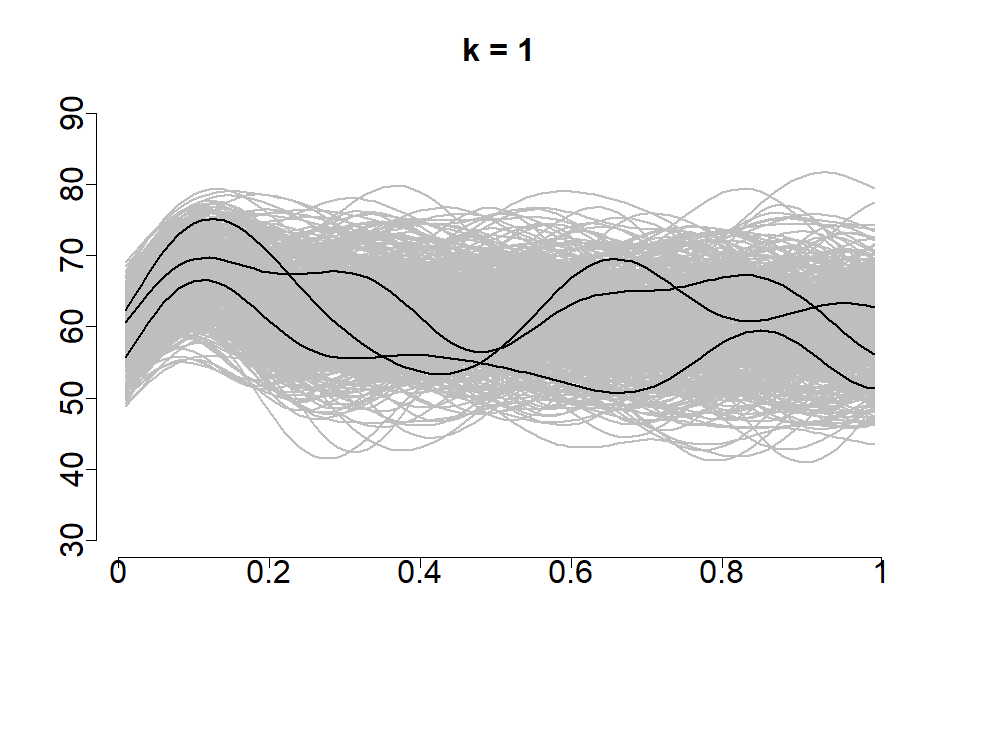} 
	\includegraphics[width=0.45\textwidth]{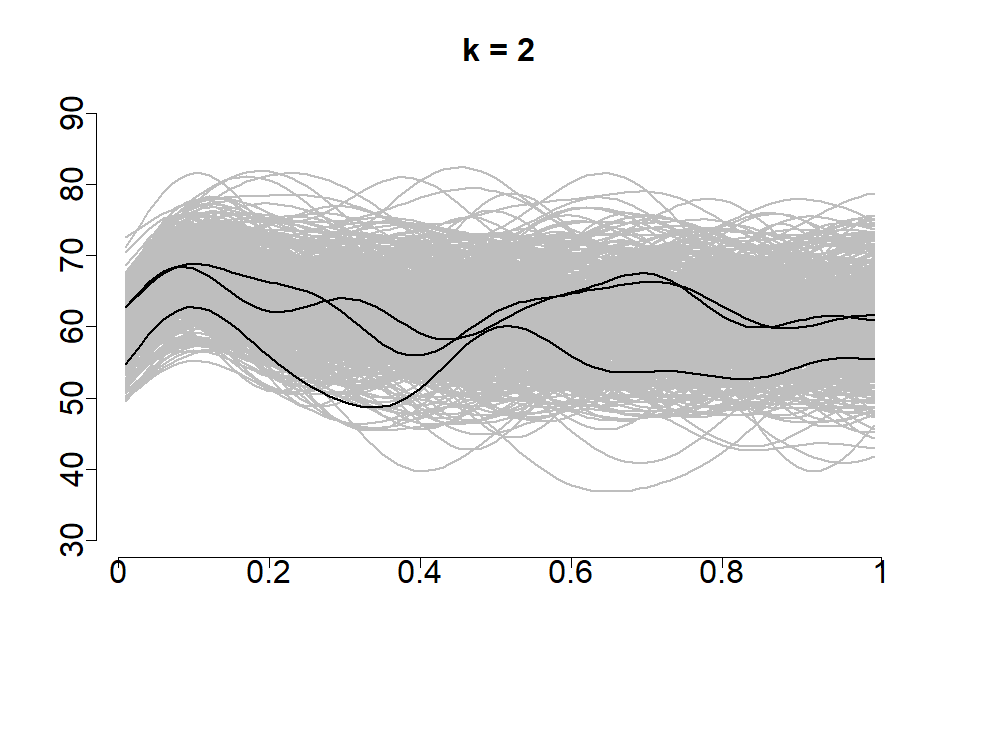} 

	\label{fun_pred} 
\end{figure}

\begin{figure}[H]
	\centering
	\caption{$\beta_{1}$ (top) and $\beta_{2}$ (bottom) for $C = \{0, 2, 5, 10\}$ from left to right respectively.}   
	\includegraphics[width=0.243\textwidth]{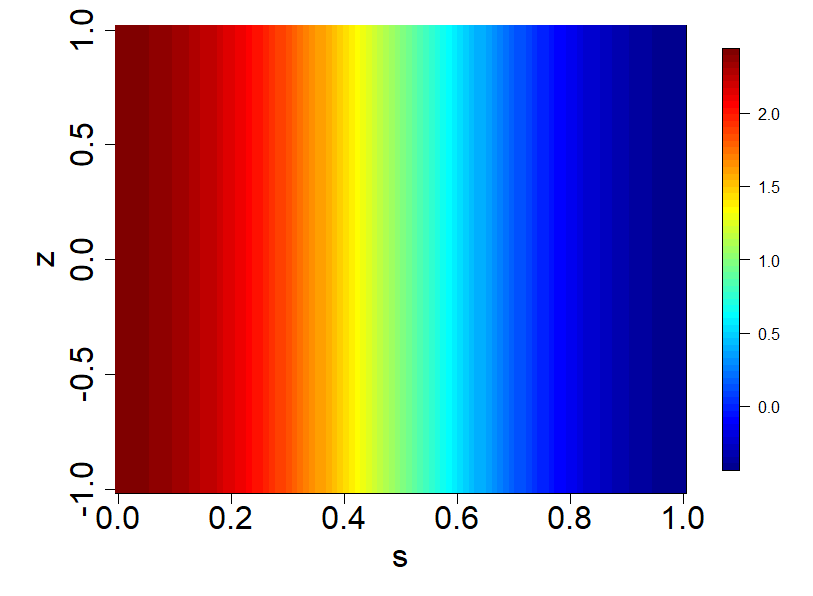} 
	\includegraphics[width=0.243\textwidth]{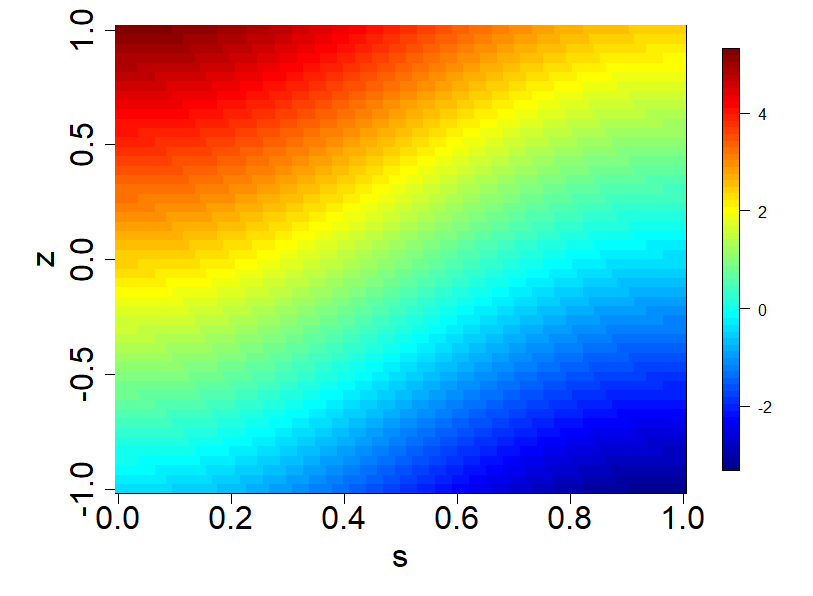} 
	\includegraphics[width=0.243\textwidth]{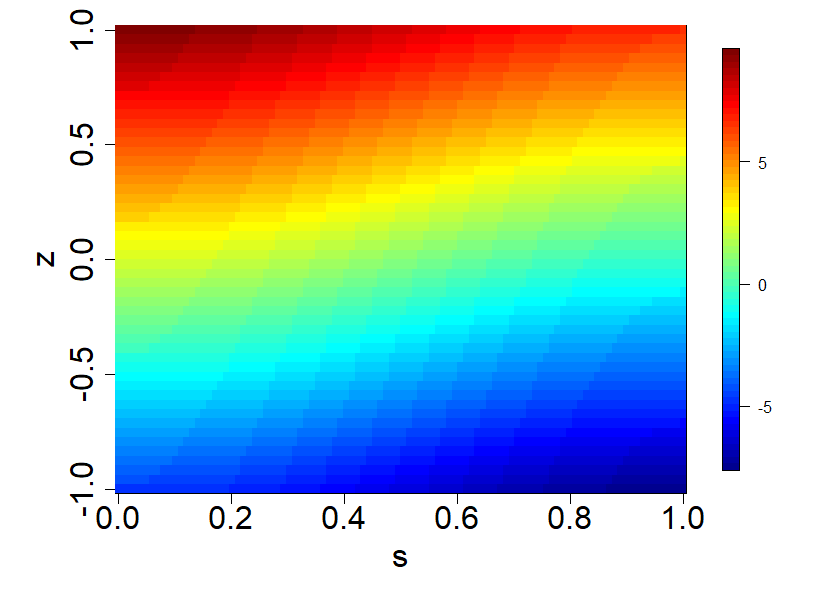}
	\includegraphics[width=0.243\textwidth]{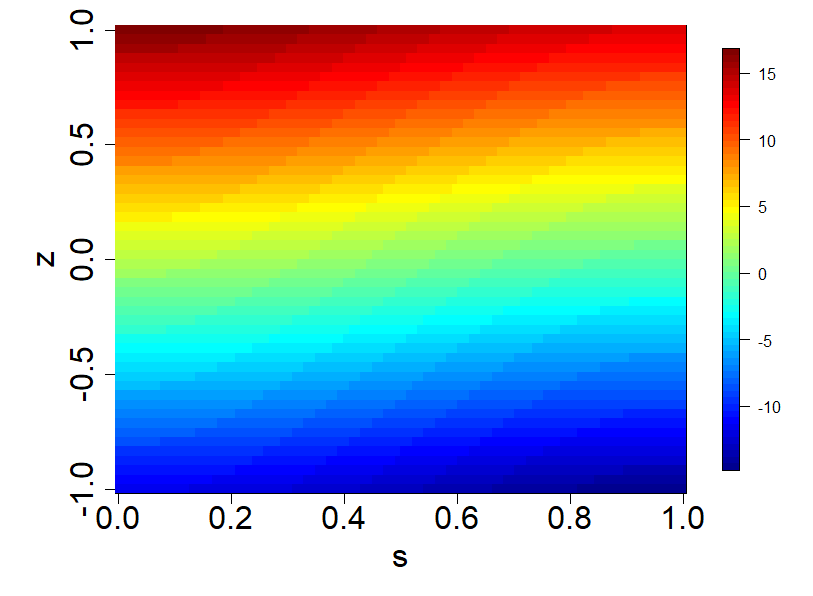}\\ 
	\includegraphics[width=0.243\textwidth]{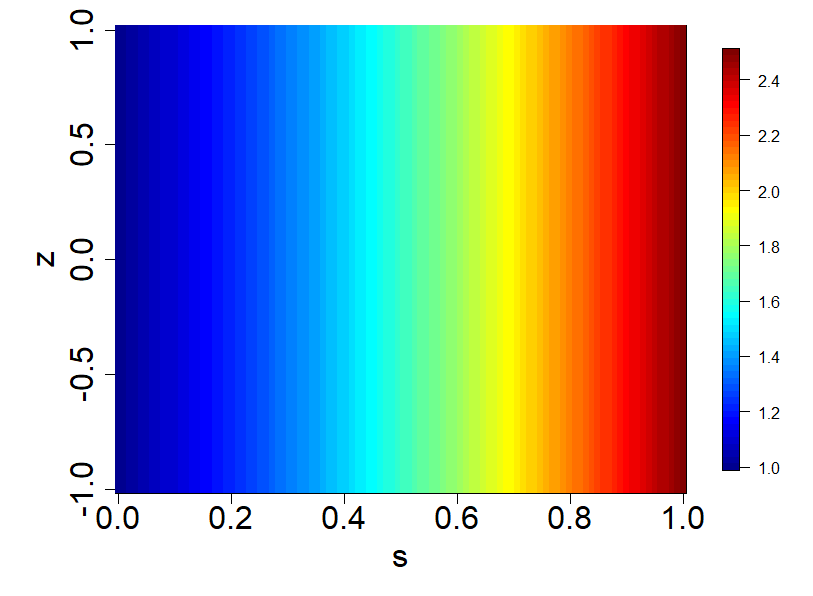} 
	\includegraphics[width=0.243\textwidth]{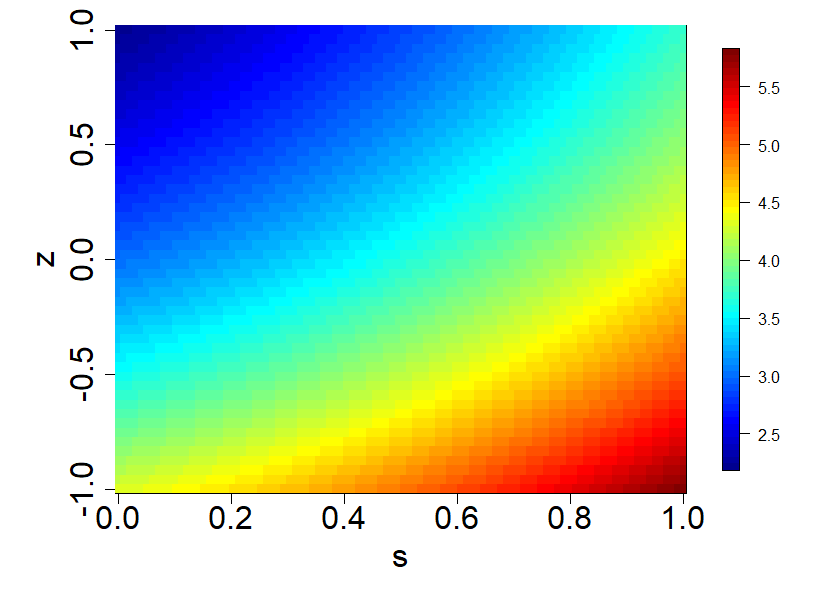} 
	\includegraphics[width=0.243\textwidth]{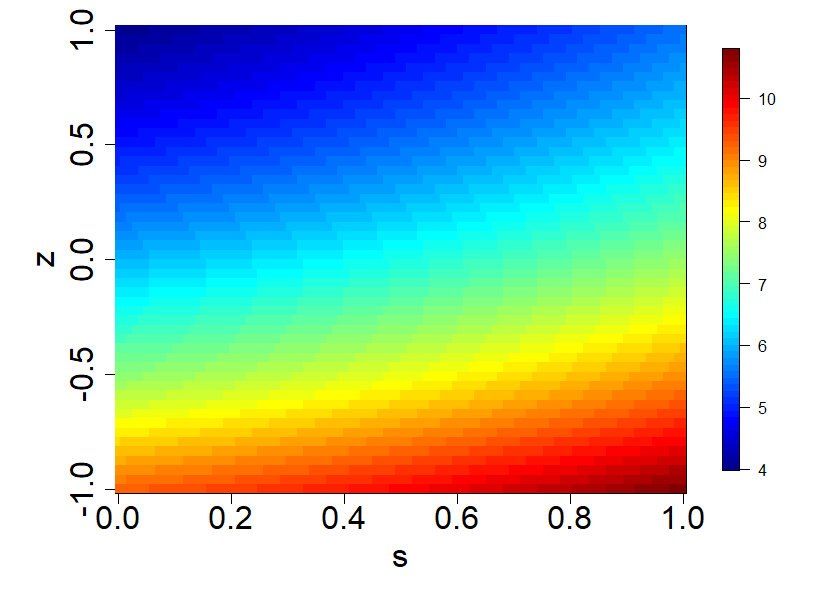} 
	\includegraphics[width=0.243\textwidth]{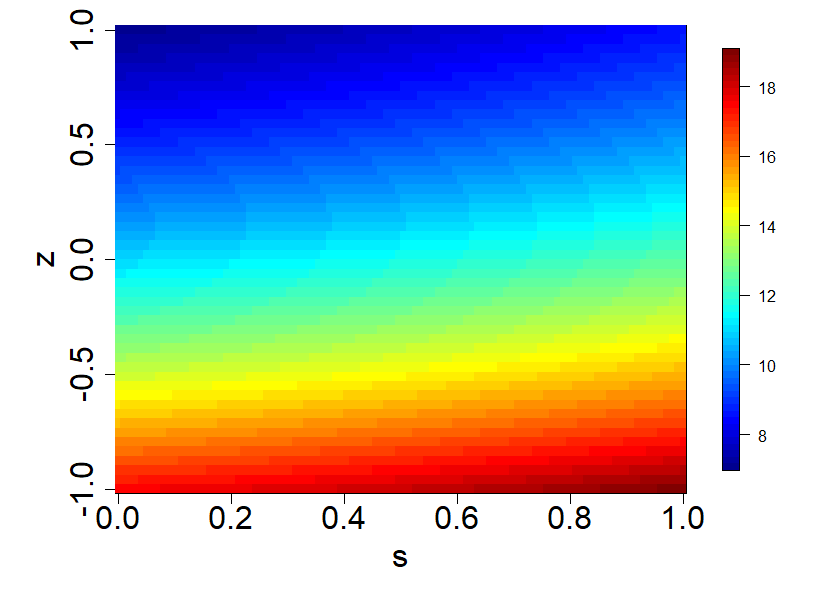}
	\label{BETA} 
\end{figure}

As described in \cite{gertheiss2013variable}, the 10 functional predictors are generated by
%
$$
\label{tuts_varying}
X_{k,i}(s) = \{\sigma(s)\}^{-1}  \sum_{r=1}^{5} [ a_{ikr} \text{sin}\{ 2 \pi s (5 - a_{ikr})\} - m_{ikr} ]. 
$$
%
Figure \ref{fun_pred} illustrates the first two functional predictors. Similar to \cite{gertheiss2013variable} and \cite{pannu2017robust}, we assume $a_{ikr} \sim U(0, 5),$ $m_{ikr} \sim U(0, 2\pi)$ and $\sigma(s)$ is defined such  that $\text{var}\{X_{k,i}(s)\} = 0.01,$ for all $s.$ 
The response is generated from
$
\label{toy_varying}
y_{i} = \alpha + \sum_{k=1}^{2} \int_{\mathcal{S}} X_{k,i}(s)  \gamma_{k}(s, z_i) ds + \epsilon_{i}, 
$
where $\epsilon_{i}$'s are assumed to be independent and identically distributed (IID) as $\epsilon_{i} \sim \pN(0, \sigma^{2}).$ The nonzero functional coefficients $\{\gamma_{k}(\cdot, \cdot); k = 1, 2\}$'s are varying over $z_i \in [-1, 1]$ and defined as $\gamma_{1}(s, z) = 1 + \sqrt{2} C z k  +  \sqrt{2} k \text{cos} (\pi s)$ and $\gamma_{2}(s, z) = 1 + C \text{exp} (-0.5 z)  +  s + 0.5 s^2$.  For example, when $C = 0,$ we have that $\gamma_{1}(s, z) = 1 + \sqrt{2} \ k  \text{cos} (\pi s)$ and $\gamma_{2}(s, z) = 1 + s + 0.5 s^2$ which are non-varying functional coefficients; for $C \neq 0,$ $\gamma_{k}(\cdot, \cdot)$'s vary over $z.$ We investigate the cases for $C = \{0, 2, 5, 10\}$. Figure \ref{BETA} depicts the effect of functional predictors at different values of $C.$ 

The steps for fitting the procedure is similar to the one described in section~5 of the main document. One difference is that we use 15 basis functions in the $s$ direction and 7 basis functions in the $z$ direction in modeling $\gamma_{k}(\cdot, \cdot)$'s. The invariant functional coefficients $\gamma_{k}(\cdot)$'s are modeled using 15 basis functions for agLASSO, LAD-gLASSO, and FAR-gSCAD approach. Additionally, the tuning parameters are selected using random CV for all competing methods.

Table \ref{math} presents the results using the performance metrics as described in section~5 of the main document, but focus only on recovery of $\mathcal{K}=\{1,2\}$ (there are no $\mathcal{K}_F$ and $\mathcal{K}_E$ here). As expected when $C = 0,$ the numerical performance of all the methods is competitive. However, as $C$ departs from 0, say at $C = 5,$ the variable selection and prediction accuracy of the competitors deteriorates.

\begin{table}[H]
	\tiny
	\centering
	\caption{Variable selection for the toy example. Reported are the SPs (\%), model size (in square brackets), TPRs (\%), and FPRs (\%) averaged across 100 simulations. Results with superscript $\dagger$ correspond to the first stage of SAFE-gLASSO.}
	\label{math}
	\noindent\makebox[\textwidth]{
		\begin{tabular}{cccccccccccccc}
			&  & \multicolumn{3}{c}{agLASSO} & \multicolumn{3}{c}{LAD-gLASSO} & \multicolumn{3}{c}{FAR-gSCAD} & \multicolumn{3}{c}{SAFE-gLASSO} \\ \cline{3-14} 
			\multicolumn{2}{c}{Combination} & SP & TPR & FPR & SP & TPR & FPR & SP & TPR & FPR & SP & TPR & FPR \\ \cline{3-14} 
			{SNR = 0.5} & C = 0 & \begin{tabular}[c]{@{}c@{}}77\\ {[}2.33{]}\end{tabular} & 100 & 4 & \begin{tabular}[c]{@{}c@{}}41\\ {[}5.93{]}\end{tabular} & 100 & 49 & \begin{tabular}[c]{@{}c@{}}80\\ {[}2.01{]}\end{tabular} & 100 & 1 & \begin{tabular}[c]{@{}c@{}}80\\ {[}2.03{]}{[}2.21{]}$^{\dagger}$\end{tabular} & \begin{tabular}[c]{@{}c@{}}100\\ {[}100{]}$\dagger$\end{tabular} & \begin{tabular}[c]{@{}c@{}}0\\ $[3]^{\dagger}$\end{tabular} \\ \cline{3-14} 
			& C = 2 & \begin{tabular}[c]{@{}c@{}}87\\ {[}2.26{]}\end{tabular} & 94 & 5 & \begin{tabular}[c]{@{}c@{}}51\\ {[}4.94{]}\end{tabular} & 96 & 38 & \begin{tabular}[c]{@{}c@{}}89\\ {[}1.09{]}\end{tabular} & 54 & 0 & \begin{tabular}[c]{@{}c@{}}80\\ {[}2.00{]}{[}2.09{]}$^{\dagger}$\end{tabular} & \begin{tabular}[c]{@{}c@{}}100\\ {[}100{]}$\dagger$\end{tabular} & \begin{tabular}[c]{@{}c@{}}0\\ $[1]^{\dagger}$\end{tabular} \\ \cline{3-14} 
			& C = 5 & \begin{tabular}[c]{@{}c@{}}81\\ {[}1.90{]}\end{tabular} & 69 & 7 & \begin{tabular}[c]{@{}c@{}}56\\ {[}4.37{]}\end{tabular} & 82 & 34 & \begin{tabular}[c]{@{}c@{}}90\\ {[}1.03{]}\end{tabular} & 50 & 0 & \begin{tabular}[c]{@{}c@{}}80\\ {[}2.03{]}{[}2.22{]}$^{\dagger}$\end{tabular} & \begin{tabular}[c]{@{}c@{}}100\\ {[}100{]}$\dagger$\end{tabular} & \begin{tabular}[c]{@{}c@{}}0\\ $[3]^{\dagger}$\end{tabular} \\ \cline{3-14} 
			{SNR = 1} & C = 0 & \begin{tabular}[c]{@{}c@{}}79\\ {[}2.11{]}\end{tabular} & 100 & 2 & \begin{tabular}[c]{@{}c@{}}35\\ {[}6.53{]}\end{tabular} & 100 & 57 & \begin{tabular}[c]{@{}c@{}}80\\ {[}2.02{]}\end{tabular} & 100 & 0 & \begin{tabular}[c]{@{}c@{}}80\\ {[}2.00{]}{[}2.02{]}$^{\dagger}$\end{tabular} & \begin{tabular}[c]{@{}c@{}}100\\ {[}100{]}$\dagger$\end{tabular} & \begin{tabular}[c]{@{}c@{}}0\\ $[0]^{\dagger}$\end{tabular} \\ \cline{3-14} 
			& C = 2 & \begin{tabular}[c]{@{}c@{}}79\\ {[}2.09{]}\end{tabular} & 100 & 2 & \begin{tabular}[c]{@{}c@{}}44\\ {[}5.63{]}\end{tabular} & 100 & 45 & \begin{tabular}[c]{@{}c@{}}86\\ {[}1.44{]}\end{tabular} & 69 & 1 & \begin{tabular}[c]{@{}c@{}}80\\ {[}2.00{]}{[}2.01{]}$^{\dagger}$\end{tabular} & \begin{tabular}[c]{@{}c@{}}100\\ {[}100{]}$\dagger$\end{tabular} & \begin{tabular}[c]{@{}c@{}}0\\ $[0]^{\dagger}$\end{tabular} \\ \cline{3-14} 
			& C = 5 & \begin{tabular}[c]{@{}c@{}}84\\ {[}1.64{]}\end{tabular} & 74 & 2 & \begin{tabular}[c]{@{}c@{}}57\\ {[}4.31{]}\end{tabular} & 89 & 32 & \begin{tabular}[c]{@{}c@{}}90\\ {[}1.03{]}\end{tabular} & 52 & 0 & \begin{tabular}[c]{@{}c@{}}80\\ {[}2.00{]}{[}2.01{]}$^{\dagger}$\end{tabular} & \begin{tabular}[c]{@{}c@{}}100\\ {[}100{]}$\dagger$\end{tabular} & \begin{tabular}[c]{@{}c@{}}0\\ $[0]^{\dagger}$\end{tabular} \\ \cline{3-14}
			{SNR = 5} & C = 0 & \begin{tabular}[c]{@{}c@{}}80\\ {[}2.00{]}\end{tabular} & 100 & 0 & \begin{tabular}[c]{@{}c@{}}44\\ {[}5.60{]}\end{tabular} & 100 & 45 & \begin{tabular}[c]{@{}c@{}}80\\ {[}2.00{]}\end{tabular} & 100 & 0 & \begin{tabular}[c]{@{}c@{}}80\\ {[}2.00{]}{[}2.00{]}$^{\dagger}$\end{tabular} & \begin{tabular}[c]{@{}c@{}}100\\ {[}100{]}$\dagger$\end{tabular} & \begin{tabular}[c]{@{}c@{}}0\\ $[0]^{\dagger}$\end{tabular} \\ \cline{3-14} 
			& C = 2 & \begin{tabular}[c]{@{}c@{}}80\\ {[}2.00{]}\end{tabular} & 100 & 0 & \begin{tabular}[c]{@{}c@{}}45\\ {[}5.50{]}\end{tabular} & 100 & 44 & \begin{tabular}[c]{@{}c@{}}80\\ {[}2.00{]}\end{tabular} & 100 & 0 & \begin{tabular}[c]{@{}c@{}}80\\ {[}2.00{]}{[}2.00{]}$^{\dagger}$\end{tabular} & \begin{tabular}[c]{@{}c@{}}100\\ {[}100{]}$\dagger$\end{tabular} & \begin{tabular}[c]{@{}c@{}}0\\ $[0]^{\dagger}$\end{tabular} \\ \cline{3-14} 
			& C = 5 & \begin{tabular}[c]{@{}c@{}}84\\ {[}1.62{]}\end{tabular} & 81 & 0 & \begin{tabular}[c]{@{}c@{}}55\\ {[}4.54{]}\end{tabular} & 97 & 33 & \begin{tabular}[c]{@{}c@{}}88\\ {[}1.22{]}\end{tabular} & 61 & 0 & \begin{tabular}[c]{@{}c@{}}80\\ {[}2.00{]}{[}2.00{]}$^{\dagger}$\end{tabular} & \begin{tabular}[c]{@{}c@{}}100\\ {[}100{]}$\dagger$\end{tabular} & \begin{tabular}[c]{@{}c@{}}0\\ $[0]^{\dagger}$\end{tabular} \\ \cline{3-14} 
		\end{tabular}
	}
\end{table}

\begin{figure}[H]
	\centering
	\caption{Reported is MSEs based on competing approaches for different SNRs at $C = 0$ (left), $C = 2$ (middle), and $C = 5$ (right). Results are based on 100 simulations. Reference lines are drawn at MSE = 0.04 (left), MSE = 0.20 (middle), and MSE = 0.75 (right) for convenience.} 
	\noindent\makebox[\textwidth]{	
		\includegraphics[width=1.1\textwidth]{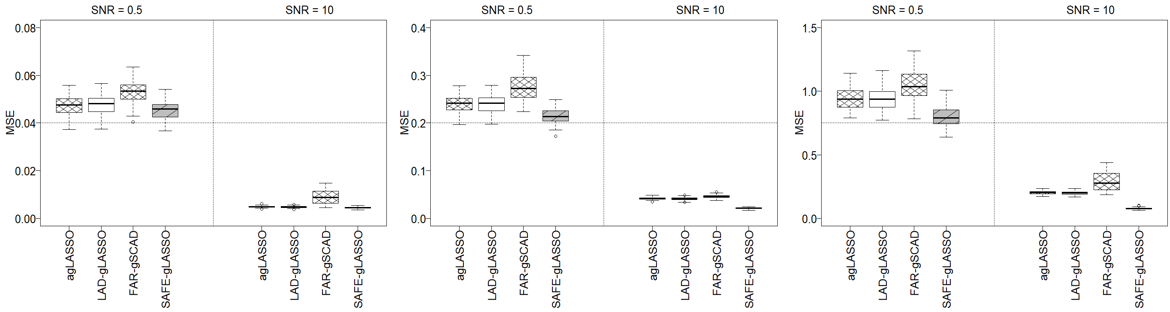}
	}
	\label{MSE_boxplot}
\end{figure}

\section{Algorithm of split conformal prediction band}\label{algorithm}

For the simplicity of exposition, denote by $\mathcal{A}$ the proposed algorithm which includes the steps of (i) variable selection and (ii) predictive modeling. In particular, we construct the split conformal prediction bands \citep{Lei2017,lei2015conformal}  for in-sample and new observations. We use the subscript \textit{in} and \textit{new} to refer to the in-sample and new data, respectively. Indeed the conformal inference offers distribution-free predictive inference in high-dimensional regression; while the method preserves the consistency properties of the estimators, it also provides the valid prediction coverage \citep{Lei2017}. Let $y_{i}$ be the scalar response, $N$ be the total number of observations, and $X_{in, i}$ be the corresponding covariates information. We adopt split conformal inference to construct prediction bands for future observations that are indexed by $i^{*}$ and use \textit{rank-one-out} (ROO) split conformal inference for in-sample observations which are indexed by $i$. For completeness, we present both algorithms as below. 
\begin{table}[H]
	\centering
	\label{algorithm5}
	\noindent\makebox[\textwidth]{ 
		\begin{tabular}{ll}
			\hline
			\multicolumn{2}{l}{$\textbf{Algorithm}$ Split Conformal Prediction} \\ \hline
			\textbf{Input:} & Data $(y_{i},  X_{in, i}), i = 1, \ldots, N;$ where $X_{in, i} = [z_i, \{X_{k,i}(s_{r}); r = 1, \cdots, R\};k = 1, \ldots, K].$\\
			&Let miscoverage level be $\alpha \in (0, 1)$ and denote the proposed SAFE-gLASSO \\ & algorithm by $\mathcal{A}$ \\
			\textbf{Output:} & Prediction intervals at future observations ($i^{*}$) with covariates $X_{new, i^{*}}; i^{*} = N+1, N+2, \cdots$\\
			& Randomly split $\{1, \cdots, N\}$ into two equal-sized subsets $\mathcal{I}_{1},$ $\mathcal{I}_{2}$ \\
			& $\widehat \mu = \mathcal{A} \big( \{ ( y_{i}, X_{in, i}) : i \in \mathcal{I}_1 \} \big)$ \\
			& $R_{i} = \big| y_{i}  - \widehat \mu(X_{in, i}) \big|$ where $i \in \mathcal{I}_2;$  \\
			& $d$ = the $p$th smallest value in $\{R_{i} : i \in \mathcal{I}_{2}\},$ where $p = \ceil{(N/2 + 1)(1 - \alpha)}$ \\
			& Return $C_{split}(X_{new, i^{*}}) = \left[ \widehat \mu(X_{new, i^{*}}) - d, \widehat \mu(X_{new, i^{*}}) + d \right],$ for all $i^{*}$ \\ \hline
		\end{tabular}
	}
\end{table}

\begin{table}[H]
	\centering
	\label{my-label}
	\noindent\makebox[\textwidth]{	
		\begin{tabular}{ll}
			\hline
			\multicolumn{2}{l}{$\textbf{Algorithm}$ Rank-One-Out (ROO) Split Conformal Prediction} \\ \hline
			\textbf{Input:} & Data $(y_{i},  X_{i}), i = 1, \ldots, N;$  where $X_{i} = [z_i, \{X_{k,i}(s_{r}); r = 1, \cdots, R\}; k = 1, \ldots, K].$\\ & Let miscoverage level be $\alpha \in (0, 1)$ and denote the proposed SAFE-gLASSO \\ 
			& algorithm by $\mathcal{A}$ \\
			\textbf{Output:} & In-sample prediction intervals at $X_{i}$ \\
			& Randomly split $\{1, \cdots, N\}$ into two equal-sized subsets $\mathcal{I}_{n},$ $n = 1, 2$ \\
			& \textbf{for} $n \in \{1, 2\}$ \textbf{do} \\
			& \quad\quad$\widehat \mu_n = \mathcal{A} \big( \{ ( y_i,  X_{i}) : i \in \mathcal{I}_n\} \big)$ \\
			& \quad\quad\textbf{for} $i \notin \mathcal{I}_{n}$ \textbf{do} \\
			& \quad\quad\quad $R_{i} = \big| y_{i}  - \widehat \mu_{n}(X_{i}) \big|$ \\
			& \quad\quad\textbf{end for} \\
			& \quad\quad\textbf{for} $i \notin \mathcal{I}_{n}$ \textbf{do} \\
			& \quad\quad\quad$d_i$ = the $m$th smallest value in $\{R_{j} : j \notin \mathcal{I}_{n}, j \neq i\},$ where $m = \ceil{N/2(1 - \alpha)}$ \\
			& \quad\quad\quad$C_{roo}(X_{i}) = [\widehat \mu_n(X_{i}) - d_{i}, \widehat \mu_n(X_{i}) + d_{i}],$ \\
			& \quad\quad\textbf{end for} \\
			& \quad\textbf{end for} \\
			& Return intervals $C_{roo}(X_{i}); i  = 1, \ldots, N$ \\ \hline
		\end{tabular}
	}
\end{table}

\end{document}